\documentclass{aa}  
\usepackage{graphicx}
\usepackage{txfonts}
\usepackage{color}
\usepackage{changes}

\usepackage{natbib}
\bibpunct{(}{)}{;}{a}{}{,}

\newcommand{\ergcm}[1]{$\times 10^{#1}$ erg cm$^{-2}$ s$^{-1}$}

\newcommand{\uergcm}{erg cm$^{-2}$ s$^{-1}$}
\newcommand{\ergs}[1]{$\times 10^{#1}$ erg s$^{-1}$}
\newcommand{\oergs}[1]{$10^{#1}$ erg s$^{-1}$}
\newcommand{\uergs}{erg s$^{-1}$}
\newcommand{\hcm}[1]{$\times 10^{#1}$ cm$^{-2}$}
\newcommand{\ohcm}[1]{$10^{#1}$ cm$^{-2}$}
\newcommand{\expo}[1]{$\times 10^{#1}$}
\newcommand{\oexpo}[1]{$10^{#1}$}

\newcommand{\nh}{N$_{\rm H}$\xspace}

\newcommand{\cts}{cts s$^{-1}$\xspace}

\newcommand{\Halpha}{H${\alpha}$\xspace}
\newcommand{\ltsima}{$\buildrel < \over \sim$}
\newcommand{\lsim}{\lower.5ex\hbox{\ltsima}}
\newcommand{\gtsima}{$\buildrel > \over \sim$}
\newcommand{\gsim}{\lower.5ex\hbox{\gtsima}}

\newcommand{\rahour}{\hbox{\ensuremath{^{\rm h}}}}
\newcommand{\ramin}{\hbox{\ensuremath{^{\rm m}}}}

\newcommand{\xspec}{\texttt{XSPEC}\xspace}
\newcommand{\eSASS}{\texttt{eSASS}\xspace}

\newcommand{\xmm}{{\it XMM-Newton}\xspace}

\newcommand{\suzaku}{{\it Suzaku}\xspace}
\newcommand{\rosat}{{\it ROSAT}\xspace}
\newcommand{\cxo}{{\it Chandra}\xspace}

\newcommand{\srg}{{\it SRG}\xspace}
\newcommand{\ero}{\mbox{eROSITA}\xspace}
\newcommand{\srget}{\mbox{SRGEt\,J052829.5$-$690345}\xspace}
\newcommand{\sxp}[1]{SXP\,{#1}\xspace}
\newcommand{\hscat}[1]{[HS16]\,{#1}\xspace}
\newcommand{\snre}{\mbox{1E\,0102.2$-$7219}\xspace}

\newcommand{\red}[1]{#1}
\newcommand{\redii}[1]{#1}
\begin{document} 

\title{eROSITA calibration and performance verification phase: High-mass X-ray binaries in the Magellanic Clouds}

\author{F.\,Haberl\inst{1} \and
        C.\,Maitra\inst{1} \and
        S.\,Carpano\inst{1} \and
        X.\,Dai\inst{2}  \and
        V.\,Doroshenko\inst{2}  \and
        K.\,Dennerl\inst{1} \and
        M.\,J.\,Freyberg\inst{1} \and
        M.\,Sasaki\inst{3} \and
        A.\,Udalski\inst{4} \and
        K.\,A.\,Postnov\inst{5,6} \and
        N.\,I.\,Shakura\inst{5,6}
       } 

\titlerunning{HMXBs in the Magellanic Clouds}
\authorrunning{Haberl et al.}

\institute{Max-Planck-Institut f{\"u}r extraterrestrische Physik, Gie{\ss}enbachstra{\ss}e 1, 85748 Garching, Germany, \email{fwh@mpe.mpg.de}
\and
Institut f{\"u}r Astronomie und Astrophysik, Sand 1, 72076 T{\"u}bingen, Germany
\and
Remeis Observatory and ECAP, Universit{\"a}t Erlangen-N{\"u}rnberg, Sternwartstra{\ss}e 7, 96049 Bamberg, Germany
\and
Astronomical Observatory, University of Warsaw, Al. Ujazdowskie 4, 00-478, 
Warszawa, Poland
\and
Moscow State University, Sternberg Astronomical Institute, 119234, Moscow, Russia
\and
Kazan Federal University, \red{Department of Astronomy and Space Geodesy}, 420008, Kazan, Russia
}

\date{Received 27 July 2021 / Accepted 25 August 2021}

\abstract
  % context heading (optional) leave it empty if necessary
   {During its performance verification phase, the soft X-ray instrument \ero aboard the Spektrum-Roentgen-Gamma (\srg) spacecraft observed large regions in the Magellanic Clouds, where almost 40 known high-mass X-ray binaries (HMXBs, including candidates) are located.}
   {We looked for new HMXBs in the \ero data, searched for  pulsations in HMXB candidates and investigated the long-term behaviour of the full sample using archival X-ray and optical data.}
   {For sources sufficiently bright, a detailed spectral and temporal analysis of their \ero data was performed. A source detection analysis of the \ero images in different energy bands provided count rates and upper limits for the remaining sources.}
   {We report the discovery of a new Be/X-ray binary in the Large Magellanic Cloud. The transient \srget was detected with a 0.2--8.0 keV luminosity of $\sim$\oergs{35} and exhibits a hard X-ray spectrum, typical for this class of HMXBs. The OGLE I-band light curve of the V$\sim$15.7\,mag counterpart shows large variations by up to 0.75\,mag, which occur quasi periodically with $\sim$511\,days. The \ero observations of the Small Magellanic Cloud covered 16 Be/X-ray binary pulsars, five of them were bright enough to accurately determine their current pulse period. The pulse periods for \sxp{726} and \sxp{1323} measured from \ero data are $\sim$800\,s and $\sim$1006\,s, respectively, far away from their discovery periods. Including archival \xmm observations we update the
   spin-period history of the two long-period pulsars which show nearly linear trends in their period evolution \red{since more than 15 years}. The corresponding average spin-down rate for \sxp{726} is 4.3\,s\,yr$^{-1}$ while \sxp{1323} shows spin-up with a rate of -23.2\,s\,yr$^{-1}$.
   We discuss the spin evolution of the two pulsars in the framework of quasi-spherical accretion.}
   {}

 \keywords{galaxies: individual: LMC, SMC --
          X-rays: binaries --
          stars: emission-line, Be -- 
          stars: neutron
         }

\maketitle   

%________________________________________________________________

\section{Introduction}
\label{sec:intro}

The Magellanic Clouds (MCs) are ideal laboratories to study X-ray source  populations in star-forming galaxies. 
They are relatively nearby with accurately known distances  \citep[50\,kpc for the Large Magellanic Cloud, LMC, and 62\,kpc for the Small Magellanic Cloud, SMC;][]{2019Natur.567..200P,2014ApJ...780...59G}, and X-ray observations are little affected by Galactic foreground absorption.
\red{However, it should be noted that in particular the SMC is significantly extended along the line of sight \citep{2021MNRAS.504.2983T,2009A&A...496..399S}.}
While the SMC was largely covered by a mosaic of pointed observations with \xmm \citep{2013A&A...558A...3S,2012A&A...545A.128H}, the large extent of the LMC on the sky allowed only to map part of it \citep[][]{2016A&A...585A.162M}.

The \xmm surveys of the MCs enlarged the number of known high-mass X-ray binaries, binary systems with a massive early-type star and a compact object, in most cases a neutron star. The majority of the HMXBs in the MCs forms the subgroup of Be/X-ray binaries  with a neutron star accreting matter from the circum-stellar disc of a Be star \citep[BeXRBs; see][for a review]{2011Ap&SS.332....1R}.
On the other hand, supergiant systems (SgXRBs), powered by accretion from the strong stellar wind, are less abundant. This is particularly true for the SMC where only one SgXRB (SMC\,X-1), but about 120 BeXRBs are known, with about half of them confirmed X-ray pulsars \citep{2016A&A...586A..81H}.
On the other hand, $\sim$60 HMXB candidates were identified in the LMC, with detected pulsations in about 20 of them. The fraction of SgXRBs is somewhat higher in the LMC as compared to the SMC \citep{2021A&A...647A...8M,2019MNRAS.490.5494M,2018MNRAS.475.3253V,2018MNRAS.475..220V} which is probably related to the different star formation history of the two Clouds \citep{2016MNRAS.459..528A,2010ApJ...716L.140A}.

\section{The \ero observations}
\label{sec:observations}

\ero, the soft X-ray instrument on the Spectrum-Roentgen-Gamma (\srg) mission \citep{2021A&A...647A...1P} 
was successfully launched from Baikonur on July 13, 2019. After the instruments were commissioned, first light observations, including an observation pointed at SN\,1987A in the LMC \citep{2021arXiv210614532M} %(Maitra et al. 2021) 
were performed, followed by a calibration and performance verification phase when pointed observations 
of selected targets were conducted. Henceforth, we call this phase
of pointed observations before the start of the all-sky survey on December 13, 2019, ``CalPV phase''.
A summary of the CalPV observations performed in the direction of LMC and SMC can be found in Table\,\ref{tabobs}. The current publication focusses on the analysis of these observations and reports first results obtained by \ero for the LMC and the SMC. 

%%%%%%%%%%%%%%%%
\begin{table*} 
\caption{Observation details} 
\begin{center}
\begin{tabular}{lccccl} 
\hline\hline\noalign{\smallskip}
Obsid\tablefootmark{a}   & TM\tablefootmark{b} & Time       & Exposure\tablefootmark{c} & Pointing  & Target  \\
        &                     & UT                          & (ks)     & R.A., Dec.&         \\
\noalign{\smallskip}\hline\noalign{\smallskip}
700016  & 3,4 & 2019-09-15 09:28:20 $-$ 2019-09-16 21:30:00 & 25002 & 83.83637,-69.31139 & SN\,1987A                  \\
700161  & 1-7 & 2019-10-18 16:54:34 $-$ 2019-10-19 15:07:54 & 71544 & 83.83638,-69.31139 & SN\,1987A FL               \\
\noalign{\smallskip}
700156  & 5-7 & 2019-10-10 01:58:14 $-$ 2019-10-10 15:18:14 & 19349 & 81.25917,-69.64417 & N\,132D                    \\
700179  & 1-7 & 2019-11-22 15:44:11 $-$ 2019-11-23 08:24:11 & 59549 & 81.25917,-69.64417 & N\,132D                    \\
700184\tablefootmark{d}  & 1-7 & 2019-11-23 21:04:51 $-$ 2019-11-24 08:11:31 & 34702 & 80.68758,-69.37765 & N\,132D off-axis           \\
700185\tablefootmark{d}  & 1-7 & 2019-11-25 08:03:01 $-$ 2019-11-25 19:09:41 & 20594 & 81.90048,-69.89292 & N\,132D off-axis           \\
700183\tablefootmark{d}  & 1-7 & 2019-11-25 19:13:51 $-$ 2019-11-26 06:20:31 & 33805 & 81.9935 ,-69.43154 & N\,132D off-axis           \\
700182  & 1-7 & 2019-11-27 06:11:41 $-$ 2019-11-27 17:18:21 & 31483 & 80.51012,-69.85367 & N\,132D off-axis           \\
\noalign{\smallskip}
700205  & 1-7 & 2019-12-07 13:00:00 $-$ 2019-12-07 21:14:00 & 18839 & 84.44442,-69.17140 & PSR\,J0537$-$6910          \\
700206  & 1-7 & 2019-12-07 21:20:00 $-$ 2019-12-08 02:50:00 & 15344 & 85.04515,-69.33173 & PSR\,J0540$-$6919          \\
700207  & 1-7 & 2019-12-08 02:56:00 $-$ 2019-12-08 13:00:00 & 22255 & 84.74217,-69.25035 & PSR\,J0540 and J0537       \\
\noalign{\smallskip}
700001  & 1-7 & 2019-11-07 17:13:18 $-$ 2019-11-08 09:53:18 & 54610 & 16.00833,-72.03194 & \snre          \\
700002  & 1-7 & 2019-11-08 09:57:18 $-$ 2019-11-09 02:37:18 & 58714 & 15.03687,-71.88838 & \snre off-axis \\
700003  & 1-7 & 2019-11-09 02:41:38 $-$ 2019-11-09 19:21:38 & 59931 & 16.99466,-72.17060 & \snre off-axis \\
700004  & 1-7 & 2019-11-09 19:25:38 $-$ 2019-11-10 10:42:18 & 55594 & 16.28235,-71.85088 & \snre off-axis \\
700005  & 1-7 & 2019-11-10 10:45:48 $-$ 2019-11-11 02:02:28 & 55672 & 16.41901,-71.76044 & \snre off-axis \\
710000\tablefootmark{e}  & 1-7 & 2020-06-18 20:00:00 $-$ 2020-06-19 06:00:00 & 37056 & 16.00833,-72.03194 & \snre          \\
\hline 
\label{tabobs} 
\end{tabular} 
\end{center}
\tablefoot{
\tablefoottext{a}{\ero observation ID.}
\tablefoottext{b}{Active telescope modules.}
\tablefoottext{c}{Maximum exposure from vignetting-corrected exposure map normalized to one TM, 
                  i.e. total exposure divided by 7.
                  This way the resulting count rates in the source detection lists are for the total of seven TMs, also when not all TMs were active.}
\tablefoottext{d}{During part of these observations the electronic chopper was set to 2, which was not handled correctly by the pipeline processing. We corrected this in the event files (see Appendix A).}
\tablefoottext{e}{Calibration observation near the end of \red{the first \ero all-sky survey (eRASS1)}.}
}
\end{table*}

To analyse the CalPV data we used the \ero Standard Analysis Software System \citep[\eSASS version {\tt eSASSusers\_201009,}][]{2021arXiv210614517B}.
% (Brunner et al. 2021).
The \eSASS pipeline processing provides energy calibrated event files which were used to 
create images, detect point sources and extract spectra and light curves of the detected objects.
While for images and light curves we used data from all seven telescope modules \citep[][]{2021A&A...647A...1P}, 
only telescope modules with on-chip optical blocking filter (TM1--TM4 and TM6) were used for spectroscopy, 
because no reliable energy calibration is available yet for TM5 and TM7 due to the optical light leak discovered soon after the start of the CalPV phase \citep{2021A&A...647A...1P}. 

We extracted \ero light curves and spectra using the \eSASS task \texttt{srctool} from circular or (at larger off-axis angles) elliptical regions around the source position and a nearby source-free background region.
Events from all valid pixel patterns (PATTERN=15) were selected.
For spectral analysis we combined the data from the five  on-chip filter cameras into a single spectrum. 
The spectra were binned to achieve a minimum of twenty counts per spectral bin to allow the use of $\chi^{2}$-statistic. 
Corresponding response files were created by \texttt{srctool}.

Spectral analysis was performed using \xspec v12.11.0k \citep{1996ASPC..101...17A}. 
To account for the photo-electric absorption by the Galactic interstellar gas we used \texttt{tbabs} in \xspec with solar
abundances following \citet{2000ApJ...542..914W} with atomic cross sections from \citet{1996ApJ...465..487V}. 
The Galactic column density was taken from  \citet{1990ARA&A..28..215D} and fixed in the fits.
For the absorption along the line of sight through the LMC/SMC and local to the source, when required we included a second absorption component with elemental abundances fixed at 0.49/0.20 solar \citep{2002A&A...396...53R,1998AJ....115..605L}, respectively, and left the column density free in the fit.
Errors were estimated at 90\% confidence intervals. 
For computing X-ray luminosities distances of 50\,kpc and 60\,kpc were assumed for LMC and SMC, respectively.
\red{Throughout the paper we provide observed fluxes and absorption-corrected luminosities.}

\section{HMXBs covered by the \ero CalPV observations }

\begin{figure*}
  \begin{center}
  \resizebox{\hsize}{!}{\includegraphics[clip=]{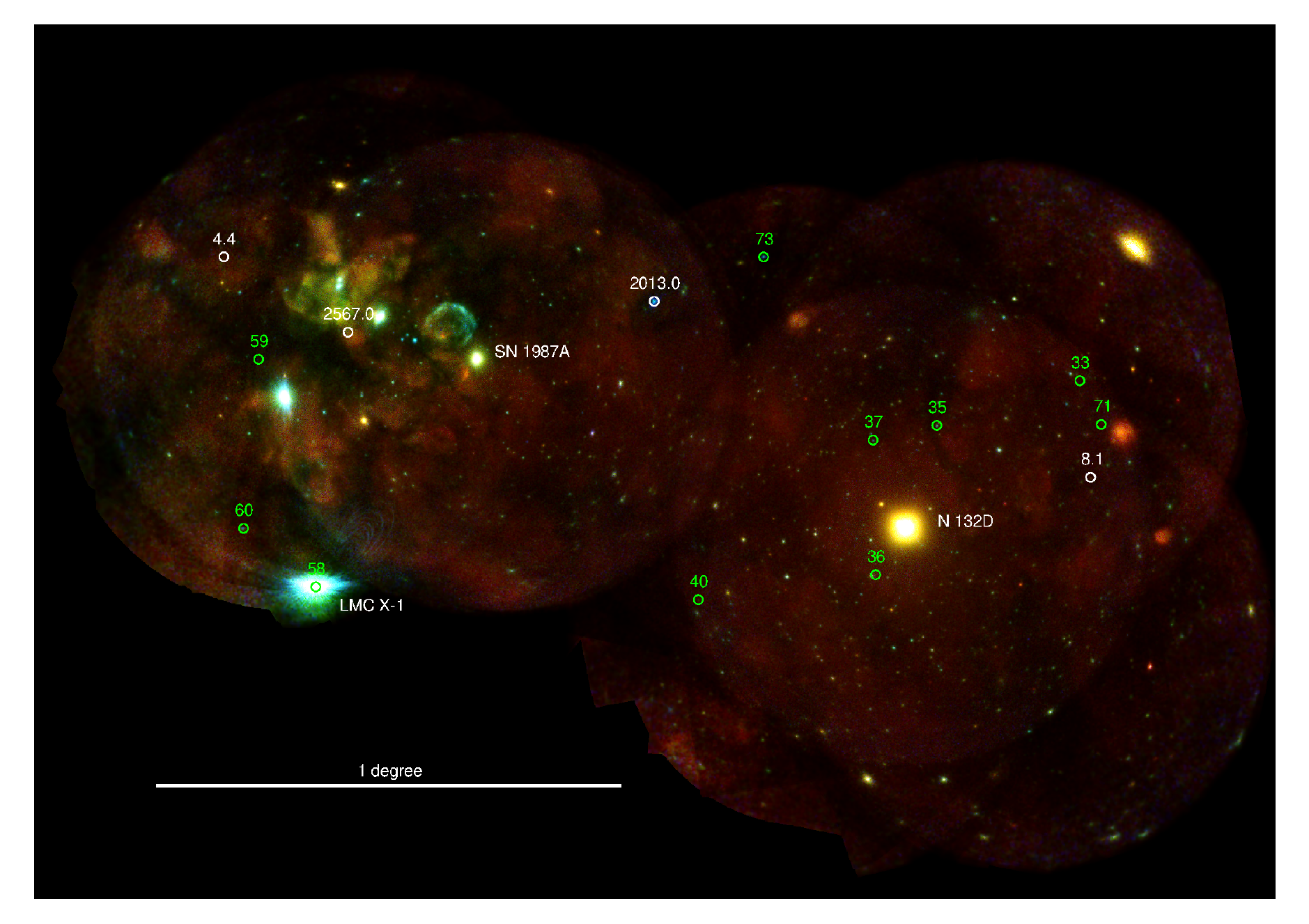}}
  \end{center}
  \caption{
    LMC mosaic from \ero observations around SN\,1987A and N\,132D.
    Red, green and blue colours represent X-ray energies in the 
    bands 0.2--1.0\,keV, 1.0--2.0\,keV and 2.0--4.5\,keV, respectively.
    The HMXB pulsars are marked with white circles and labeled with their 
    pulse period measured near the time of their discovery. 
    HMXBs and candidates without known pulse period are indicated by green circles and 
    the entry number from Table\,\ref{tab:lmc_hmxbs}.
  }
  \label{fig:imalmc}
\end{figure*}

\begin{figure*}
  \begin{center}
  \resizebox{\hsize}{!}{\includegraphics[clip=]{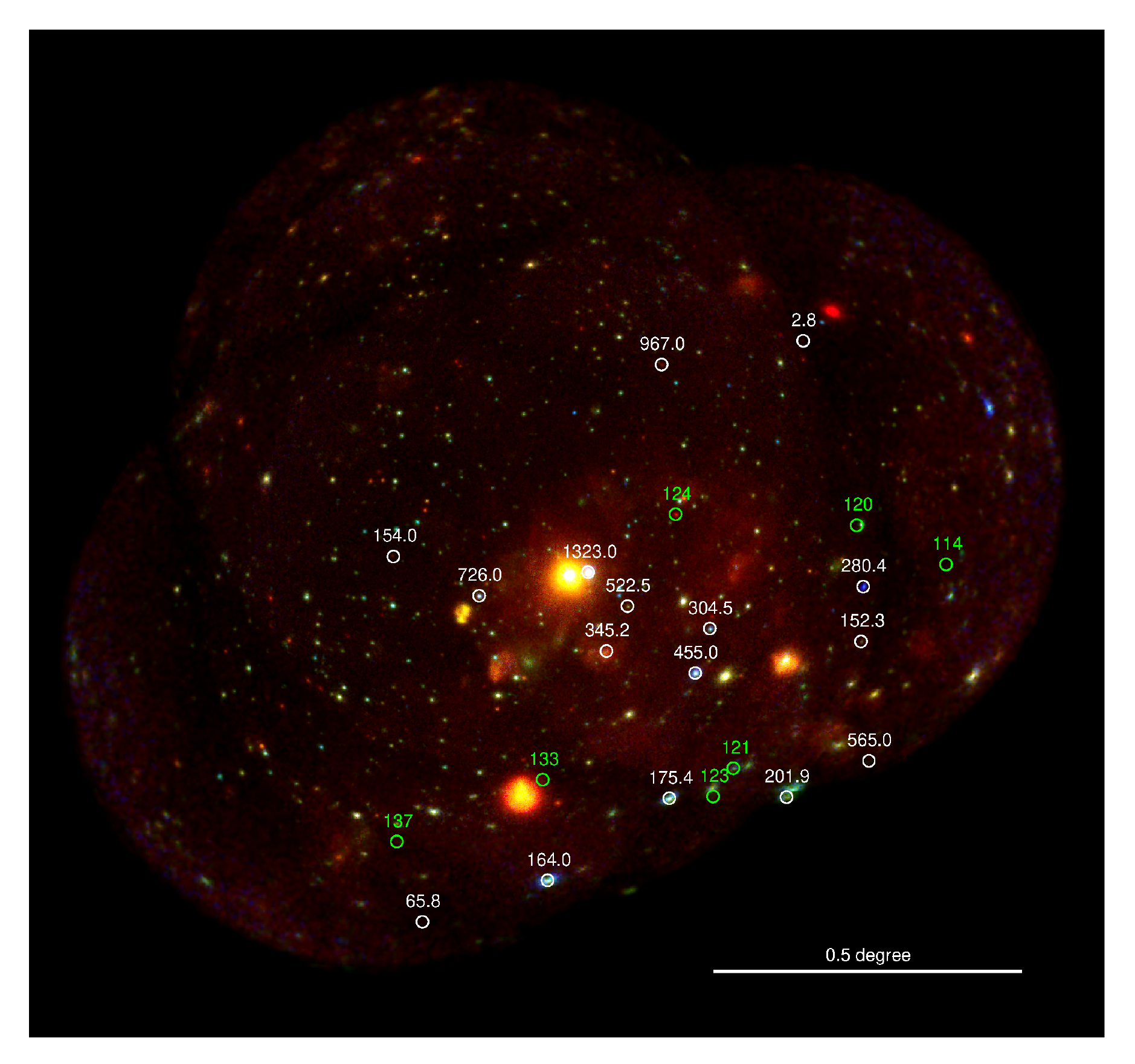}}
  \end{center}
  \caption{
    SMC mosaic from \ero observations around the supernova remnant \snre, 
    the brightest source near the image centre. Image colours and labels for the HMXB pulsars are as in Fig.\,\ref{fig:imalmc}. 
    HMXBs without known pulse period are indicated by green circles and their entry number from HS16 (see Table\,\ref{tab:smc_hmxbs}). 
  }
  \label{fig:imasmc}
\end{figure*}

The large ($\sim$1\degr\ diameter) field of view (FoV) of the \ero telscopes implies that CalPV observations pointed towards the LMC around SN\,1987A and N\,132D overlap and cover a contiguous area 
of 4.4 square degrees. 
Therefore, we produced mosaic images from the merged calibrated event file (combining data from all available cameras and observations) in three energy bands (red: 0.2--1.0\,keV, green: 1.0--2.0\,keV and blue: 2.0--4.5\,keV). 
The images were adaptively smoothed and divided by vignetting-corrected exposure maps to obtain count rate images. 
The corresponding RGB mosaic image is shown in Fig.\,\ref{fig:imalmc}.
The detector background was not subtracted yet from the images
because more data with filter wheel closed are still to be collected to construct a reliable background model.
This leads to a slight over-correction of the detector-background component in the images towards the edges of the FoV, \red{but does not affect our analysis of point sources where a local full background subtraction is applied.}
Similarly, the observations of the SMC supernova remnant \snre
(placed at different off-axis angles) cover an area of 2.5 square degrees and the 
RGB image is presented in Fig.\,\ref{fig:imasmc}.

%--------------------------
%--------------------------
%--------------------------
%--------------------------

For each observation, the \ero pipeline system performs source detection, 
based on a maximum likelihood Point Spread Function (PSF) fitting algorithm, 
simultaneously on the images from the available TMs in the three energy bands 
0.2--0.6\,keV, 0.6--2.3\,keV and 2.3--5.0\,keV \citep{2021arXiv210614517B}.
%(Brunner et al. 2021).
We correlated the source detection lists with the catalogues of HMXBs (see below) with 
a matching radius of 10\arcsec\ to check which of the known HMXBs were detected at which 
brightness. 
Upper limits were derived for the non-detections from the sensitivity maps using
the \eSASS task \texttt{apetool}. 
Tables\,\ref{tab:lmc_rates} and \ref{tab:smc_rates} summarize the detections and upper limits for all HMXBs observed in the LMC and SMC, respectively. Count rates are reported for the 0.2--5\,keV energy band and all rates given in subsections describing the sources refer to this band, unless otherwise noted.

The available data cover several epochs which also allows to investigate variability of detected sources on several time scales.
The observations of the LMC were irregularly distributed in time over nearly three months, whereas five of the SMC observations were performed consecutively, followed by an observation centered on \snre about 7 months later, providing two epochs for the HMXBs located within 0.5\degr\ of the supernova remnant.

\section{The LMC field}

The CalPV observations of the LMC cover fourteen (candidate) HMXBs which include nine well confirmed 
cases (four X-ray pulsars). 
These objects are listed in Table\,\ref{tab:lmc_hmxbs} and marked in the image 
presented in Fig.\,\ref{fig:imalmc}. \red{We used the numbering scheme from an HMXB catalogue of the LMC in preparation, similar to that of the SMC \citep{2016A&A...586A..81H}.}
For two cases the identification of the optical counterpart is still uncertain 
(No. 37 and 60), while the three remaining candidates for BeXRBs (No. 33, 35, 36) did not show significant
\Halpha emission during optical spectroscopy reported by  \citet{2018MNRAS.475.3253V}.
As can be already seen from the image, apart from the persistently bright black hole system
LMC\,X-1, only one source was sufficiently bright for a detailed spectroscopic and temporal
analysis. This is the supergiant fast X-ray transient (SFXT) XMMU\,J053108.3$-$690923 
(marked with its pulse period of 2013\,s in the image), 
for which the \ero results are already presented in \citet{2021A&A...647A...8M}.
One new BeXRB was discovered in this work (\srget, source 73 in Table\,\ref{tab:lmc_hmxbs}
and Fig.\,\ref{fig:imalmc}), which is described in detail in the following section.

\begin{table*}
\caption{HMXBs in the LMC}
\begin{center}
\begin{tabular}{rllcrc}
\hline\hline\noalign{\smallskip}
No. &Type    & X-ray source names                 & Spin (s) & AZ16 & References  \\
\noalign{\smallskip}\hline\noalign{\smallskip}
 71 & Be/SSS & XMMU\,J052016.0$-$692505           &      -   &  -   & KHP06       \\
  5 & Be/X   & RX\,J0520.5$-$6932                 &   8.04   & 14   & VHS14       \\
 33 & Be/X?  & XMMU\,J052049.1$-$691930 Halpha?   &      -   &  -   & [VBM18]7    \\
 35 & Be/X?  & XMMU\,J052417.1$-$692533 Halpha?   &      -   &  -   & [VBM18]8    \\
 36 & Be/X?  & XMMU\,J052546.4$-$694450 Halpha?   &      -   &  -   & [VBM18]9    \\
 37 & Be/X?? & XMMU\,J052550.7$-$692729 ID?       &      -   &  -   & [VBM18]18   \\
 73 & Be/X   & \srget                             &      -   &  -   & This work   \\
 40 & Be/X   & XMMU\,J053010.8$-$694755           &      -   &  -   & [VBM18]11   \\
 43 & SFXT   & XMMU\,J053108.3$-$690923           &   2013   &  -   & VMH18, MHV20\\
 18 & Be/X?  & CXOU\,J053833.4$-$691158           &   2567   & 46   & CBB15 \\
 58 & SG/BH  & LMC\,X$-$1                         &      -   & 37   &           \\
 59 & Be/X   & XMMU\,J054045.4$-$691452           &      -   &  -   & [VBM18]19 \\
 60 & SG?    & RX\,J0541.4$-$6936 ID?             &      -   & 39   & SHP00     \\
  3 & Be/X   & IGR\,J05414$-$6858                 &   4.42   & 38   & RSG10, SHR12 \\
\noalign{\smallskip}\hline\noalign{\smallskip}
\end{tabular}
\end{center}
\tablefoot{
HMXBs covered by the eROSITA calibration and performance verification observations of the LMC. References: 
AZ16: \citet{2016MNRAS.459..528A},
CBB15: \citet{2015A&A...579A.131C},
KHP06: \citet{2006A&A...458..285K},
MHV21: \citet{2021A&A...647A...8M},
RSG10: \citet{2010ATel.2704....1R},
SHP00: \citet{2000A&AS..143..391S},
SHR12: \citet{2012A&A...542A.109S},
VBM18: \citet{2018MNRAS.475.3253V},
VHS14: \citet{2014A&A...567A.129V},
VMH18: \citet{2018MNRAS.475..220V}.
}
\label{tab:lmc_hmxbs}
\end{table*}

\begin{table*}
\caption{HMXBs in the LMC: Variability between observations}
\begin{center}
\begin{tabular}{rcccccc}
\hline\hline\noalign{\smallskip}
 No.  & 700016                  & 700161                  & 700156                  & 700179                  & 700184                  & 700185                  \\
      & 2019-09-15              & 2019-10-18              & 2019-10-10              & 2019-11-22              & 2019-11-23              & 2019-11-25              \\
\noalign{\smallskip}\hline\noalign{\smallskip}                                                                                                
 71   &   --                    &  --                     &  $<$4.6\expo{-3}        &  $<$5.2\expo{-3}        &  $<$2.0\expo{-3}        & --                      \\
  5   &   --                    &  --                     &  $<$5.8\expo{-3}        &  5.40$\pm$1.28 \expo{-3}&  6.79$\pm$0.93 \expo{-3}& --                      \\
 33   &   --                    &  --                     &  $<$2.9\expo{-3}        &  $<$6.6\expo{-3}        &  $<$1.6\expo{-3}        & --                      \\
 35   &   --                    &  --                     &  1.14$\pm$0.15 \expo{-2}&  1.51$\pm$0.09 \expo{-2}&  1.33$\pm$0.10 \expo{-2}& --                      \\
 36   &   --                    &  --                     &  $<$1.8\expo{-3}        &  $<$2.9\expo{-3}        &  $<$1.0\expo{-2}        & $<$3.0\expo{-3}         \\
 37   &   --                    &  --                     &  $<$2.7\expo{-3}        &  5.59$\pm$0.62 \expo{-3}&  6.81$\pm$1.21 \expo{-3}& $<$5.7\expo{-3}         \\
 73   &   --                    &  --                     &  --                     &  --                     &  --                     & --                      \\
 40   &   --                    &  --                     &  1.47$\pm$0.31 \expo{-2}&  1.05$\pm$0.09 \expo{-1}&  --                     & 7.34$\pm$1.29 \expo{-3}\\
 43   &  1.27$\pm$0.04 \expo{-1}&  5.61$\pm$0.81 \expo{-3}&  --                     &  --                     &  --                     & --                      \\
 18   &  $<$8.3\expo{-3}        &  $<$7.4\expo{-3}        &  --                     &  --                     &  --                     & --                      \\
 59   &   --                    &  $<$4.5\expo{-3}        &  --                     &  --                     &  --                     & --                      \\
 60   &   --                    &  --                     &  --                     &  --                     &  --                     & --                      \\
  3   &   --                    &  --                     &  --                     &  --                     &  --                     & --                      \\
\noalign{\smallskip}\hline\noalign{\smallskip}                                                                                                
 No.  & 700183                  & 700182                  & 700205                  & 700206                  & 700207                  \\
      & 2019-11-25              & 2019-11-27              & 2019-12-07              & 2019-12-07              & 2019-12-08              \\
\noalign{\smallskip}\hline\noalign{\smallskip}                                                                                                
 71   &  --                     &  $<$4.4\expo{-3}        &  --                     & --                      & --                      \\
  5   &  --                     &  7.09$\pm$1.45 \expo{-3}&  --                     & --                      & --                      \\
 33   &  --                     &  --                     &  --                     & --                      & --                      \\
 35   &  $<$3.8\expo{-3}        &  $<$8.9\expo{-3}        &  --                     & --                      & --                      \\
 36   &  $<$8.3\expo{-3}        &  $<$9.2\expo{-3}        &  --                     & --                      & --                      \\
 37   &  5.61$\pm$0.81 \expo{-3}&  --                     &  --                     & --                      & --                      \\
 73   &  3.99$\pm$0.24 \expo{-2}&  --                     &  --                     & --                      & --                      \\
 40   &  7.21$\pm$1.77 \expo{-3}&  --                     &  --                     & --                      & --                      \\
 43   &  1.69$\pm$0.05 \expo{-1}&  --                     &  --                     & --                      & --                      \\
 18   &  --                     &  --                     &  $<$2.7\expo{-3}        & $<$2.6\expo{-3}         & $<$2.2\expo{-3}         \\
 59   &  --                     &  --                     &  $<$1.5\expo{-3}        & $<$4.2\expo{-3}         & $<$1.5\expo{-3}         \\
 60   &  --                     &  --                     &  1.73$\pm$0.09 \expo{-1}& $<$6.6\expo{-3}         &  2.11$\pm$0.11 \expo{-1}\\
  3   &  --                     &  --                     &  $<$5.7\expo{-3}        & $<$9.0\expo{-3}         & $<$5.2\expo{-3}         \\
\noalign{\smallskip}\hline\noalign{\smallskip}
\end{tabular}
\end{center}
\tablefoot{
Background-subtracted, PSF and vignetting-corrected source count rates (\cts) are obtained from source detection and are given for the 0.2--5.0\,keV band normalized to seven telescope modules.
We exclude LMC\,X-1 here due to its brightness which causes strong photon pile-up effects.
Numbers without error \red{and marked with ``$<$''} denote upper limits (3$\sigma$) for undetected sources in the FoV, derived from sensitivity maps produced by the \eSASS task \texttt{apetool}.
}
\label{tab:lmc_rates}
\end{table*}

\subsection{\srget, a new BeXRB in the LMC}
\label{sec:srget}

\begin{figure}
  \begin{center}
  \resizebox{\hsize}{!}{\includegraphics[clip=]{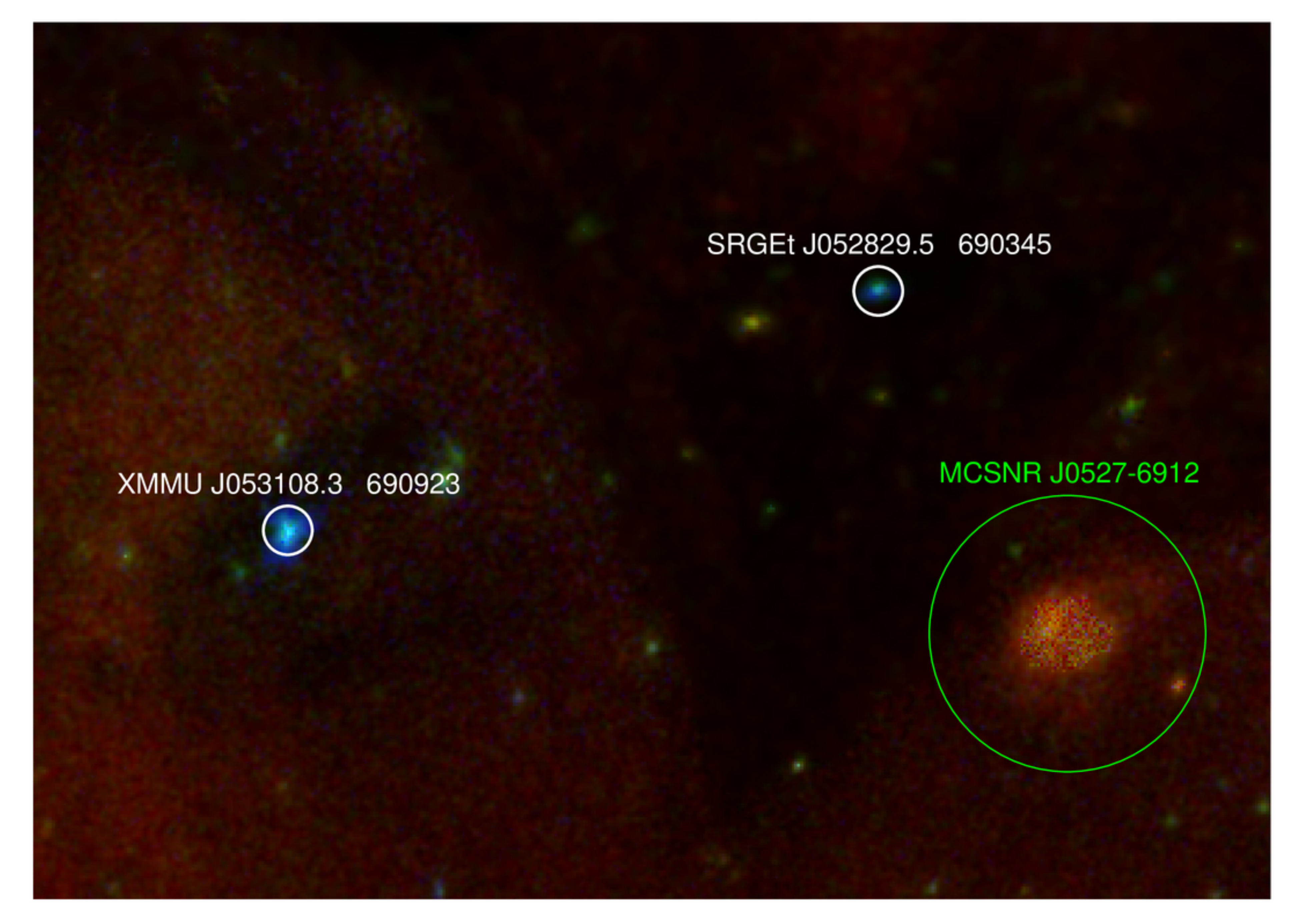}}
  \end{center}
  \caption{
    \ero colour image (red: 0.2--1.0\,keV, green: 1.0--2.0\,keV, blue: 2.0--4.5\,keV) of the region around \srget.
    The transient source appears similar in colours to the SFXT XMMU\,J053108.3$-$690923 \citep{2021A&A...647A...8M}, which is located at an angular distance of $\sim$15\arcmin. The soft extended X-ray emission from the supernova remnant MCSNR\,J0527-6912 \citep[e.g.][]{2016A&A...585A.162M} is also marked.
  }
  \label{fig:eroima}
\end{figure}

\paragraph{Source position and optical counterpart:}
The faint transient \srget was discovered in the \ero data of the calibration observation 700183 of the 
supernova remnant N132D performed in Nov. 2019 (Table\,\ref{tabobs}).
\srget appears as a blue source on the \ero colour image (Fig.\,\ref{fig:eroima}), indicating a hard X-ray spectrum.
Source detection resulted in an X-ray source position of R.A. (2000) = 05\rahour\,28\ramin\,29\fs1, Dec. = -69\degr\,03\arcmin\,51\arcsec\ 
with a statistical 1\,$\sigma$ uncertainty of 1.2\arcsec. 
The \ero position is thereby 7\arcsec\ from a star with V = 15.70\,mag, B = 15.90\,mag, U = 15.08\,mag and R = 16.66\,mag \citep{2002ApJS..141...81M} with coordinates R.A. = 05\rahour\,28\ramin\,29\fs34, Dec. = -69\degr\,03\arcmin\,44\farcs7, as listed in the Gaia DR2 catalogue \citep{2018A&A...616A..12G}. 
The brightness and colours of this star are typical for an early B star in the LMC, suggesting a new BeXRB.

%--------------------------
%--------------------------

The observation also covers the supergiant fast X-ray transient (SFXT) XMMU\,J053108.3$-$690923
\citep{2021A&A...647A...8M} at a similar off-axis angle. The SFXT is securely identified with a supergiant in the LMC. Also for this source the \ero position is shifted by a similar value
(6.6\arcsec)\footnote{\redii{This shift is consistent with the values obtained from other \ero observations \citep[][Lamer et al. in preparation]{2021arXiv210614517B}. These papers describe  statistical methods to determine astrometric corrections using large samples of objects. In the case of the Magellanic Clouds this is more difficult as only few sources in the field of view have secure optical counterparts \citep[see][]{2019A&A...622A..29M,2013A&A...558A...3S}.}} and in the same direction with respect to the optical counterpart.
Therefore, we used the SFXT for astrometric correction and obtained R.A. = 05\rahour\,28\ramin\,29\fs57, Dec. = -69\degr\,03\arcmin\,45\farcs5 for the refined coordinates of the new BeXRB candidate. The corrected X-ray position is 1.5\arcsec\ from that of the proposed optical counterpart, compatible within the errors. We use these coordinates for the source name following the \srg/\ero convention for a new transient: \srget.

The position of the transient \srget was covered by two \xmm pointed observations (observation IDs 0690743501/3601) 
at large off-axis angles and during many slews. 
The observations were performed one after the other between 2012-09-04 16:55 and 2012-09-05 11:25 as part of the \xmm LMC survey (PI Haberl).
The source was not detected and we derived 3\,$\sigma$ upper limits in the 0.2--4.5\,keV energy band of 1.6\expo{-3} \cts, 1.6\expo{-3} \cts and 3.4\expo{-3} \cts from the EPIC-MOS1, MOS2 and pn sensitivity maps, respectively. The upper limits during slew are typically more than a factor 100 higher and not constraining.

\paragraph{Spectral analysis:}
As described above, we extracted \ero spectra \red{(with total net exposure of 28.4\,ks)} from circular regions around the source position (radius 25\arcsec)\footnote{\red{The half-energy width of the telescope point spread function is about 16\arcsec\ at 1.5\,keV \citep{2021A&A...647A...1P}.}} and a nearby source-free background region (radius 50\arcsec). 
An absorbed power-law model yields an acceptable fit (reduced $\chi^2 = 0.8$ for 13 degrees of freedom, dof) to the \ero spectrum. The power-law index of 0.66$^{+0.74}_{-0.54}$, although with large error is within the range observed for BeXRBs in the Magellanic Clouds \citep[e.g.][]{2016A&A...586A..81H} and the LMC absorption, required in addition to the Galactic foreground (fixed at 6.2\hcm{20}), is 1.0$^{+1.0}_{-0.6}$\hcm{22}. The observed 0.2--8.0\,keV flux is 3.4\ergcm{-13}, which results in soft X-ray source luminosity of 1.2\ergs{35}.

Using WebPIMMS\footnote{\url{https://heasarc.gsfc.nasa.gov/cgi-bin/Tools/w3pimms/w3pimms.pl}} and the \ero best-fit parameters for a power law with a single column density, the expected 0.2--4.5\,keV count rates for EPIC-MOS and pn are 1.3\expo{-2} \cts and 3.3\expo{-2} \cts, about a factor of 10 higher than the upper limits derived from the \xmm observations. Therefore, considering the non-detection by \xmm, we conclude that the source is a transient as expected for a BeXRB.

Considering that most known BeXRBs host accreting neutron stars, we also carried out a search for possible X-ray pulsations. Unfortunately, the search did not reveal any significant signal because of the low statistics. From the extraction region a total of 330 counts is available, with $\sim$100 counts background contribution, which also does not allow to put meaningful limits on the amplitude of possible pulsations.  

%--------------------------
\begin{figure}
  \begin{center}
  \resizebox{0.95\hsize}{!}{\includegraphics[clip=]{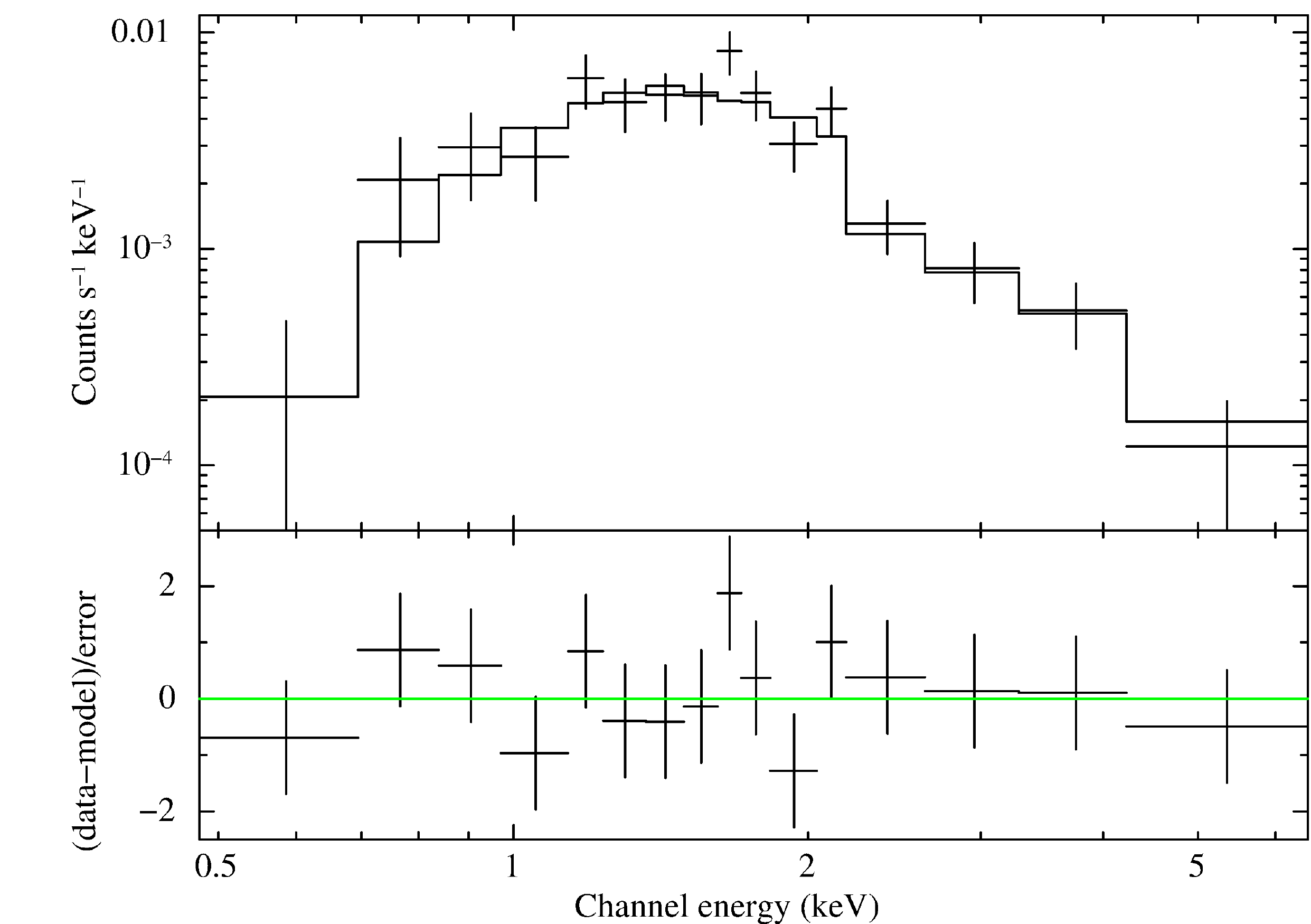}}
  \end{center}
  \caption{
    Combined \ero spectrum of \srget from the cameras with on-chip filter (TM1--4 and TM6).
    The best-fit absorbed power-law model is shown as histogram, 
    with the residuals plotted in the bottom panel.
  }
  \label{fig:erospec}
\end{figure}
%--------------------------

\paragraph{OGLE long-term monitoring:}
The region around the new BeXRB candidate \srget was monitored regularly in the I and V bands with the  
the Optical Gravitational Lensing Experiment \citep[OGLE;][]{2008AcA....58...69U,2015AcA....65....1U}. 
Images were taken in V and I filter bands, while photometric magnitudes are calibrated to the standard VI system.

\begin{figure*}
  \resizebox{\hsize}{!}{\includegraphics[clip=]{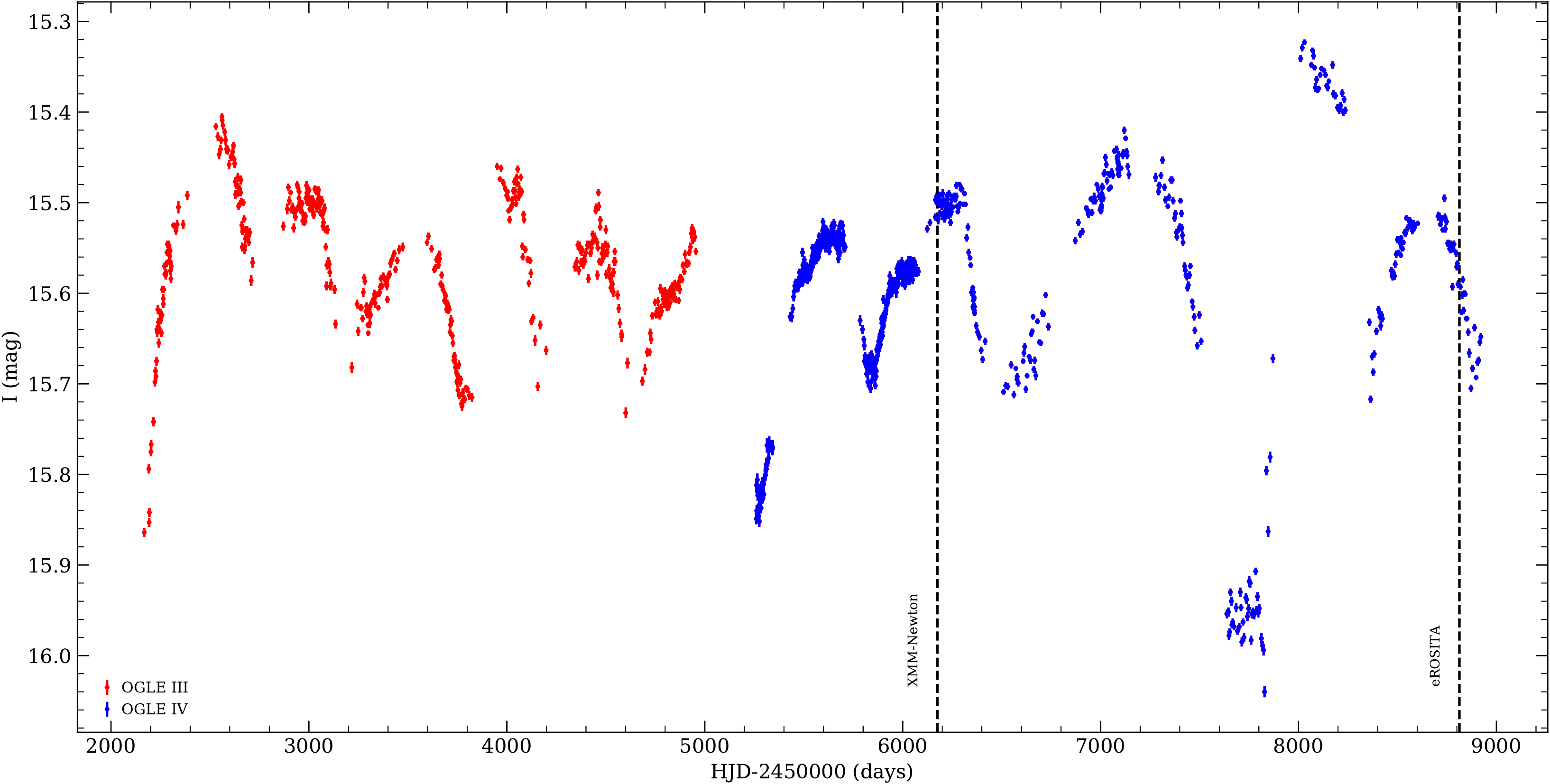}}
  \caption{
    OGLE I-band light curve of \srget between 2001-09-15 and 2020-03-13. 
    The vertical dashed lines indicate the times of the \ero and archival \xmm observations.
  }
  \label{fig:oglelci}
\end{figure*}

\begin{figure}
  \resizebox{\hsize}{!}{\includegraphics[clip=]{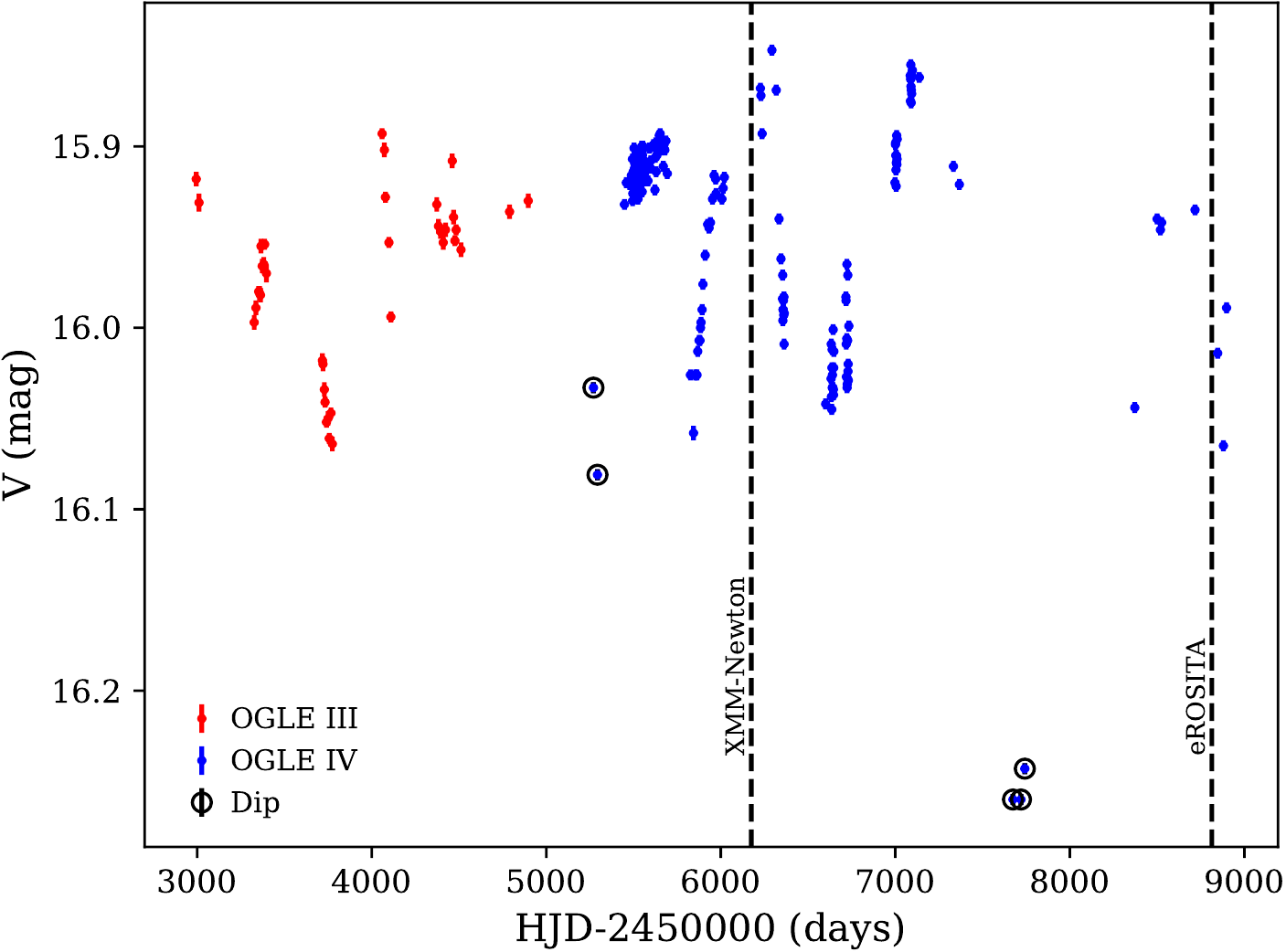}}
  \caption{
    OGLE V-band light curve of \srget. See also Fig.\,\ref{fig:oglelci}. 
    \red{The data points marked with additional black circles indicate two deep dips in the light curve, which show extreme colours (see Fig.\,\ref{fig:oglevi}).}
  }
  \label{fig:oglelcv}
\end{figure}

The optical counterpart of \srget was covered for about 18.5 years during OGLE phases III \red{(star lmc161.4.31169)} and IV \red{(lmc517.06.71858)} as shown in Figs.\,\ref{fig:oglelci} and \ref{fig:oglelcv}.
Both light curves exhibit remarkable variations by up to 0.75\,mag in the I-band.
A Lomb-Scargle (LS) analysis of the I-band light curve reveals that these variations occur quasi-periodically, repeating 
every $\sim$511\,days. The outbursts show different durations, which lead to phase shifts, a behaviour very similar to 
the optical counterpart of another BeXRB: RX\,J0529.8$-$6556 \citep{2021MNRAS.503.6187T}. The authors suggest this to be resulting from a misalignment of 
the Be star disc with respect to the orbital plane, which causes precession of the disc.

%--------------------------
%--------------------------
%--------------------------
%--------------------------
%--------------------------
\begin{figure}
  \resizebox{\hsize}{!}{\includegraphics[clip=]{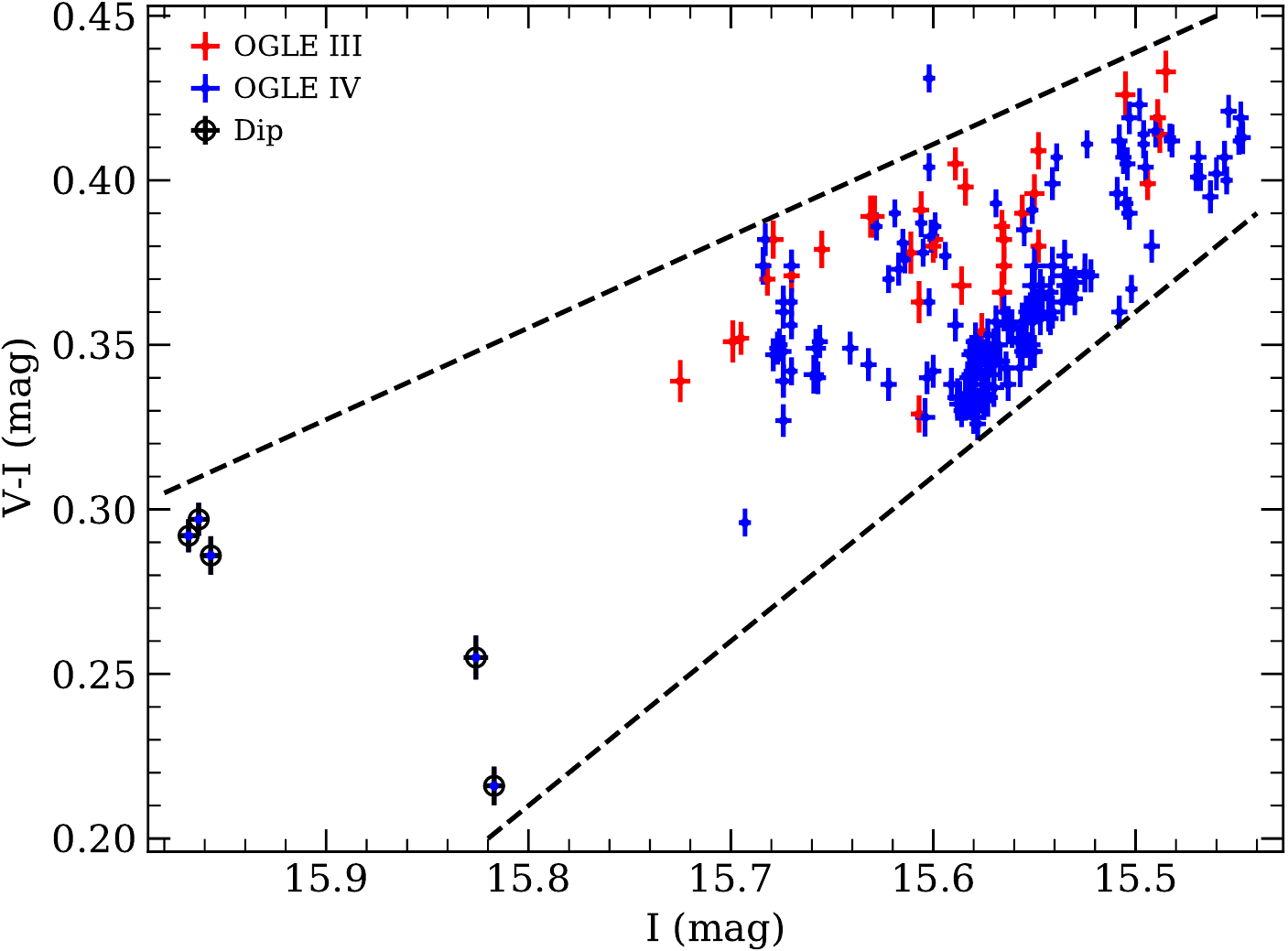}}
  \caption{
    OGLE V-I colour index vs. I (interpolated linearly to the times of the V-band measurements).
    \red{The data points marked with additional black circles originate from two deep dips in the light curve (see also Fig.\,\ref{fig:oglelcv}). V-I is correlated with I in a well-defined region of the parameter space as indicated by the dashed lines.}
  }
  \label{fig:oglevi}
\end{figure}
%--------------------------

The OGLE V-band observations provided a less frequent coverage of the BeXRB system than the I-band observations.
To compute the I$-$V colour index we chose the I-band magnitude nearest in time to the V-band measurement, disregarding those more than 2 days apart.
The colour index I$-$V is shown as a function of the I-band magnitude in Fig.\,\ref{fig:oglevi}. Large colour variations with brightness (redder when brighter) become apparent especially during the deep dips seen in the light curves.

\subsection{\red{XMMU\,J052417.1$-$692533, a BeXRB without significant \Halpha emission?}}

\red{\citet{2018MNRAS.475.3253V} selected 19 candidate BeXRBs from \xmm  observations of the LMC based on X-ray hardness ratio criteria.
They present the results from optical spectroscopy and the analysis of the OGLE data of the proposed counterparts. Six of their candidates were covered by the \ero observations (reference [VBM18] in Table\,\ref{tab:lmc_hmxbs}) and the proposed counterparts of three of these did not exhibit significant \Halpha emission}
(sources No. 33, 35 and 36),
\red{questioning their BeXRB identification}. 
XMMU\,J052049.1$-$691930 (No.\,33) and XMMU\,J052546.4$-$694450 (No.\,36) were not detected in the \ero data, \red{not providing further information on their nature.}

The position of XMMU\,J052417.1$-$692533 (No.\,35) was covered by three observations (700156, 700179 and 700184) from which similar count rates 
were obtained and two observations with no detection (see Table\,\ref{tab:lmc_rates}). 
The upper limits indicate variability of at least a factor of four on  a time scale of 6 weeks.
The eROSITA positions obtained from the three detections 
\red{have angular separations of 4.4\arcsec, 3.0\arcsec\ and 6.5\arcsec\ from} the optical position of the V = 16\,mag early-type star which is proposed as optical counterpart \red{\citep[for more information on the star see][]{2018MNRAS.475.3253V}}. \red{Given the current uncertainties in the astrometry of the CalPV observations (see Sect.\,\ref{sec:srget}) the X-ray position obtained from \xmm is more reliable \citep[see][]{2018MNRAS.475.3253V}.}

\begin{figure}
  \begin{center}
  \resizebox{0.95\hsize}{!}{\includegraphics[clip=]{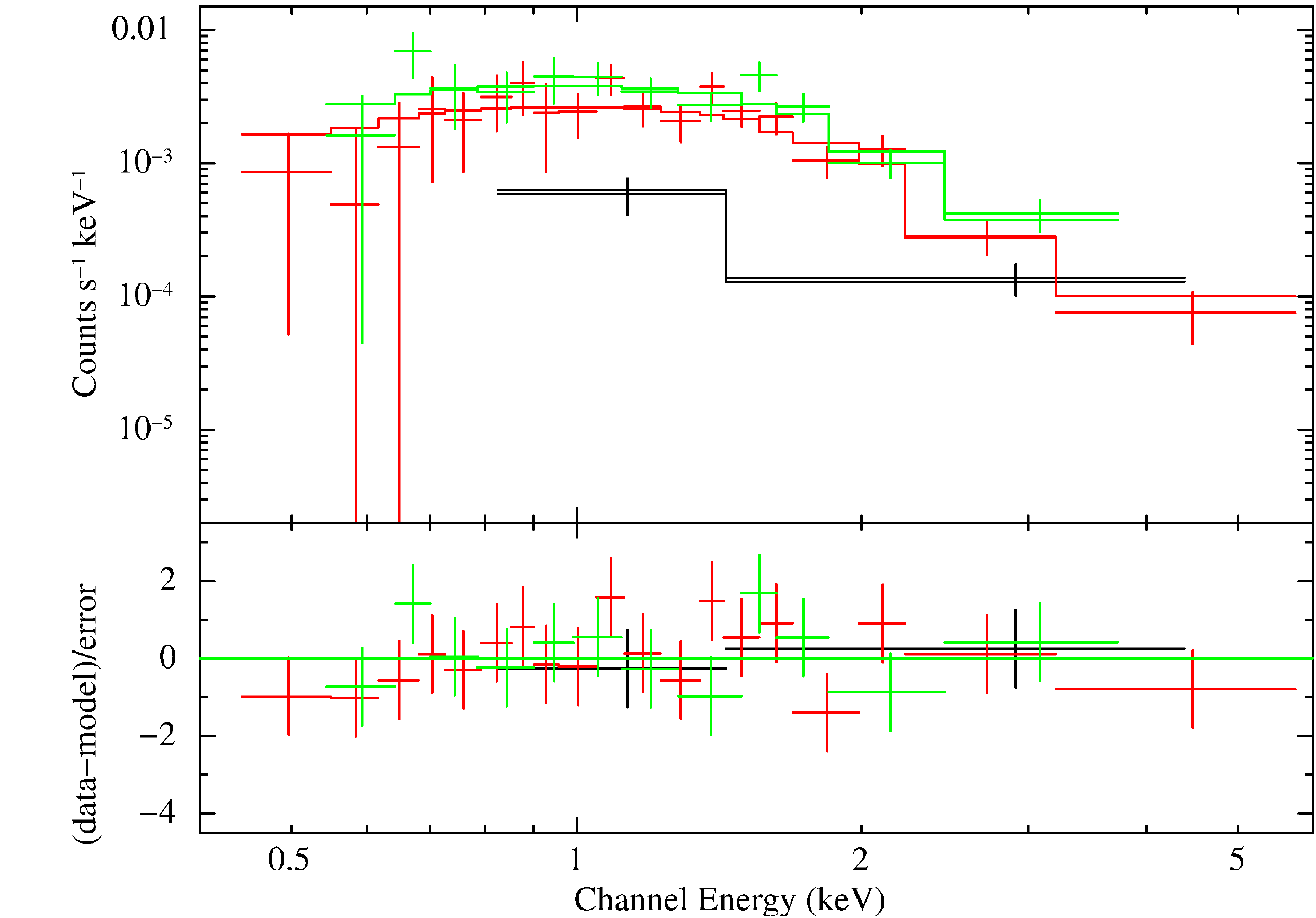}}
  \end{center}
  \caption{
    Combined \ero spectra of XMMU\,J052417.1$-$692533 from the cameras with on-chip filter (TM1--4 and TM6).
    The best-fit absorbed power-law model is shown as histogram, 
    with the residuals plotted in the bottom panel.
    The different colours indicate the observations 700156 (black), 700179 (red) and 700184 (green).
    The spectrum from observation 700156 includes data from TM6 only (see Table\,\ref{tabobs}).
  }
  \label{fig:erospecLMCcand}
\end{figure}

We extracted and analysed the eROSITA spectra of XMMU\,J052417.1$-$692533 in a similar way as for \srget (Sect.\,\ref{sec:srget}). Because of a nearby source, the extraction regions were chosen smaller (15\arcsec\ and 25\arcsec\ for source and background regions, respectively). The spectra from the three observations \red{(with total net exposure of 135\,ks)} were fitted simultaneously with a power-law model. Absorption (foreground \nh fixed at 6.2\hcm{20}) and photon index were assumed to be constant between the observations, only the power-law normalization was free in the fit to follow the flux variations. The spectra with the best-fit model are shown in Fig.\,\ref{fig:erospecLMCcand}. 
The model formally yields an acceptable fit \red{(reduced $\chi^2$ = 0.73 for 29 dof) to the \ero spectra. The power-law index of 0.92$\pm$0.22} is typical for BeXRBs in the Magellanic Clouds \citep[e.g.][]{2016A&A...586A..81H}, excluding a background active galactic nucleus and supporting the identification of the X-ray source with the early-type star.
For the LMC absorption no significant \nh was required in the fit, however, the systematic residuals might indicate a more complex spectrum at energies below $\sim$1\,keV. 
The 0.2--8.0\,keV flux might vary slightly between observations, but is consistent within the 90\% errors: 
7.9$^{+3.0}_{-2.6}$\,\ergcm{-14}, 
6.3$^{+1.4}_{-1.5}$\,\ergcm{-14} and 
7.2$^{+1.5}_{-1.7}$\,\ergcm{-14} (ordered in time), which corresponds to an average X-ray luminosity of 2.2\ergs{34}.

%--------------------------
%--------------------------

\section{The SMC field}

\ero performed five calibration observations consecutively in Nov. 2019, one centered on \snre, the other four having the supernova remnant offset into different directions (Table\,\ref{tabobs}). This results in different off-axis angles for the sources covered by several observations and does not allow us to immediately handle the five observations as one long one (duration $\sim$3.4 days). However, we concatenated light curves after they were background-subtracted and corrected for vignetting and PSF losses. Similarly, spectra were extracted for each observation and analysed simultaneously. A sixth observation in June 2020 was again centred on \snre, allowing further long-term studies of sources located within $\sim$30\arcmin\ of the supernova remnant. 

\begin{table*}
\caption{HMXBs in the SMC}
\begin{center}
\begin{tabular}{llcrc}
\hline\hline\noalign{\smallskip}
Type    & X-ray source names     & Spin (s)   & HS16 & References\\
\noalign{\smallskip}\hline\noalign{\smallskip}
  Be/X   & XMMU\,J005605.8$-$720012                                               &       - & 114 & NLM11, MGB17 \\
  Be/X   & CXOUJ005736.2-721934                                                   &   564.8 &  54 & MFL03, CEG05 \\
  Be/X   & RX\,J0057.8$-$7202 = AXJ0058$-$72.0                                    &   280.3 &  43 & TIY99, SPH03 \\
  Be/X   & RX\,J0057.8$-$7207 = CXOU\,J005750.3$-$720756                          &   152.3 &  33 & HS00, SPH03, MFL03 \\
  Be/X?  & RX\,J0057.9$-$7156                                                     &       - & 120 & HS00 \\
  Be/X   & RX\,J0059.2$-$7138                                                     &   2.76  &   4 & H94, KYK00 \\
  Be/X   & RX\,J0059.3$-$7223 = XMMU\,J005921.0-722317                            &   201.9 &  38 & KPF99, SPH03, MLM04 \\
  Be/X   & XMMU\,J010030.2$-$722035                                               &       - & 121 & SPH03 \\
  Be/X   & XMMU\,J010055.8$-$722320                                               &       - & 123 & SHP13 \\
  Be/X   & RX\,J0101.0$-$7206                                                     &     304 &  46 & KP96, MFL03 \\
  Be/X   & RX\,J0101.3$-$7211                                                     &     455 &  51 & SHK01 \\
  Be/SSS & Be+WD? XMMU\,J010147.5$-$715550                                        &       - & 124 & SHP12 \\
  Be/X   & RXJ\,0101.8$-$7223 = AX\,J0101.8$-$7223 = XMMU\,J010152.4$-$722336     &     175 &  37 & HS00, YIT03, TDC11 \\
  Be/X   & CXOU\,J010206.6$-$714115                                               &     967 &  60 & HEP08 \\
  Be/X   & 2XMM\,J010247.4$-$720449 = Suzaku\,J0102$-$7204                        &     522 &  53 & WTE13 \\
  Be/X   & SAX\,J0103.2$-$7209 = AX\,J0103$-$722                                  &   345.2 &  50 & ICC00, YK98 \\
  Be/X   & RX\,J0103.6$-$7201                                                     &    1323 &  62 & HS00, HP05, CHS17 \\
  Be/X   & XMMU\,J010429.4$-$723136                                               &     164 & 132 & MSH13, MGB17, HCM19 \\
  Be/X   & RX\,J0104.5$-$7221 = XMMU\,J010435.4$-$722147                          &       - & 133 & HS00, SHP13 \\
  Be/X   & RX\,J0105.9$-$7203 = AX\,J0105.8$-$7203                                &     726 &  57 & HS00, YIT03, EH08 \\
  Be/X   & CXOU\,J010712.6$-$723533 = RX\,J0107.1$-$7235 = AX\,J0107.2$-$7234     &    65.8  &  24 & MCS07 \\
  Be/X   & XMMU\,J010743.1$-$715953                                               &     154 &  34 & CHS12  \\
  Be/X   & CXOU\,J010744.51$-$722741.7 = Swift\,J010745.0$-$722740                &       - & 137 & MSH14, VHM17 \\
\noalign{\smallskip}\hline\noalign{\smallskip}
\end{tabular}
\end{center}
\tablefoot{
HMXBs covered by the eROSITA calibration observations of the supernova remnant 1E\,0102.2$-$7219, 
extracted from \citet[][\red{HS16}]{2016A&A...586A..81H}. References: 
CEG05: \citet{2005MNRAS.356..502C}, 
CHS12: \citet{2012MNRAS.424..282C}, CHS17: \citet{2017A&A...602A..81C}, EH08: \citet{2008A&A...485..807E}, 
H94: \citet{1994ApJ...427L..25H},   HCM19: \citet{2019ATel13312....1H}, HEP08: \citet{2008A&A...489..327H}, 
HP05: \citet{2005A&A...438..211H},  HS00: \citet{2000A&A...359..573H},  ICC00: \citet{2000ApJ...531L.131I},
KP96: \citet{1996A&A...312..919K},  KPF99: \citet{1999A&AS..136...81K}, KYK00: \citet{2000PASJ...52..299K},
MCS07: \citet{2007MNRAS.376..759M}, MFL03: \citet{2003ApJ...584L..79M}, MGB17: \citet{2017MNRAS.467.1526M}, 
MLM04: \citet{2004ApJ...609..133M}, MSH13: \citet{2013ATel.5674....1M}, MSH14: \citet{2014ATel.5778....1M}, 
NLM11: \citet{2011A&A...532A.153N}, SHK01: \citet{2001A&A...369L..29S}, SHP12: \citet{2012A&A...537A..76S}, 
SHP13: \citet{2013A&A...558A...3S}, SPH03: \citet{2003A&A...403..901S}, TDC11: \citet{2011ATel.3311....1T}, 
TIY99: \citet{1999PASJ...51L..21T}, VHM: \citet{2017ATel10253....1V},   WTE13: \citet{2013PASJ...65L...2W},
YIT03: \citet{2003PASJ...55..161Y}, YK: \citet{1998IAUC.7009....3Y}
}
\label{tab:smc_hmxbs}
\end{table*}

In total, the observations of the SMC cover 23 HMXBs which include 20 well confirmed cases of Be X-ray binaries (BeXRBs) 
and 16 X-ray pulsars (including XMMU\,J010429.4$-$723136, see below). 
The (candidate) HMXBs are summarized in Table\,\ref{tab:smc_hmxbs} and marked in the RGB image which is presented in Fig.\,\ref{fig:imasmc}. 
As can be seen from the image, one of the candidate HMXBs \citep[XMMU\,J010147.5$-$715550, number 124;][]{2012A&A...537A..76S} 
is characterised by a red colour, which indicates a supersoft X-ray source (SSS). SSSs with a Be star as optical counterpart
are promising candidates for Be systems with a white dwarf (WD) as compact object \citep[see also][]{2006A&A...458..285K,2020MNRAS.497L..50C}. 
As can also be seen, apart from the persistently bright supergiant system SMC\,X-1, 
there were several other ``classical'' BeXRBs (with hard X-ray spectrum), which were sufficiently bright for a detailed spectroscopic and temporal analysis. 
All these are known to be BeXRB pulsars: 
RX\,J0101.0$-$7206 a.k.a. \sxp{304}, 
RX\,J0101.3$-$721 a.k.a. \sxp{455}, 
AX\,J0103$-$722 a.k.a. \sxp{522}, 
RX\,J0105.9$-$720 a.k.a. \sxp{726}, 
and RX\,J0103.6$-$7201 a.k.a. \sxp{1323}. 
The results from the HMXB XMMU\,J010429.4$-$723136, from which 164\,s pulsations were discovered in the \ero CalPV data 
\citep{2019ATel13312....1H}, are presented in \citet{2021arXiv210614536C}, % Carpano et al. (2021),
the results from the other sources are described below.

\begin{table*}
\caption{HMXBs in the SMC: Variability between observations}
\begin{center}
\begin{tabular}{rcccccc}
\hline\hline\noalign{\smallskip}
 HS16 & 700001                     & 700002                   & 700003                    & 700004                    & 700005                    & 710000     \\
      & 2019-11-07                 & 2019-11-08               & 2019-11-09                & 2019-11-09                & 2019-11-10                & 2020-06-18 \\
\noalign{\smallskip}\hline\noalign{\smallskip}
 114  &       --                  &  $<$2.3\expo{-3}          &         --                &        --                 &      --                  & --                      \\
  54  &        --                 &  1.14$\pm$1.76 \expo{-2}  &       --                  &      --                   &      --                  & --                      \\
  43  &  1.90$\pm$0.20 \expo{-2}  &  2.38$\pm$1.07 \expo{-2}  &       --                  &     --                    &      --                  & --                      \\
  33  &  4.50$\pm$1.64 \expo{-3}  &  7.00$\pm$7.90 \expo{-3}  &       --                  &     --                    &      --                  & --                      \\
 120  &  $<$2.6\expo{-3}          &  $<$2.3\expo{-3}          &       --                  &      --                   &      --                  & --                      \\
   4  &  $<$3.3\expo{-3}          &  2.32$\pm$4.24 \expo{-3}  &       --                  &  $<$1.4\expo{-3}          & $<$2.0\expo{-3}          & --                      \\
  38  &  --                       &  2.36$\pm$9.53 \expo{-1}  &       --                  &      --                   &       --                 & --                      \\
 121  &  2.40$\pm$0.17 \expo{-2}  &  2.72$\pm$1.96 \expo{-2}  &       --                  &       --                  &       --                 & $<$3.2\expo{-3}         \\
 123  &  5.11$\pm$1.37 \expo{-3}  &       --                  &       --                  &       --                  &       --                 & $<$5.5\expo{-3}         \\
  46  &  3.21$\pm$0.12 \expo{-2}  &  3.41$\pm$1.18 \expo{-2}  &       --                  &  2.41$\pm$0.17 \expo{-2}  & $<$4.0\expo{-3}          & 1.56$\pm$0.13 \expo{-2} \\
  51  &  8.68$\pm$0.20 \expo{-2}  &  1.10$\pm$2.37 \expo{-1}  &  4.82$\pm$0.37 \expo{-2}  &  9.27$\pm$0.30 \expo{-2}  &       --                 & 1.57$\pm$0.04 \expo{-1} \\
 124  &  6.42$\pm$0.55 \expo{-3}  &  7.36$\pm$4.94 \expo{-3}  &       --                  &  8.59$\pm$0.70 \expo{-3}  & 1.00$\pm$0.10 \expo{-2}  & 6.89$\pm$0.89 \expo{-3} \\
  37  &  1.74$\pm$0.03 \expo{-1}  &        --                 &       --                  &       --                  &      --                  & 2.07$\pm$0.05 \expo{-1} \\
  60  &  2.84$\pm$0.73 \expo{-3}  &  2.00$\pm$3.79 \expo{-3}  &       --                  &  2.60$\pm$0.58 \expo{-3}  & $<$1.7\expo{-3}          & 1.70$\pm$0.19 \expo{-2} \\
  53  &  5.34$\pm$0.48 \expo{-3}  &  5.16$\pm$6.79 \expo{-3}  &  4.17$\pm$1.11 \expo{-3}  &  8.56$\pm$0.88 \expo{-3}  & 1.31$\pm$0.14 \expo{-2}  & 3.70$\pm$0.59 \expo{-3} \\
  50  &  $<$1.4\expo{-3}          &  $<$4.8\expo{-3}          &  $<$4.2\expo{-3}          &  $<$4.3\expo{-3}          & $<$6.6\expo{-3}          & $<$1.9\expo{-3}         \\
  62  &  6.80$\pm$1.21 \expo{-1}  &  5.19$\pm$3.13 \expo{-1}  &  4.85$\pm$0.59 \expo{-1}  &  5.13$\pm$0.42 \expo{-1}  & 4.55$\pm$0.14 \expo{-1}  & 8.86$\pm$2.76 \expo{-2} \\
 132  &  1.36$\pm$0.06 \expo{-1}  &       --                  &  1.36$\pm$0.03 \expo{-1}  &        --                 &        --                & 1.96$\pm$0.05 \expo{-1} \\
 133  &  $<$2.4\expo{-3}          &       --                  &  $<$2.4\expo{-3}          &       --                  &       --                 & $<$3.0\expo{-3}         \\
  57  &  1.49$\pm$0.07 \expo{-2}  &  9.33$\pm$1.53 \expo{-3}  &  1.63$\pm$0.08 \expo{-2}  &  1.41$\pm$0.08 \expo{-2}  & 1.65$\pm$0.11 \expo{-2}  & 1.34$\pm$0.12 \expo{-1} \\
  24  &       --                  &       --                  &  3.71$\pm$0.81 \expo{-2}  &       --                  &       --                 &  --                     \\
  34  &  1.35$\pm$0.39 \expo{-3}  &       --                  &  $<$5.7\expo{-4}          &  $<$1.0\expo{-3}          & 2.49$\pm$0.57 \expo{-3}  & 5.55$\pm$0.68 \expo{-3} \\
 137  &  $<$2.3\expo{-3}          &       --                  &  $<$1.8\expo{-3}          &       --                  &       --                 & 1.12$\pm$0.17 \expo{-2} \\
\noalign{\smallskip}\hline\noalign{\smallskip}
\end{tabular}
\end{center}
\tablefoot{
For a description of count rates and upper limits (numbers without error, \red{marked with ``$<$''}) see Table\,\ref{tab:lmc_rates}.}
\label{tab:smc_rates}
\end{table*}

\subsection{RX\,J0101.0$-$7206 = \sxp{304} = \hscat{46}}
\label{sec:304}

\sxp{304} was discovered in \rosat pointed observations as a highly variable source and was proposed to be a BeXRB \citep{1996A&A...312..919K}. Pulsations at 304\,s were discovered using \cxo, confirming its nature as a BeXRB pulsar \citep{2003ApJ...584L..79M}.

In Nov. 2019 \sxp{304} was detected in the \ero  observations 700001, 700002 and 700004 and
had decreased in flux by more than a factor of two in June 2020 (obsid 710000). 
The source was not covered during observation 700003 and was not detected during observation 700005, when it was located very close to the edge of the FoV (see Table\,\ref{tab:smc_rates}). For timing analysis we used only the two observations with best statistics while for the spectral analysis we analysed the spectra from all observations in which the source was detected.
In order to look for periodic signals in the X-ray light curves of this and other sources, Lomb-Scargle periodogram analysis \citep{1976Ap&SS..39..447L,1982ApJ...263..835S} was performed.  
To determine the spin periods more precisely, an epoch-folding technique was applied \citep{1987A&A...180..275L}.

\begin{figure}
  \begin{center}
  \resizebox{0.9\hsize}{!}{\includegraphics[clip=]{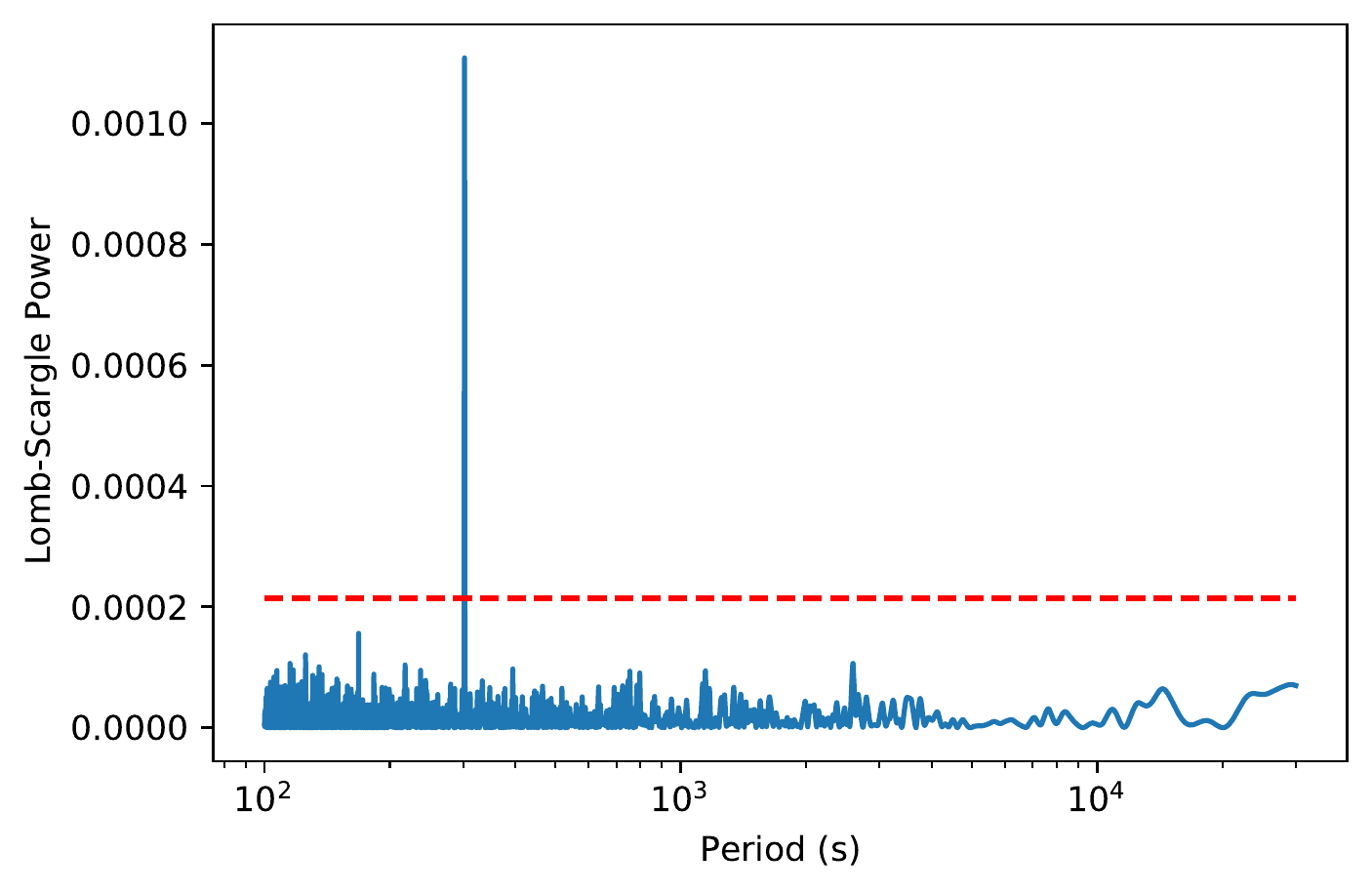}}
  \resizebox{\hsize}{!}{\includegraphics[clip=]{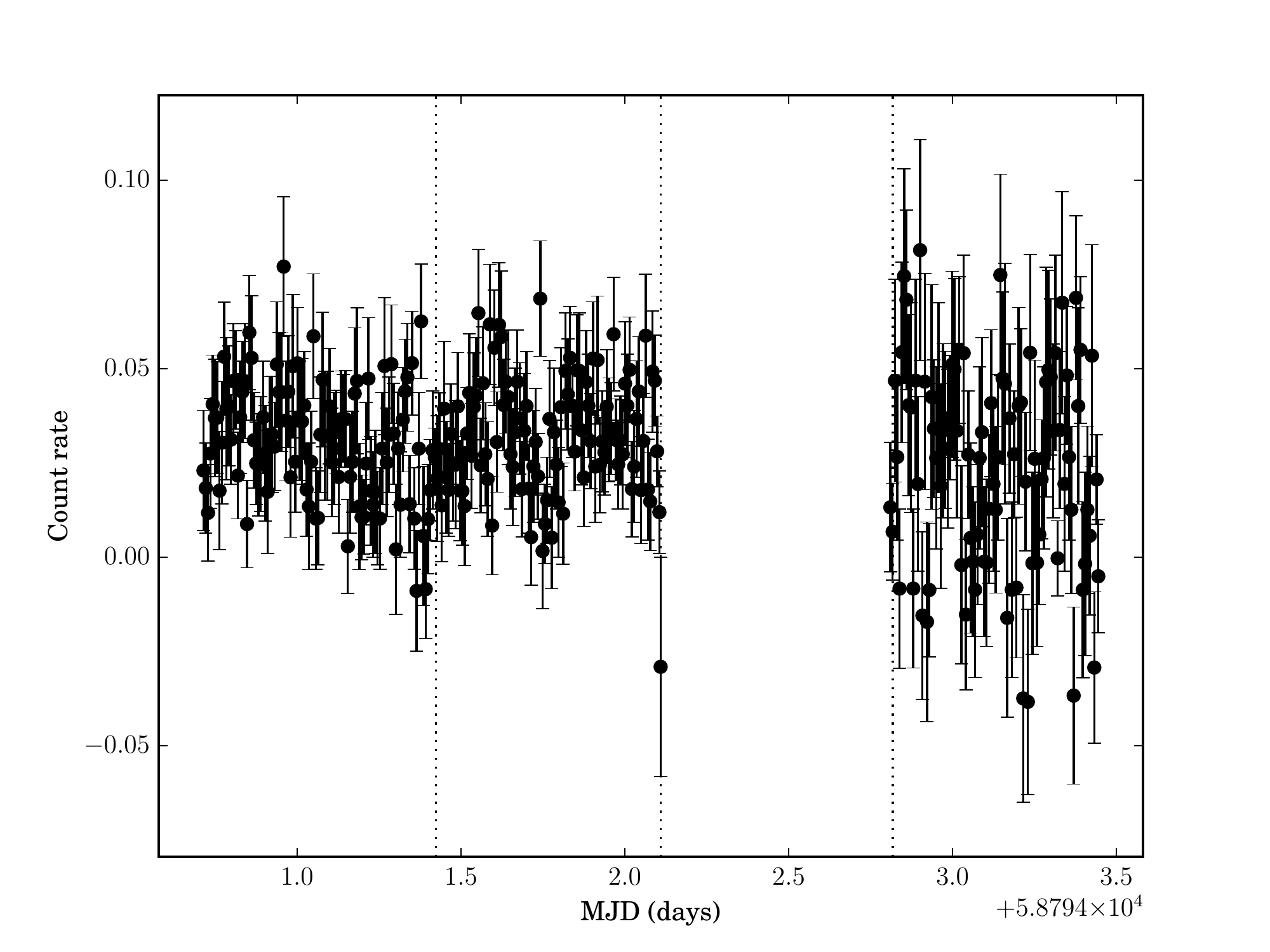}}
  \end{center}
  \caption{Top: Lomb-Scargle periodogram from the light curve of \sxp{304}, combining observations 700001 and 700002. 
           The red dashed line indicates the 99\% confidence level.
           Bottom: Background subtracted and vignetting-corrected light curve of \sxp{304} 
           (binned at twice the pulse period and including obsid 700004) in the energy band of 0.2--8\,keV. 
           The dashed vertical lines separate the intervals corresponding to observations 700001--4. \red{During observation 700003 the source was not in the FoV and during observation 700005 it was not detected.}}
  \label{figperiodsxp304lc}
\end{figure}

\begin{figure}
  \begin{center}
  \resizebox{0.8\hsize}{!}{\includegraphics[clip=,angle=-90]{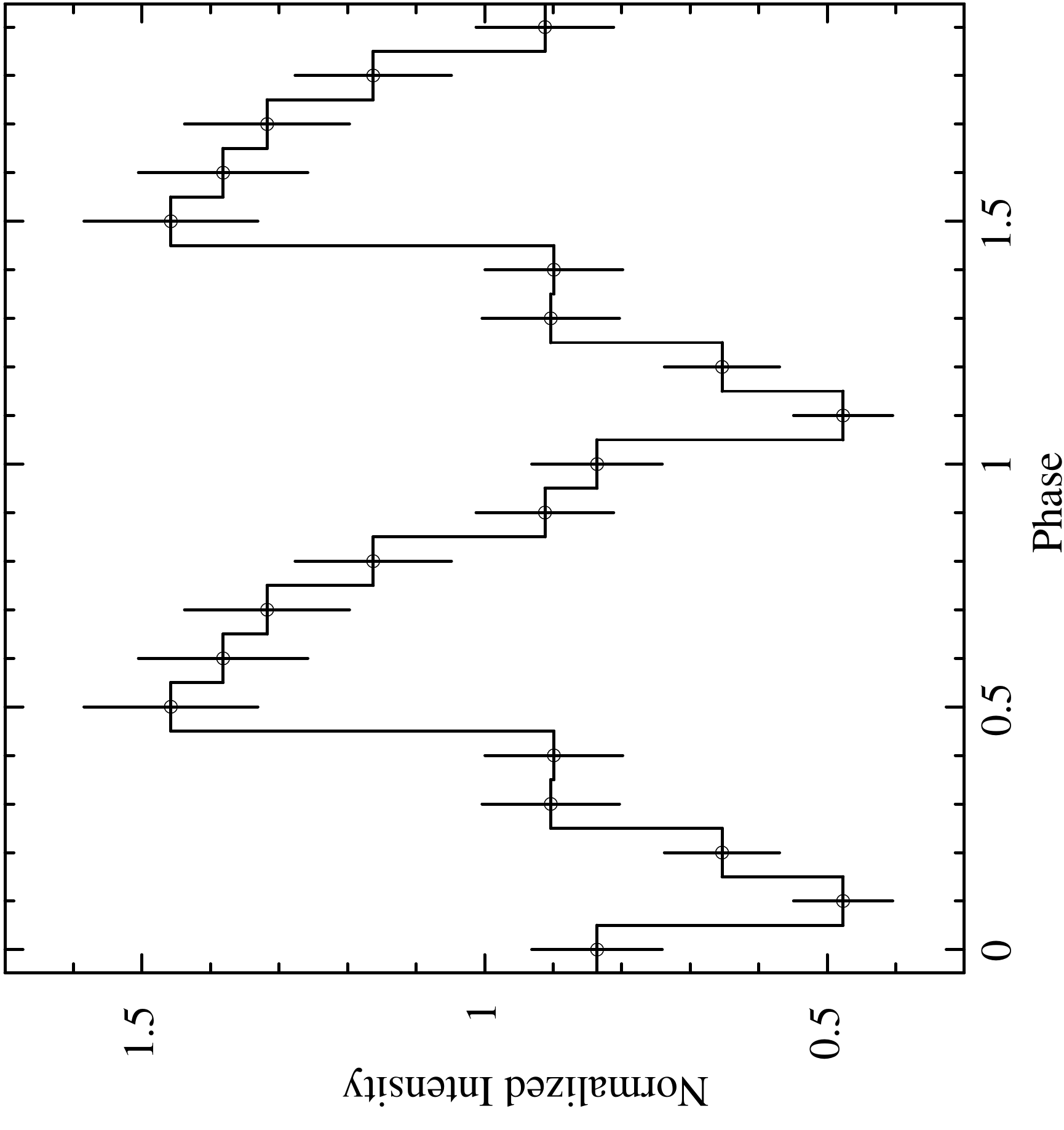}}
  \end{center}
  \caption{Background subtracted pulse profile of \sxp{304} combined over 
           observations 700001-2 in the energy range of 0.2--8\,keV. }
   \label{figsxp304pulse}
\end{figure}

\red{The lower panel of Fig.\,\ref{figperiodsxp304lc} shows the light curve of \sxp{304} obtained from the three observations with detection.
For the LS periodogram (upper panel of Fig.\,\ref{figperiodsxp304lc}) we used only data from the first two observations with better statistical quality. A} strong periodic signal at $\sim$302\,s is  visible. From the epoch-folding technique the spin period and corresponding 1$\sigma$ errors for Obsids 700001 and 700002 and combining the two were determined to  $302.21\pm0.55$, $302.33\pm 0.60$ s and $302.29\pm0.27$, respectively, and are consistent with each other. 
This period is consistent with the shortest value of $302.6\pm0.4$\,s measured from \xmm data in Nov./Dec. 2005 \citep{2008A&A...491..841E}, demonstrating that the pulsar shows little long-term period change on a time scale of 15 years.
The pulse profiles are single peaked and exhibit no change in shape between the \ero observations. The pulse profile in the energy range of 0.2--10\,keV combining both observations is shown in Fig.\,\ref{figsxp304pulse}, with a pulsed fraction of $51\pm11$\%.
 
\begin{figure}
  \begin{center}
  \resizebox{0.9\hsize}{!}{\includegraphics[clip=,angle=-90]{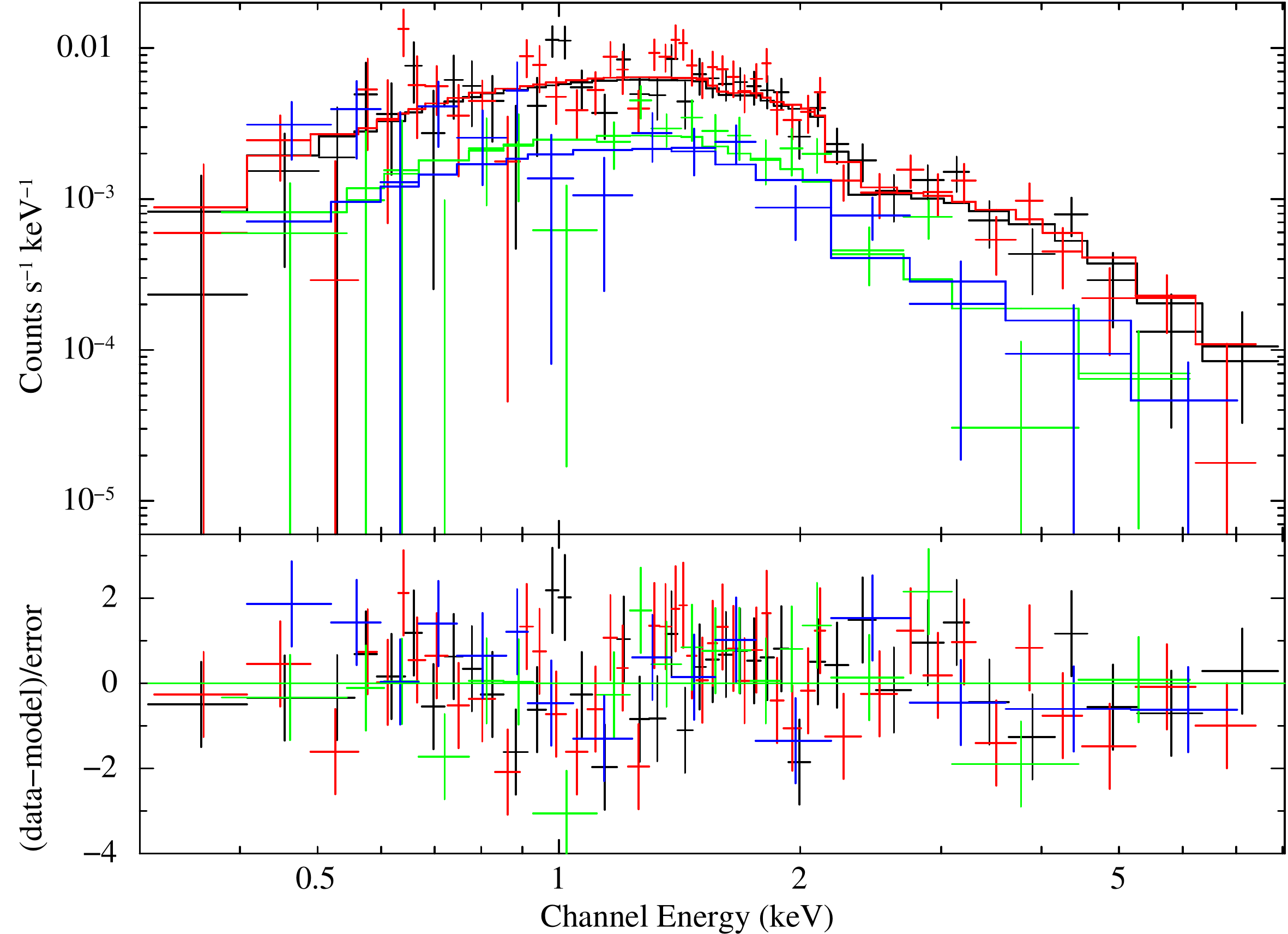}}
  \end{center}
  \caption{Simultaneous spectral fit of \sxp{304} spectra from observations 700001 (black), 700002 (red), 700004 (green) and 710000 (blue), showing the best-fit model as histogram and the residuals in the bottom panel. Each spectrum was obtained by combining the data from TM1--4 and 6.}
   \label{figsxp304spec}
\end{figure}

The spectra can be modelled with a simple absorbed power law. 
The absorption component was consistent with the Galactic foreground column density. 
No additional SMC component was required and the \nh was fixed at 6\hcm{20}.
%For this purpose we used two absorption components: one to describe the Galactic foreground absorption and another to account for the column density of both the interstellar medium of the LMC and the intrinsic absorption corresponding to the source.
%For this absorption component, the abundances were set to 0.2 solar for elements heavier than helium. For the Galactic photo-electric absorption, we used a fixed column density \citep{1990ARA&A..28..215D} with abundances taken from \cite{2000ApJ...542..914W}. 
From the simultaneous fit to the spectra of observations 700001, 700002, 700004 and 710000 a power-law index of $\Gamma=0.31\pm0.11$ was derived. 
The observed flux (0.2--10\,keV) varied from $\sim$3.6\ergcm{-13} in the first two observations to 3.2\ergcm{-13} (700004) and 1.5\ergcm{-13} (710000), 
corresponding to a luminosity between 6.5\ergs{34} and 1.6\ergs{35}. The spectra with best-fit model are shown in Fig.\,\ref{figsxp304spec}.

%%%%%%%%%%%%%%%%%%%%%%%%%%%% figures SXP304 %%%%%%%%%%%%%%%%%%%%%%%%%%%%%

%--------------------------------

\subsection{RX\,J0101.3$-$7211 = \sxp{455} = \hscat{51}}
\label{sec:455}

\sxp{455} was discovered as a BeXRB pulsar using \xmm EPIC-pn observations \citep{2001A&A...369L..29S}. 
The source was detected in the FoV of observations 700001--4 and 710000 (see  Table\,\ref{tab:smc_rates}).
Note that in the case of 700003 we did not use  the flag to select the circular FoV (as is applied to the released data) because the source was located just outside the selection mask and provides useful data for our temporal analysis.
In Nov. 2019 pulsations were detected significantly only from observations 700001 and 700002.  
The spin periods and corresponding 1$\sigma$ errors measured for the two observations using epoch-folding technique are 
\red{447.9$\pm1.0$\,s and 449.0$\pm1.9$\,s}, respectively.
These values are well within the range measured from the \rosat and \xmm observations reported by \citet{2008A&A...491..841E} and as in the case of \sxp{304} indicate little long-term spin period variations since 1993.

\begin{figure}
  \resizebox{0.95\hsize}{!}{\includegraphics[clip=]{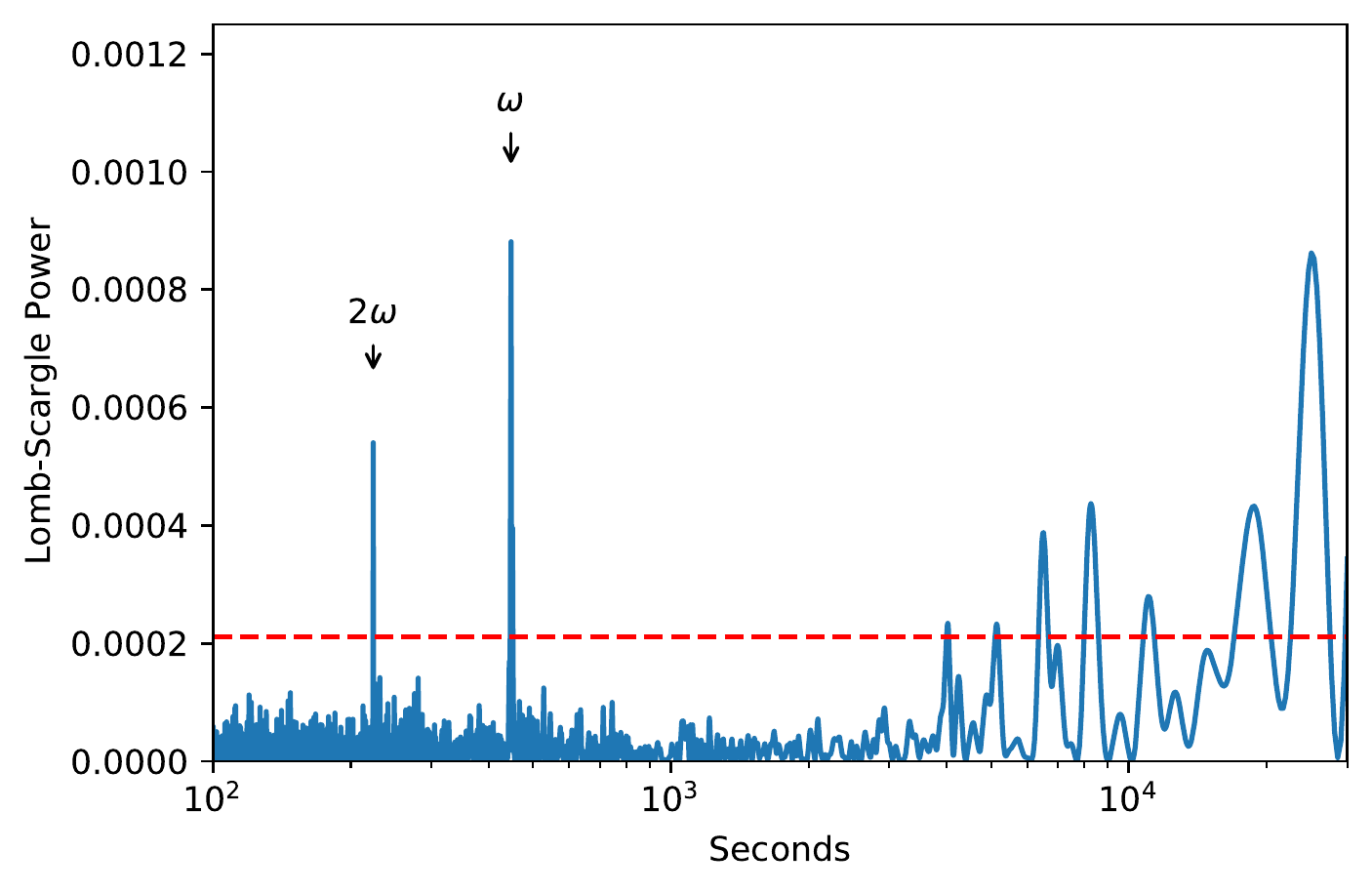}}
  \resizebox{\hsize}{!}{\includegraphics[clip=]{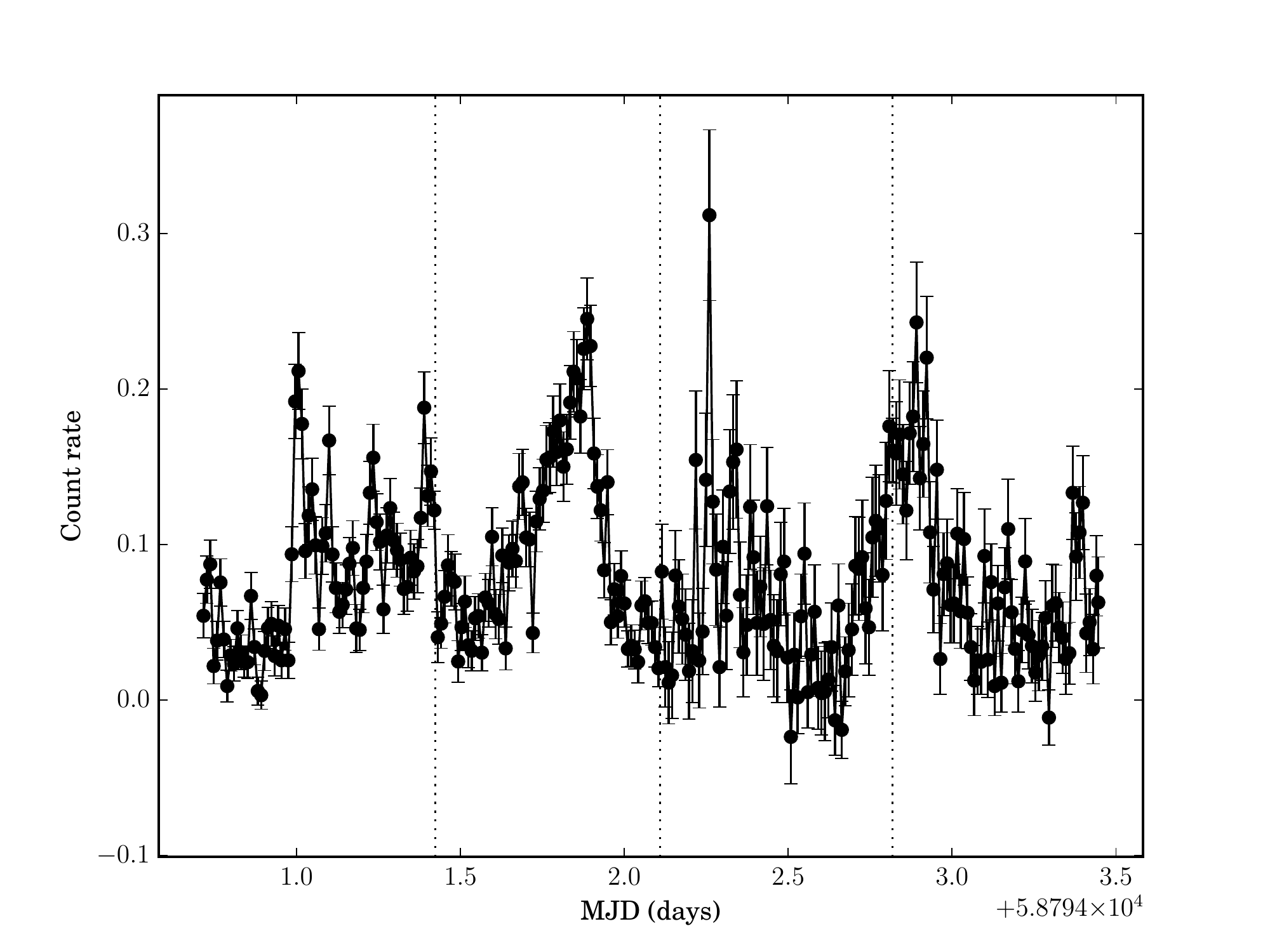}}
  \caption{Top: LS periodogram of \sxp{455} from the combined light curve of observations 700001--2.
           The red dashed line indicates the 99\% confidence level.
           Bottom: Background subtracted and vignetting-corrected light curve of \sxp{455} (binned at twice the pulse period) in the energy band of 0.2--8\,keV. 
           The dashed vertical lines separate the intervals corresponding to observations 700001--4. \red{During observation 700005 the source was not in the FoV.}}
  \label{figperiodsxp455}
\end{figure}

\begin{figure}
  \centering
  \resizebox{0.75\hsize}{!}{\includegraphics[clip=,angle=-90]{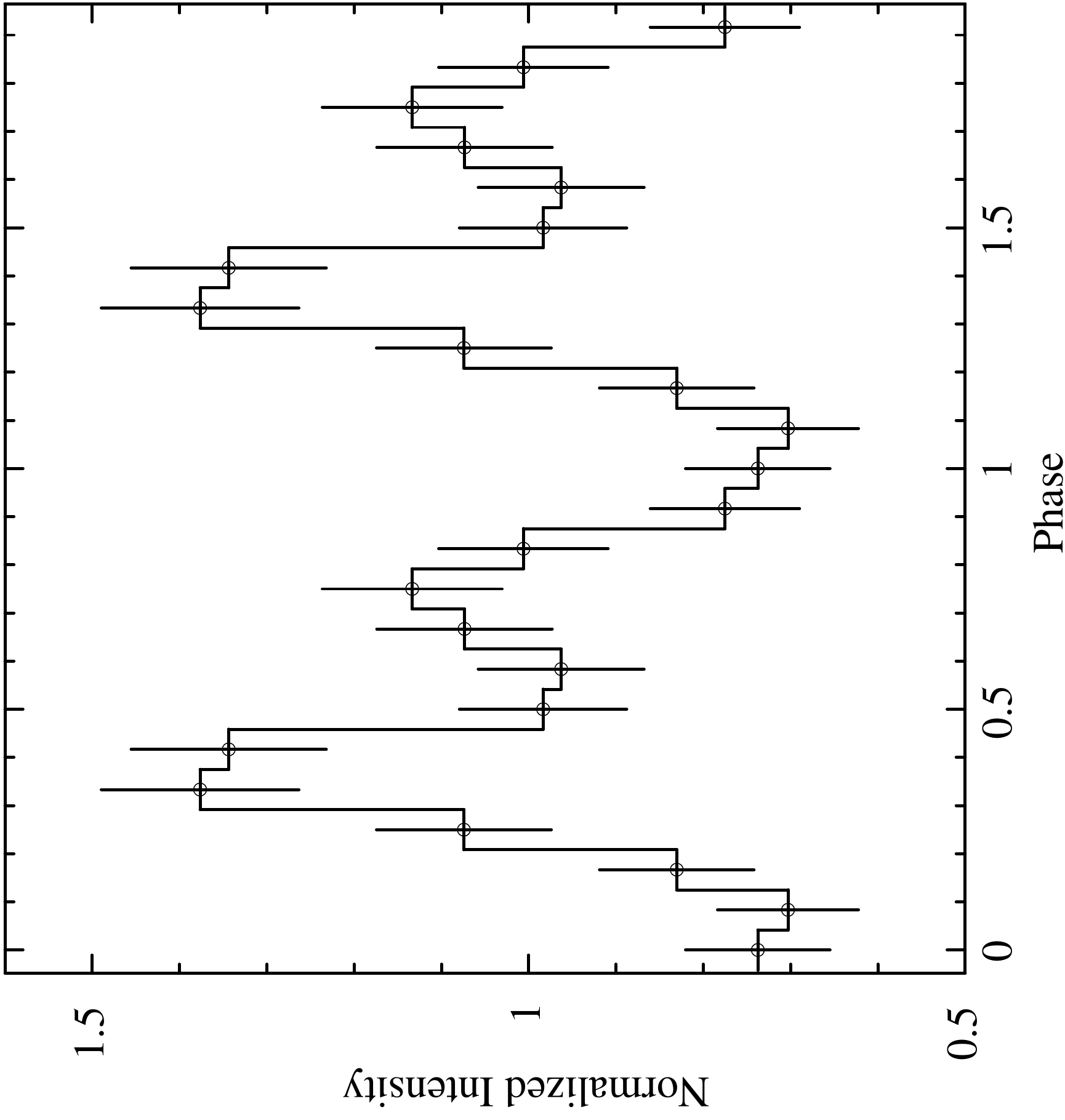}}
  \resizebox{0.75\hsize}{!}{\includegraphics[clip=,angle=-90]{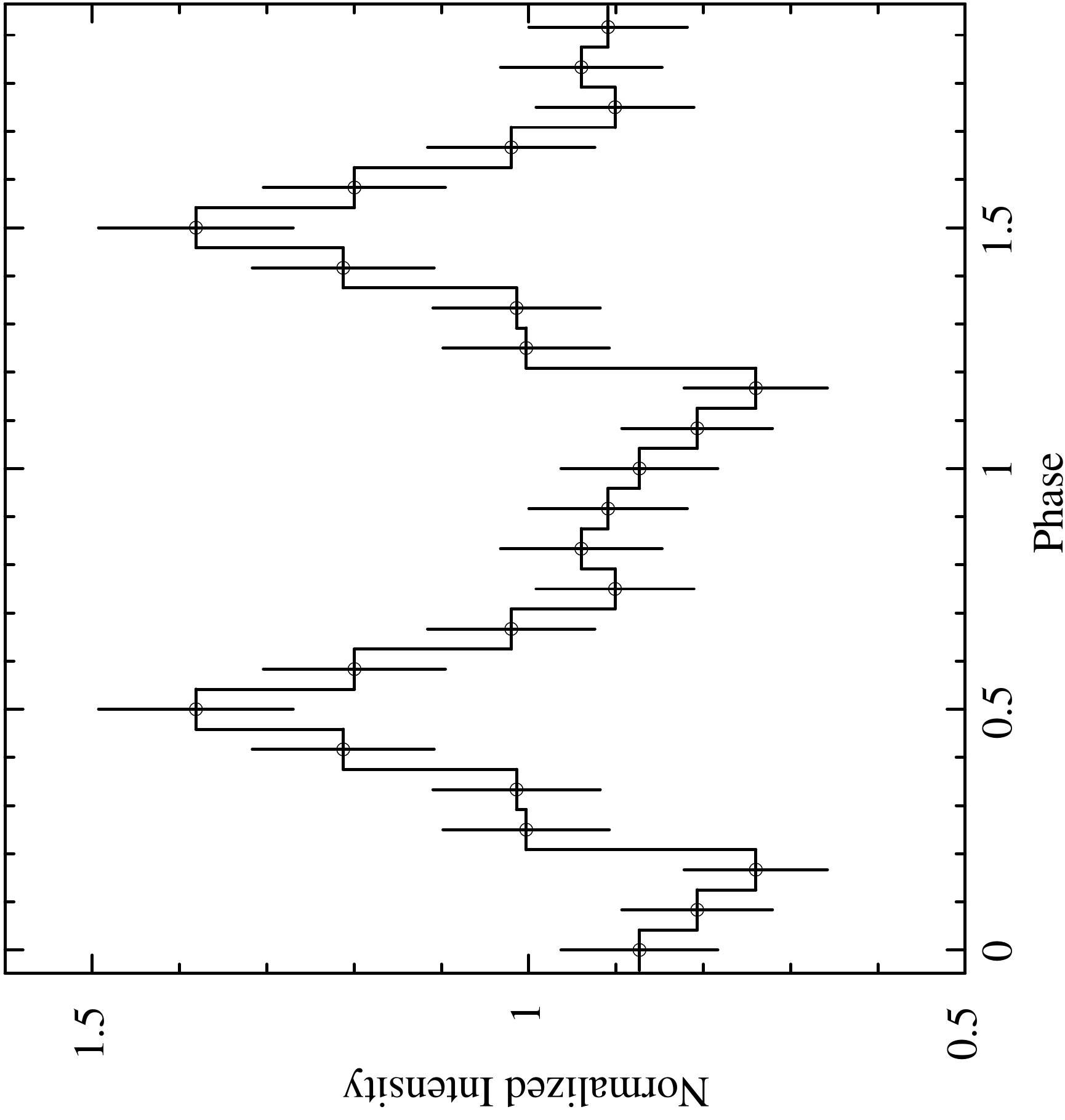}}
  \caption{Normalized pulse profiles of \sxp{455} in the energy range of 0.2--8\,keV (top: 700001 and bottom: 700002).}
   \label{figsxp455pp}
\end{figure}

The combined light curve demonstrates the flaring behaviour of the source (Fig.\,\ref{figperiodsxp455}). The pulse profiles also exhibit evidence for a change in pulse shape between observations 700001 and 700002 as shown in Fig.\,\ref{figsxp455pp}, which might be related to the flaring activity.

\begin{figure}[h!]
  \centering
  \resizebox{0.9\hsize}{!}{\includegraphics[clip=,angle=-90]{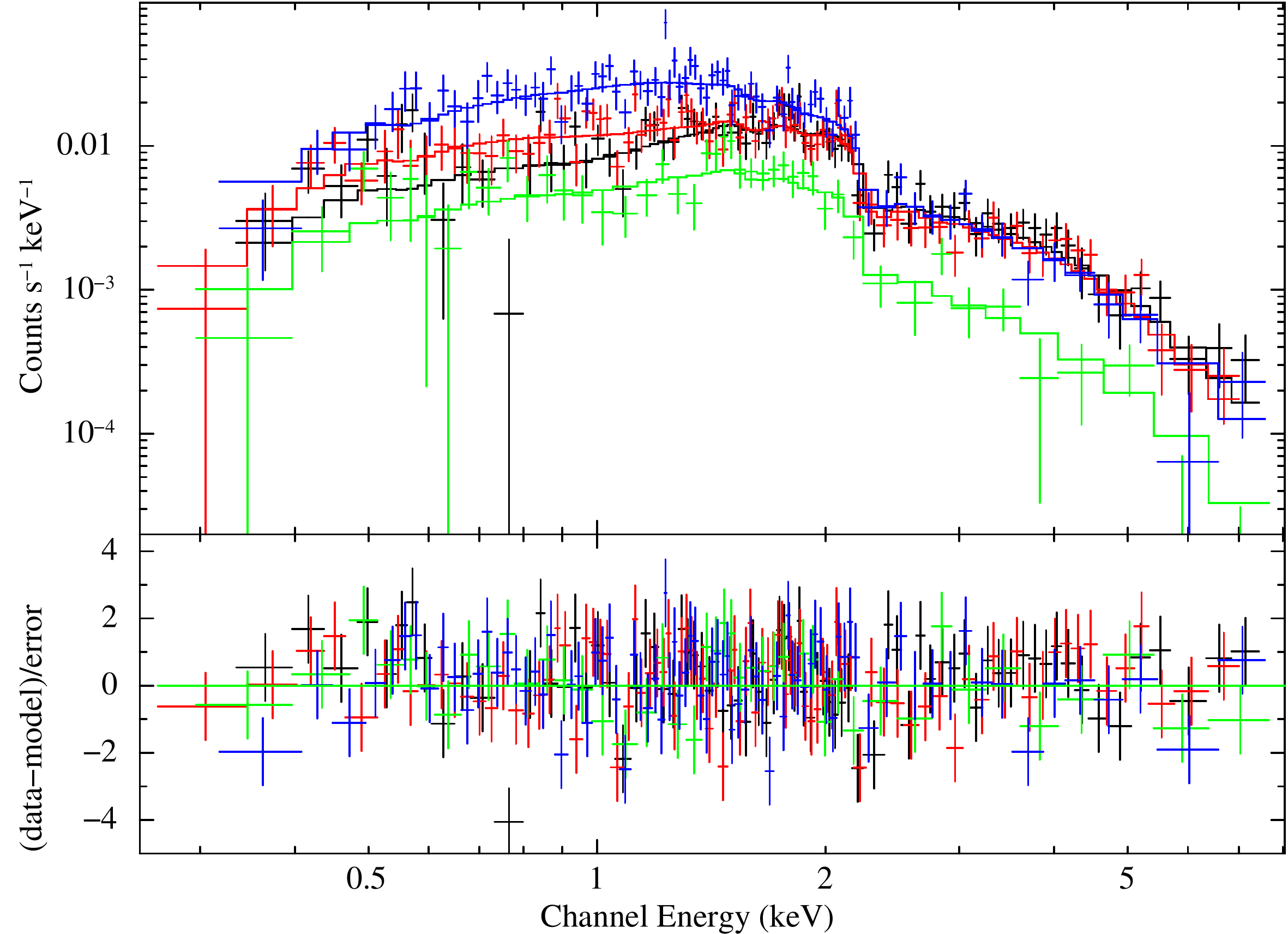}}
  \caption{Simultaneous analysis of the spectra of \sxp{455} using a power-law model with a partially covering absorber 
          (observation 700001: black, 700002: red, 700004: green, and 710000: blue).
          The best-fit model is plotted as histogram and residuals are shown in the lower panel. 
          The spectrum from each observation is obtained by combining data from TM1--4 and 6.}
   \label{figsxp455spec}
\end{figure}

Because of the much reduced statistics due to the large off-axis angle of the source, 
we excluded observation 700003 from the spectral analysis. 
The spectra from observations 700001, 700002, 700004 and 710000 were fitted simultaneously. 
A  simple absorbed power law did not provide an adequate fit to the data with an excess seen at low energies. 
Including an additional black-body component (with  $kT\sim$1.7) or alternately a partial covering  absorbing  component  improved the spectral fit. 
As the partial absorbing component provided a better statistical fit (339.7/290 dof w.r.t. 387.5/295 dof), it was used as the final model for the spectral analysis. 
For the simultaneous fitting only the power-law normalization and the parameters of the partial covering absorber were left free. 
The obtained power-law index is $0.99\pm0.16$.
The corresponding values of the partial absorber and the covering fraction (CF) for the four observations are 
\red{summarized in Table\,\ref{tab:pcmodel}, together with observed fluxes and luminosities.} 
The \ero spectra with the best-fit model are presented in Fig.\,\ref{figsxp455spec}.
As can be seen from the figure, the spectra show variations mainly at energies below $\sim$1.5\,keV, which is also reflected in the parameters of the partial covering absorber. 
In particular the spectrum from June 2020 shows a lower absorption column density and a lower covering fraction leading to the highest flux at low energies.

%--------------------------
\begin{table*}
\caption[]{\red{\sxp{455}: Results for the partial covering absorber model.}}
\begin{tabular}{lcccc}
\hline\hline\noalign{\smallskip}
\multicolumn{1}{l}{Obsid} &
\multicolumn{1}{c}{N$_{\rm H}$} &
\multicolumn{1}{c}{CF} &
\multicolumn{1}{c}{F$_{\rm observed}$\tablefootmark{a}} &
\multicolumn{1}{c}{L\tablefootmark{a}} \\
\multicolumn{1}{c}{} &
\multicolumn{1}{c}{(\ohcm{22})} &
\multicolumn{1}{c}{} &
\multicolumn{1}{c}{(\uergcm)} &
\multicolumn{1}{c}{(\uergs)} \\
\noalign{\smallskip}\hline\noalign{\smallskip}
  700001 & 2.24$\pm$0.52          & 0.85$\pm$0.05          & 8.2\expo{-13} & 4.4\expo{35} \\ 
  700002 & 2.68$^{+0.66}_{-0.57}$ & 0.76$\pm$0.07          & 9.9\expo{-13} & 4.3\expo{35} \\ 
  700004 & 2.14$^{+0.80}_{-0.62}$ & 0.77$\pm$0.09          & 8.1\expo{-13} & 3.5\expo{35} \\ 
  710000 & 0.81$^{+0.57}_{-0.50}$ & 0.51$^{+0.12}_{-0.17}$ & 8.9\expo{-13} & 3.9\expo{35} \\ 
\noalign{\smallskip}\hline
\end{tabular}
\tablefoot{
\tablefoottext{a}{0.2--10\,keV.}
}
\label{tab:pcmodel}
\end{table*}
%--------------------------

%%%%%%%%%%%%%%%%%%%%%%%%%%%%%%%%%%%%%%%%%%% figures SXP455 %%%%%%%%%%%%%%%%%%%%

%%%%%%%%%%spectra%%%%%%%%%%%%%%%%%%%%%%%%
%%%%%%%%%%%%%%%%%%%%%%%%%%%%%%%%%%%%%%%%

\subsection{Suzaku\,J0102$-$7204 = \sxp{522} = \hscat{53}}
\label{sec:522}
\begin{figure}
  \centering
  \resizebox{\hsize}{!}{\includegraphics[clip=]{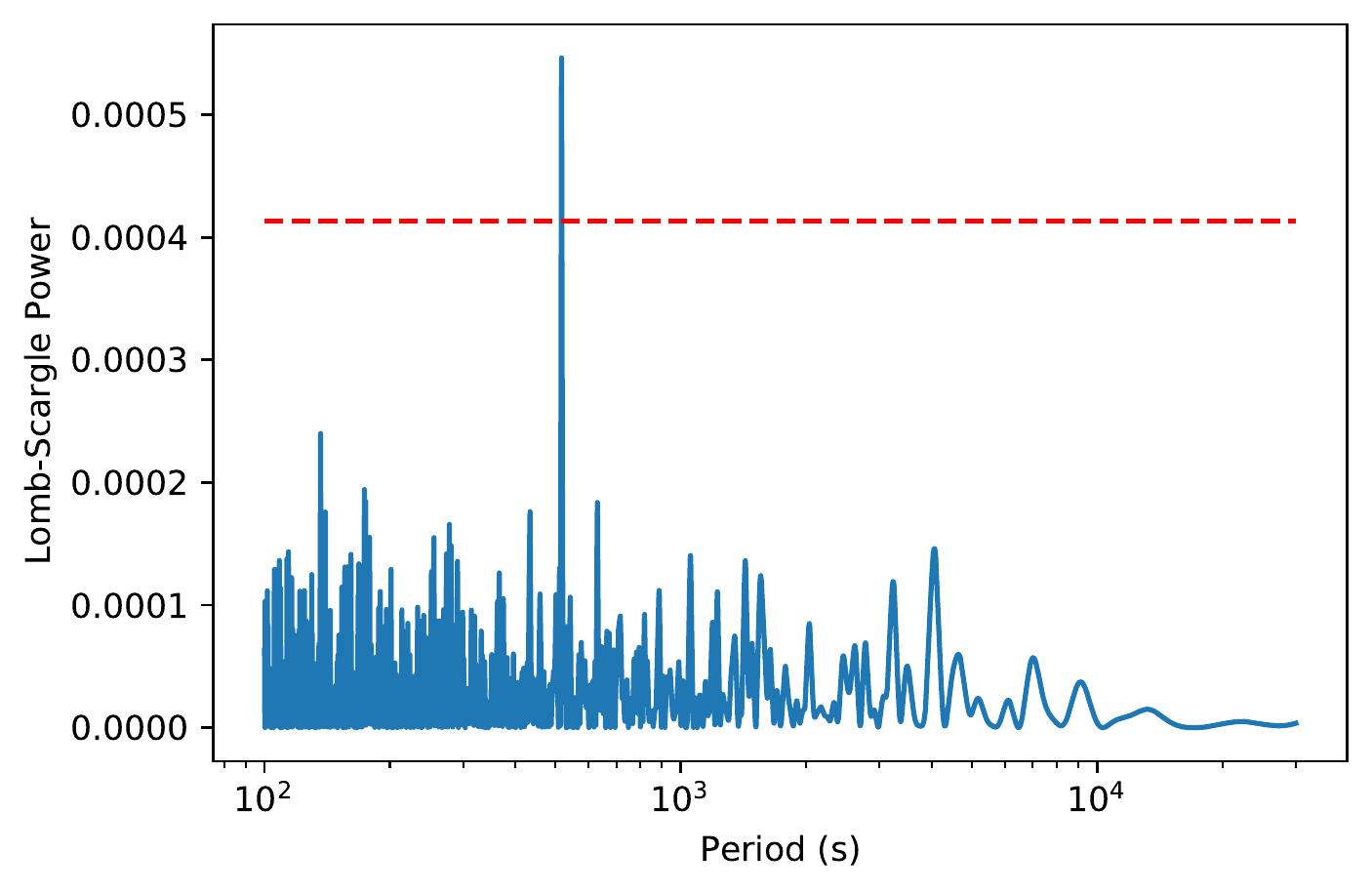}}
  \resizebox{1.1\hsize}{!}{\includegraphics[clip=]{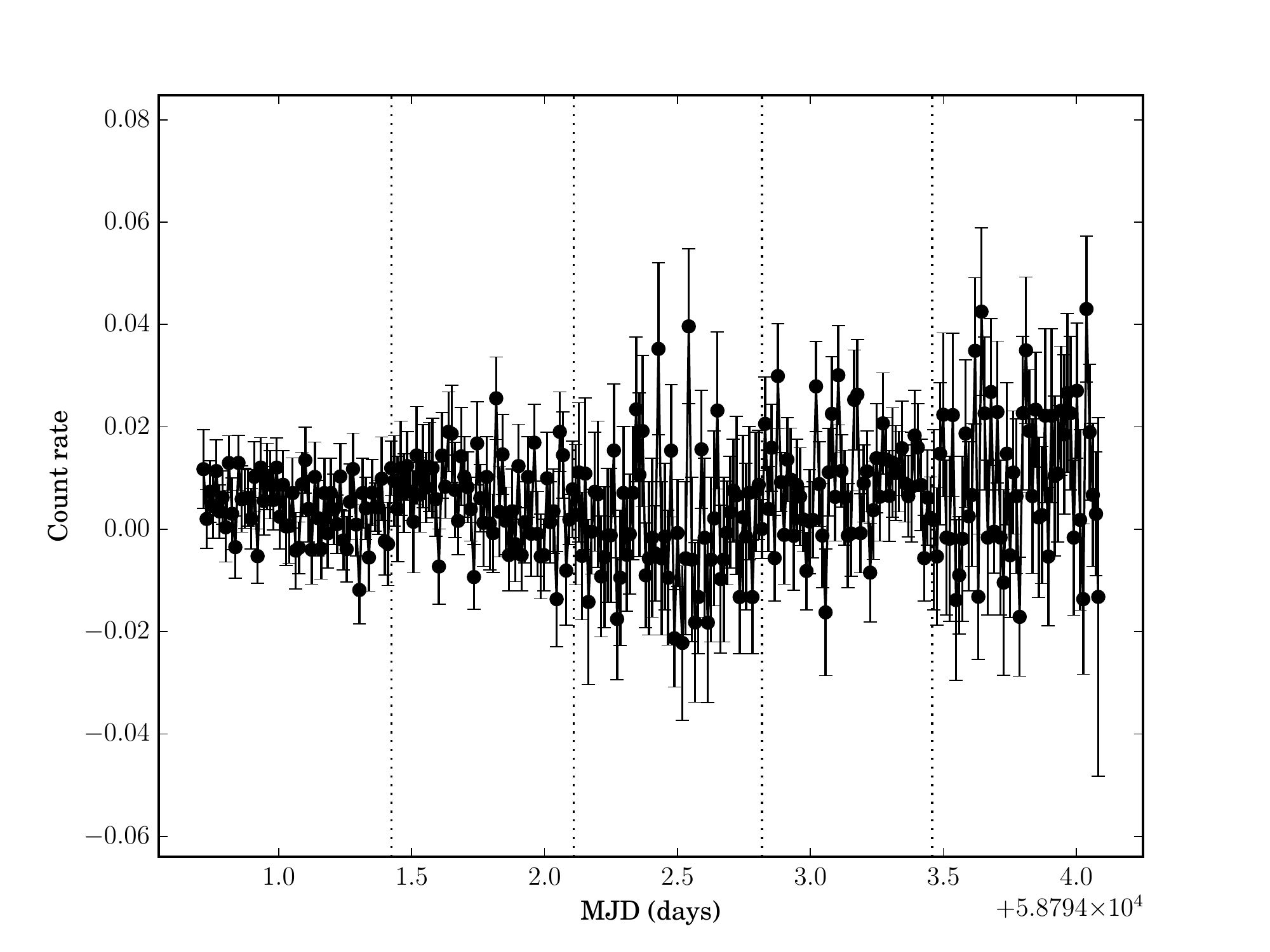}}
  \caption{Top: LS periodogram from Obsid 700001 for \sxp{522}.
           The red dashed line indicates the 99\% confidence level.
           Bottom: Background subtracted and vignetting-corrected light curve of \sxp{522} (binned at twice the pulse period) in the energy band of 0.2--8\,keV for the five Nov. 2019 observations.}
   \label{figperiodsxp522lc}
\end{figure}

\begin{figure}
  \begin{center}
  \resizebox{0.8\hsize}{!}{\includegraphics[clip=,angle=-90]{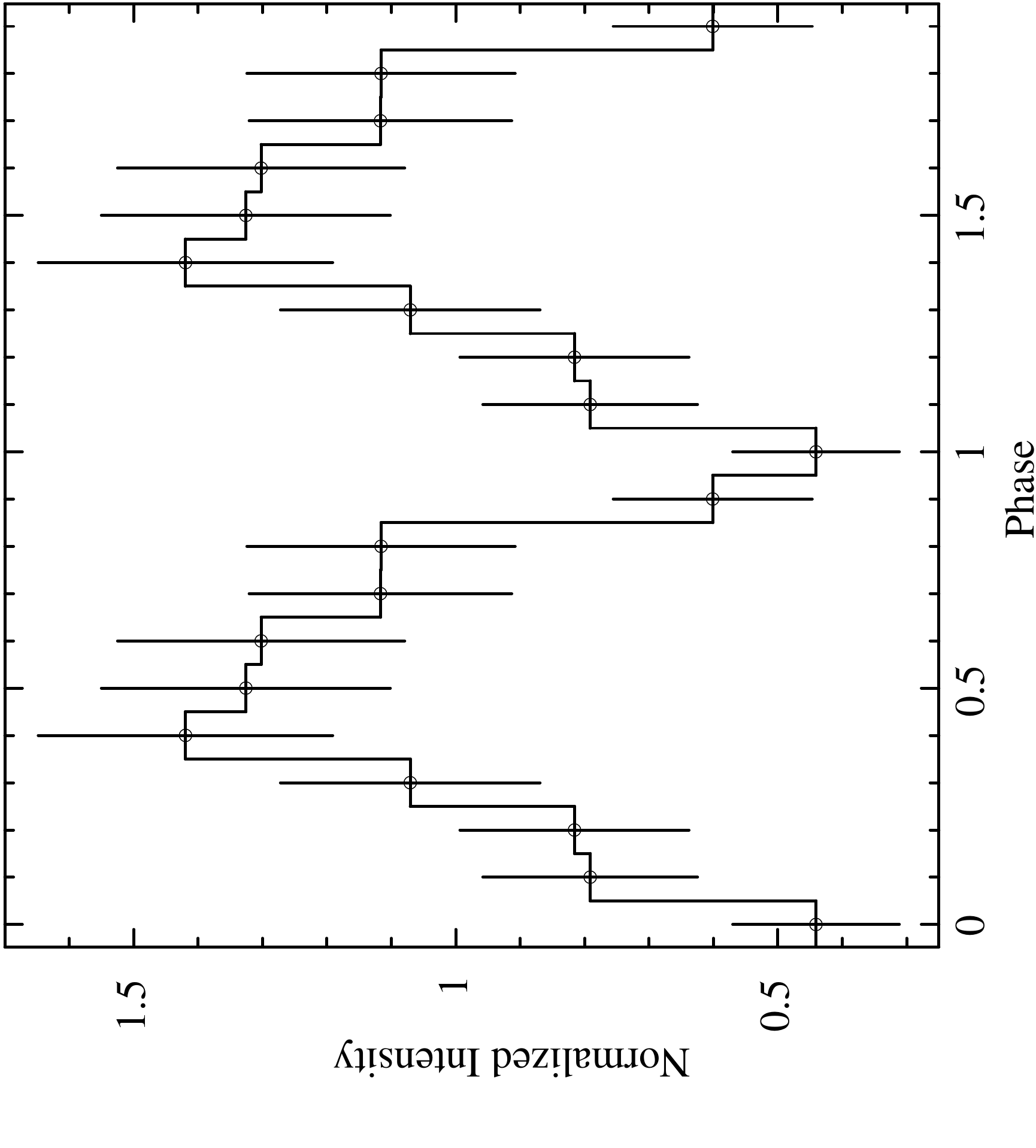}}
  \end{center}
  \caption{Background subtracted pulse profile of \sxp{522} in the energy range of 0.2--8\,keV.}
   \label{figsxp522pulse}
\end{figure}

\sxp{522} was detected in the Nov. 2019 observations 700001--5 with increasing flux during the last two observations while it was again weaker in June 2020 (see Table\,\ref{tab:smc_rates}).
The source was fainter than the other BeXRB pulsars studied in this work, and only data from observation 700001 (source was detected at smallest off-axis angle) were suitable for performing timing analysis. Pulsations are detected with a peak at $\sim$517\,s in the Lomb-Scargle periodogram as shown in Fig.\,\ref{figperiodsxp522lc}. 
The precise spin period with 1$\sigma$ error is $516.8\pm2.0$\,s. The profile exhibits a simple sinusoidal profile (Fig.\,\ref{figsxp522pulse}).
The measured spin period indicates that \sxp{522} experienced an episode of slight long-term spin-up since the discovery of the pulsations in Oct. 2012 \citep[522.5$\pm$0.5\,s;][]{2012ATel.4628....1W,2013PASJ...65L...2W} and an \xmm detection in Dec. 2012 \citep[521.42$\pm$0.21\,s;][]{2013ATel.4719....1S}.
To our knowledge, the new \ero spin measurement is the first reported since 2012.

\begin{figure}
  \begin{center}
  \resizebox{0.9\hsize}{!}{\includegraphics[angle=-90]{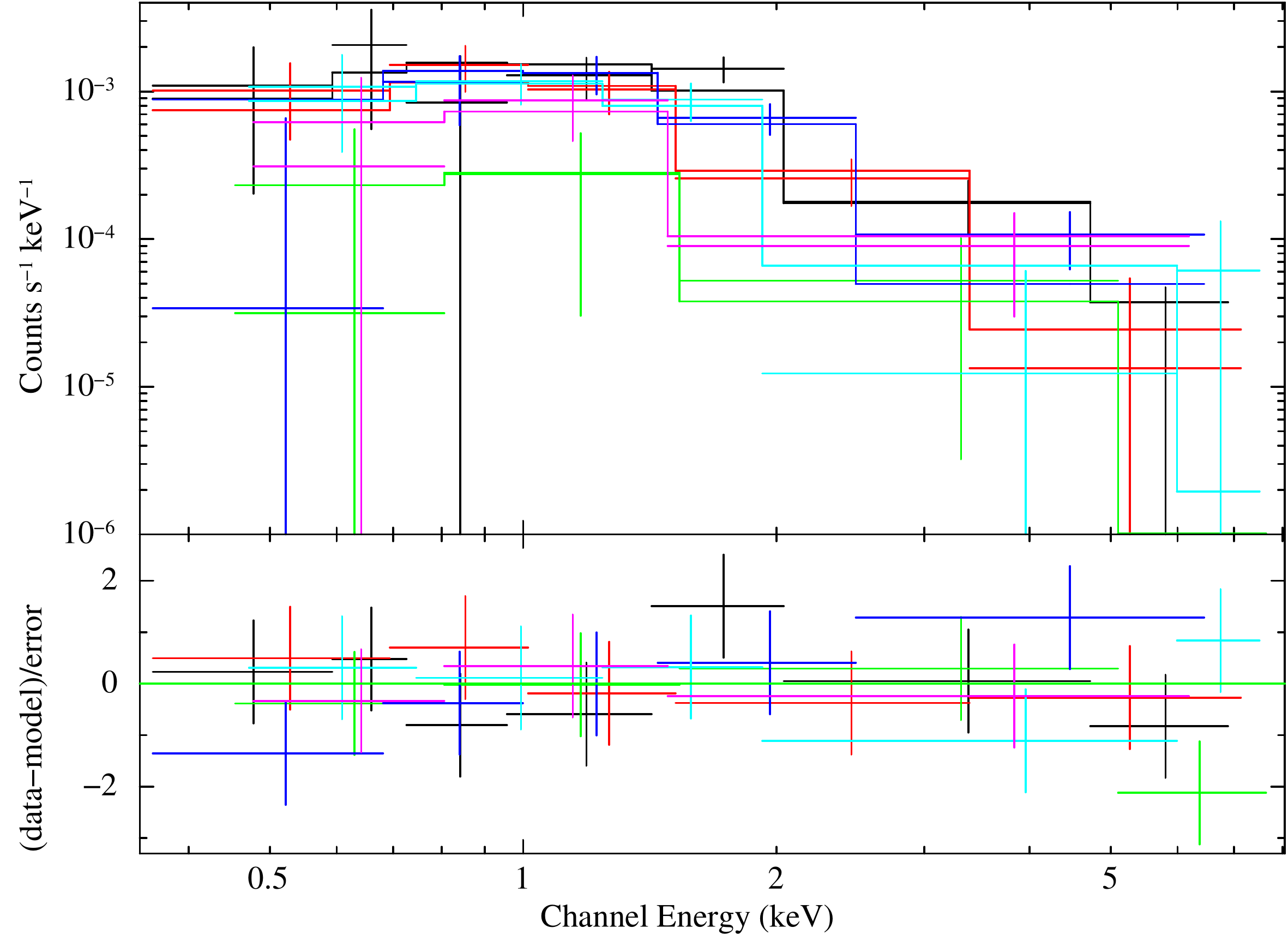}}
  \end{center}
  \caption{\ero spectra of \sxp{522} obtained from observations 700001--5 marked in black, red, green, blue, and cyan, respectively,
           and 710000 (magenta). Each spectrum includes data from TM1--4 and 6.
           The histograms show the best-fit model and the residuals are drawn in the lower panel.}
   \label{figsxp522spec}
\end{figure}

The six spectra from all \ero observations can be modelled with a simple absorbed power law. The absorption column density was consistent with the Galactic foreground absorption and fixed at 6\hcm{20}. The power-law index derived from a simultaneous fit is $0.93\pm0.35$. 
\red{The low statistical quality of the spectra does not allow us to draw firm conclusions about variability between the observations. 
The observed average fluxes (0.2--10\,keV) of (2.7$^{+1.5}_{-0.9}$, 3.1$^{+1.8}_{-1.3}$, 1.4$^{+1.7}_{-1.4}$, 4.1$^{+2.4}_{-1.6}$, 5.4$^{+3.7}_{-2.6}$, and 1.4$^{+1.0}_{-0.8}$) \ergcm{-14} for the six observations are consistent within the errors and correspond to luminosities} of (1.2, 1.4, 0.63, 1.8, 2.4, and  0.63) \ergs{34}. The spectra with the best-fit model are shown in Fig.\,\ref{figsxp522spec}.

%%%%%%%%%%%%%%%%%%%%%%%%%%%%%%%%%%%%%%%%%%%%%%%%%%sxp 522%%%%%%%%%%%%%%%%%%%%

%%%%%%%%%%%%%%%%%%%%%%%%%%%%%%%%%%%%%%%%%%%%%%%%

\subsection{RX\,J0105.9$-$7203 = \sxp{726} = \hscat{57}}

RX\,J0105.9$-$7203 was proposed as BeXRB by \citet{2000A&A...359..573H} and finally confirmed 
by the discovery of pulsations by \citet{2008A&A...485..807E} of $\sim$726\,s.
Seven detections of the period in \xmm observations between Apr. 2001 and Nov. 2003 
indicated a long-term spin-down rate of $\sim$3.2\,s\,yr$^{-1}$, although of low significance
due to large uncertainties in the individual period measurements.

\begin{figure}
  \begin{center}
  \resizebox{0.95\hsize}{!}{\includegraphics[]{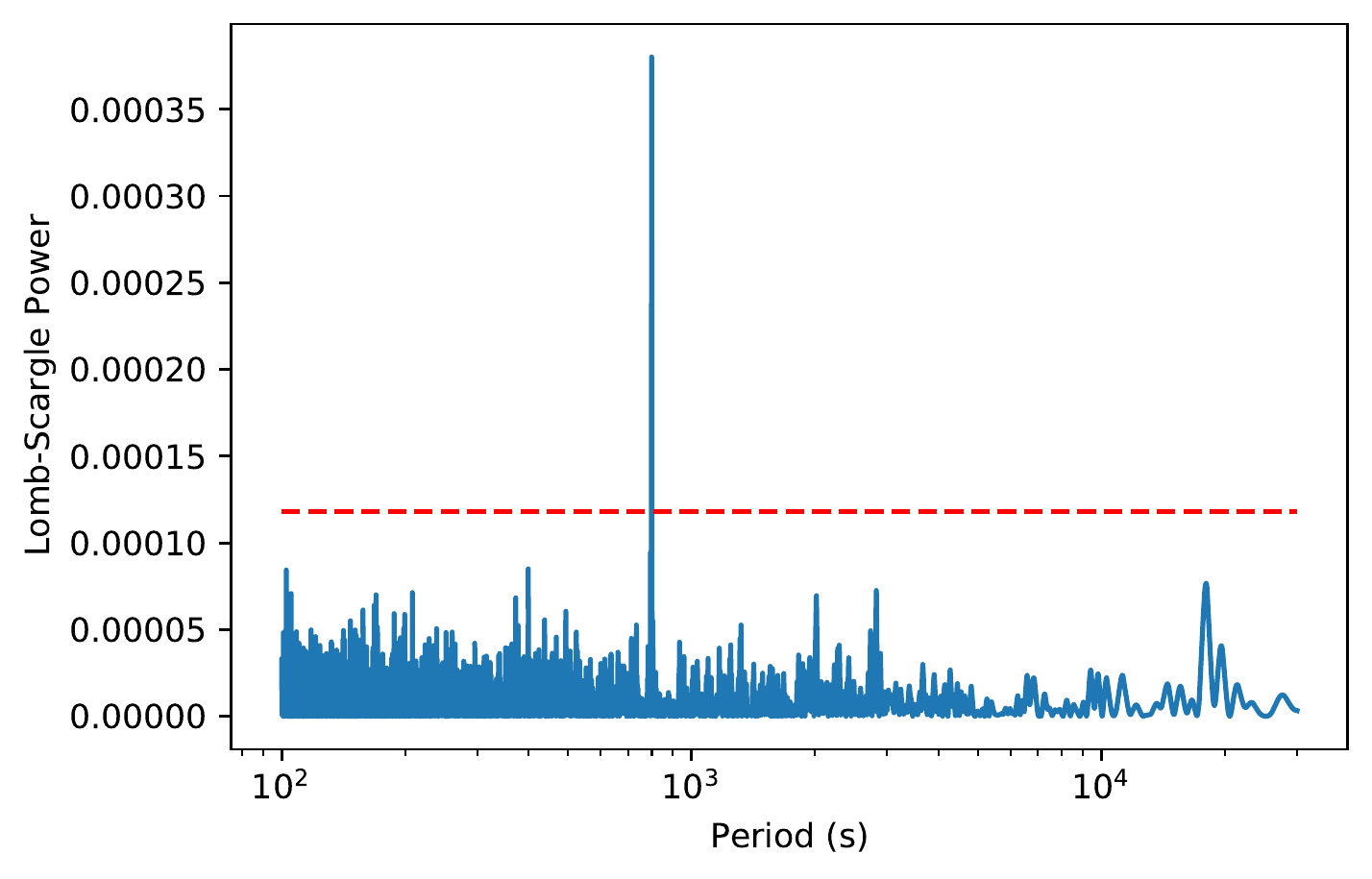}}
  \resizebox{0.95\hsize}{!}{\includegraphics[]{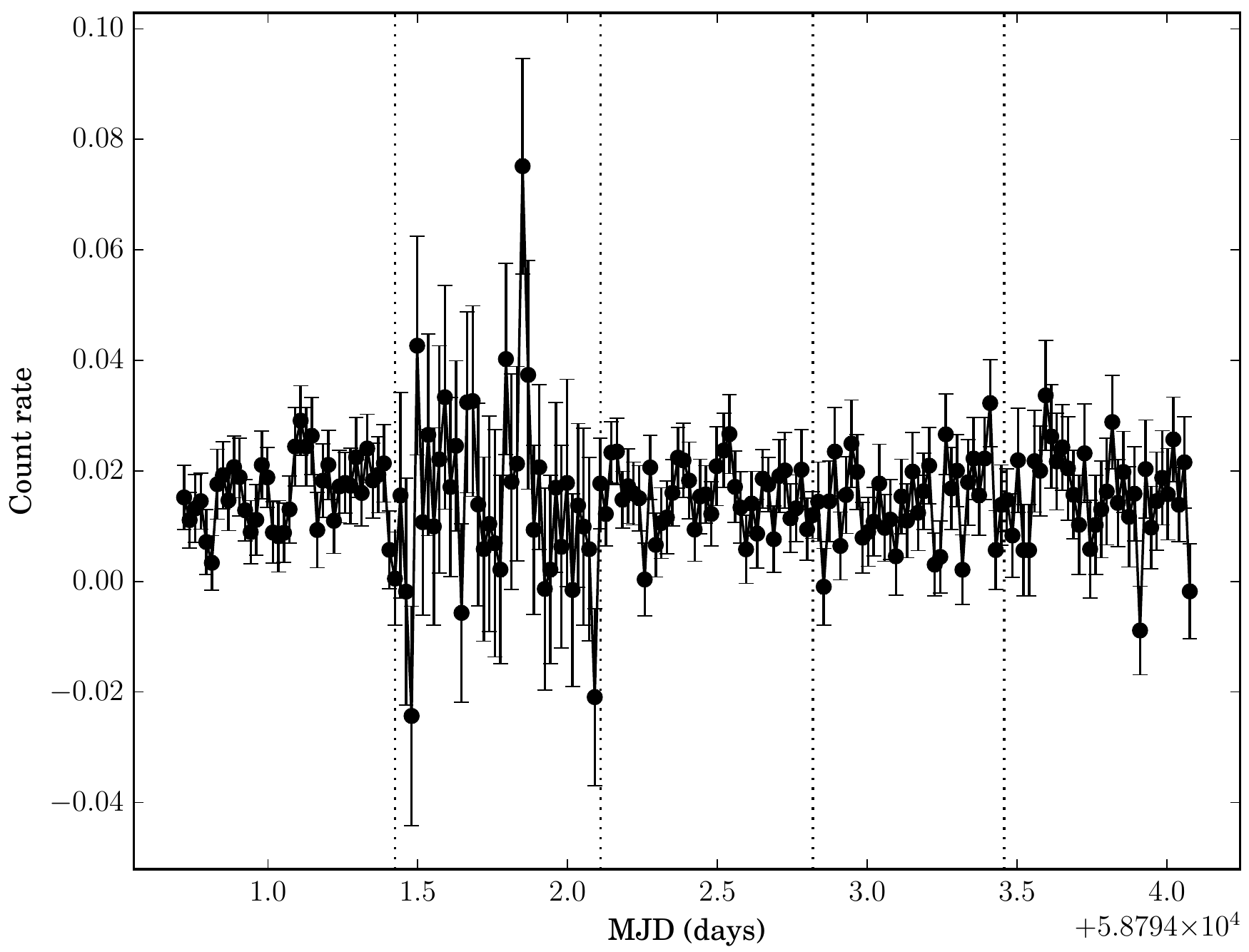}}
  \end{center}
  \caption{Top: LS periodogram obtained from the combined light curve of \sxp{726} shown below.
           The red dashed line indicates the 99\% confidence level.
           Bottom: Background subtracted and vignetting-corrected light curve of \sxp{726} 
           (binned at twice the pulse period) in the energy band of  0.2--8\,keV.}
   \label{figperiodsxp726lc}
\end{figure}

\sxp{726} was detected during the Nov. 2019 observations 700001--5 at a similar flux level, while half a year later it was a factor of $\sim$7 brighter (see Table\,\ref{tab:smc_rates}).
Pulsations are detected from all observations except 700002 where the source was located at a very large off-axis angle. 
The LS periodogram in Fig.\,\ref{figperiodsxp726lc} shows a strong periodic signal with the fundamental at $\sim$799\,s, demonstrating that the pulsar has exhibited a significant spin-down by nearly 80\,s since the time of the discovery of the period. 
The light curve exhibits a nearly constant count rate (Fig.\,\ref{figperiodsxp726lc}). 
The precise spin periods with 1$\sigma$ error from the Nov. 2019 observations 700001, 700003, 700004, 700005 and from the merged data from these observations are $798.8\pm4.0$, $800.0\pm3.8$, $797.6\pm6.6$, $801.5\pm4.4$ and $799.9\pm3.2$\,s respectively. During observation 710000 in June 2020, the period had further increased to $803.0\pm5.5$\,s.

\begin{figure}
  \begin{center}
     \resizebox{0.8\hsize}{!}{\includegraphics[angle=-90]{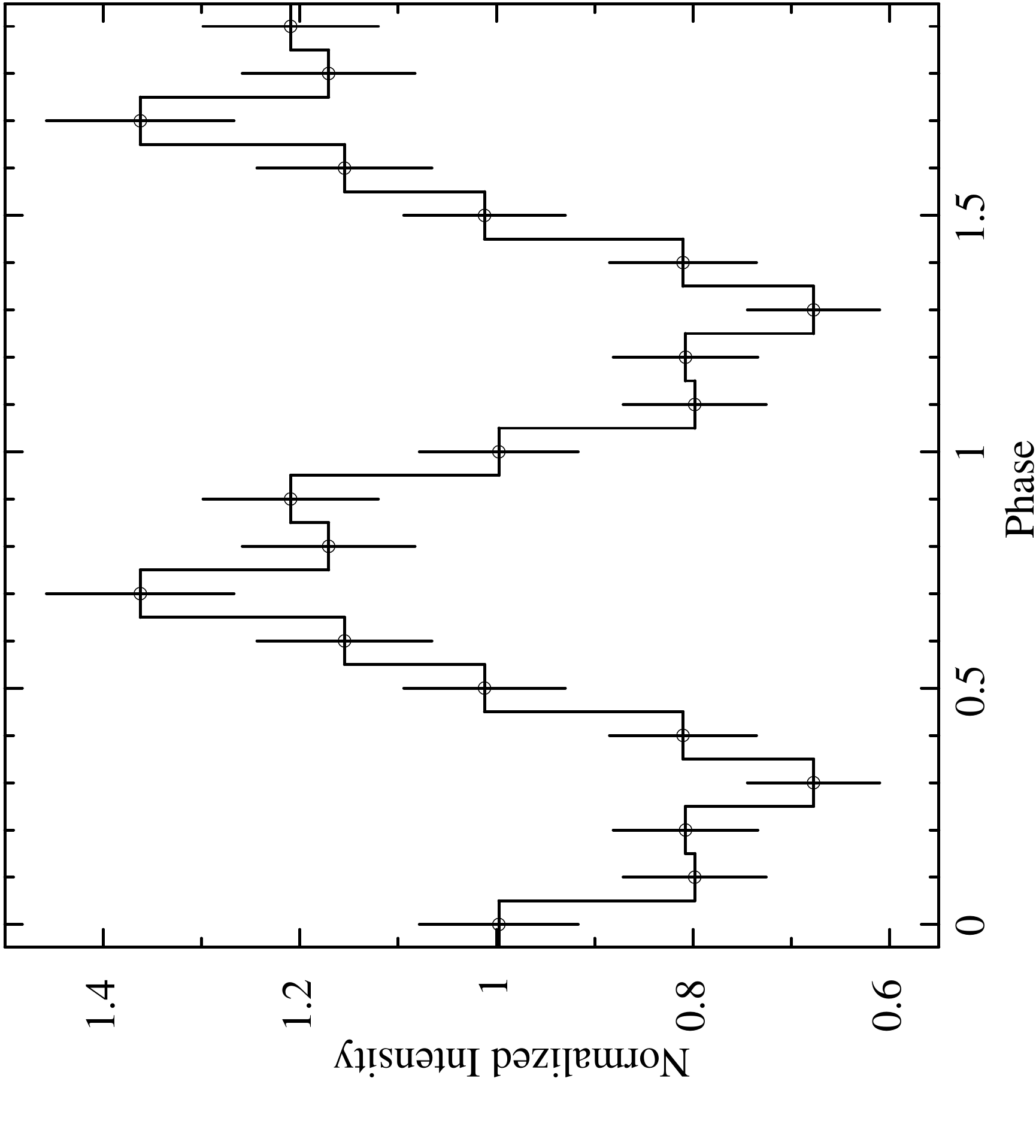}}
     \resizebox{0.8\hsize}{!}{\includegraphics[angle=-90]{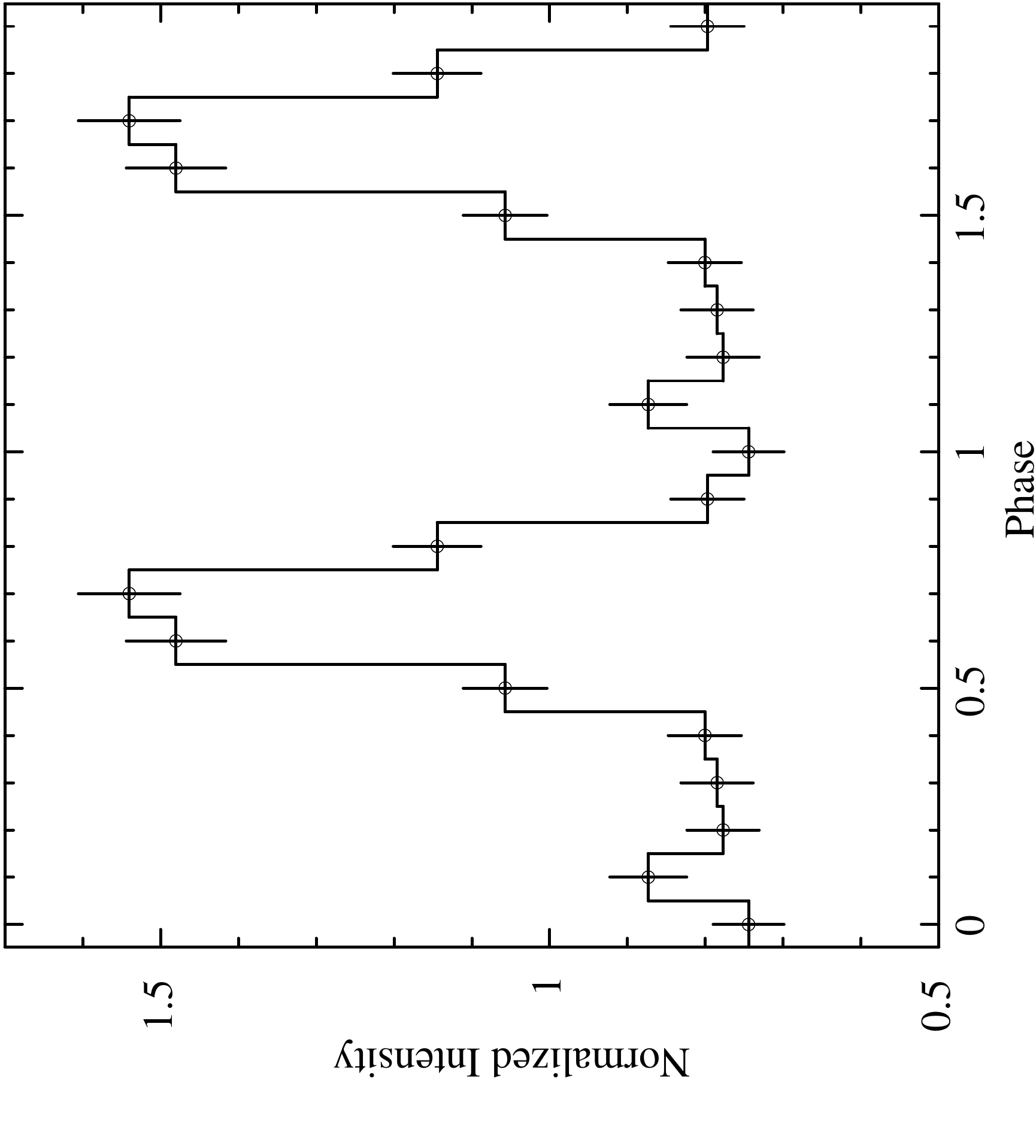}}
  \end{center}
  \caption{Background subtracted pulse profile of \sxp{726} in the energy range of 0.2--8\,keV
           averaged over all the observations in which pulsations were detected (top) and for Obsid 710000 (bottom).}
  \label{figsxp726pulse}
\end{figure}

No significant change in the spin period or the pulse shape is detected between the Nov. 2019 observations and the combined pulse profile is shown in Fig.\,\ref{figsxp726pulse} (top). 
The average profile is single peaked and the pulsed fraction (0.2--8\,keV) is $30\pm8$\%. The pulse profile from observation 710000 shows a narrower peak with a pulsed fraction of $35\pm5$\% (Fig.\,\ref{figsxp726pulse}, bottom). 

%%%%%%%%%%%%%%%%%%%%%%%%%%%%%%%%%%%%%%%% figures SXP726 %%%%%%%%%%%%%%%%%%%%

 \begin{figure}
     \centering
     \includegraphics[width=\columnwidth]{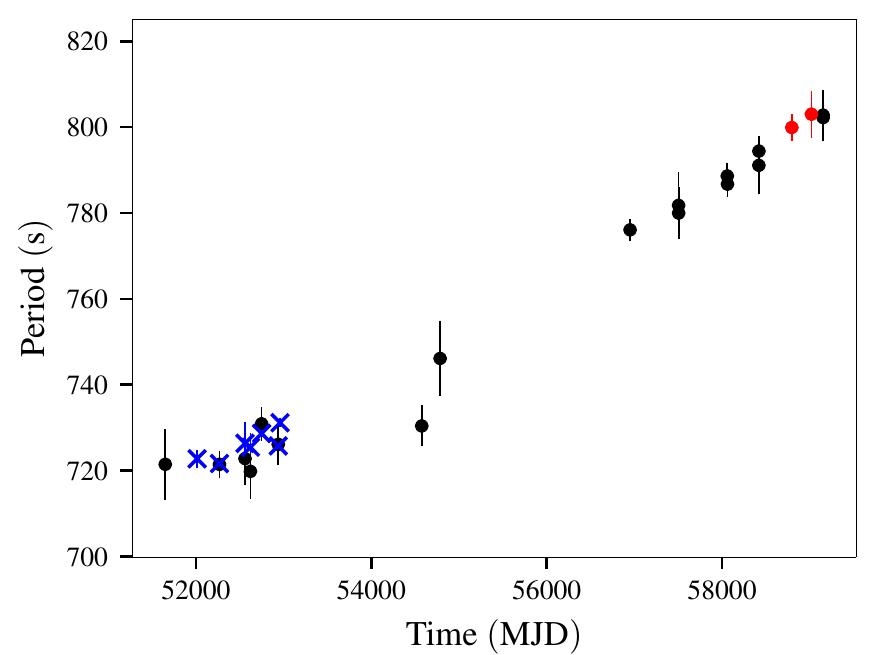}
     \caption{Spin period evolution of \sxp{726} measured in this work using the \xmm (black) and \ero (red) data. The values reported by \cite{2008A&A...485..807E} for some of the \xmm observations are also plotted for reference (blue crosses).}
     \label{fig:sxp726xmmpspin}
 \end{figure}

\begin{figure}
  \begin{center}
     \resizebox{0.9\hsize}{!}{\includegraphics[angle=-90]{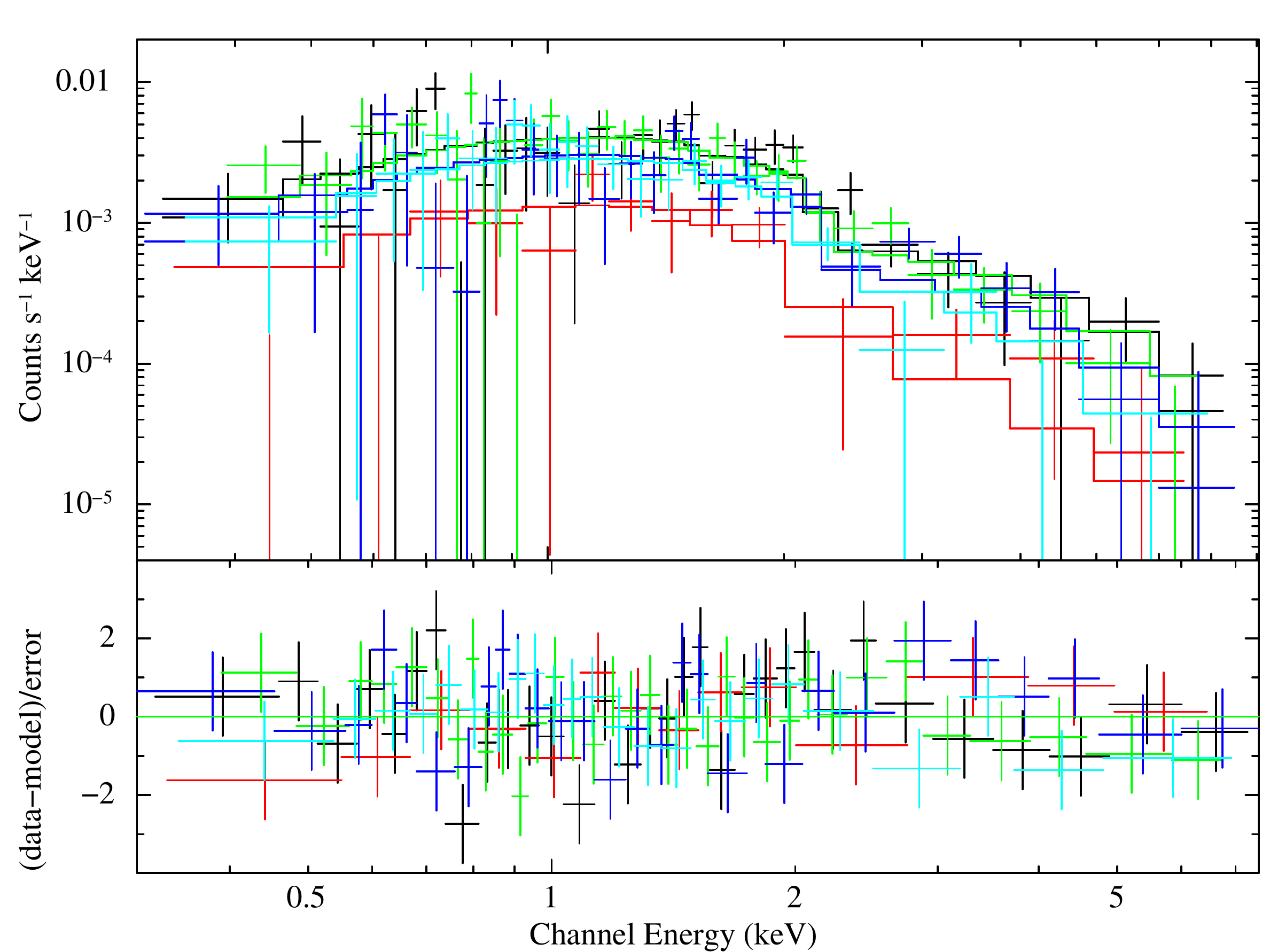}}
     \resizebox{0.9\hsize}{!}{\includegraphics[angle=-90]{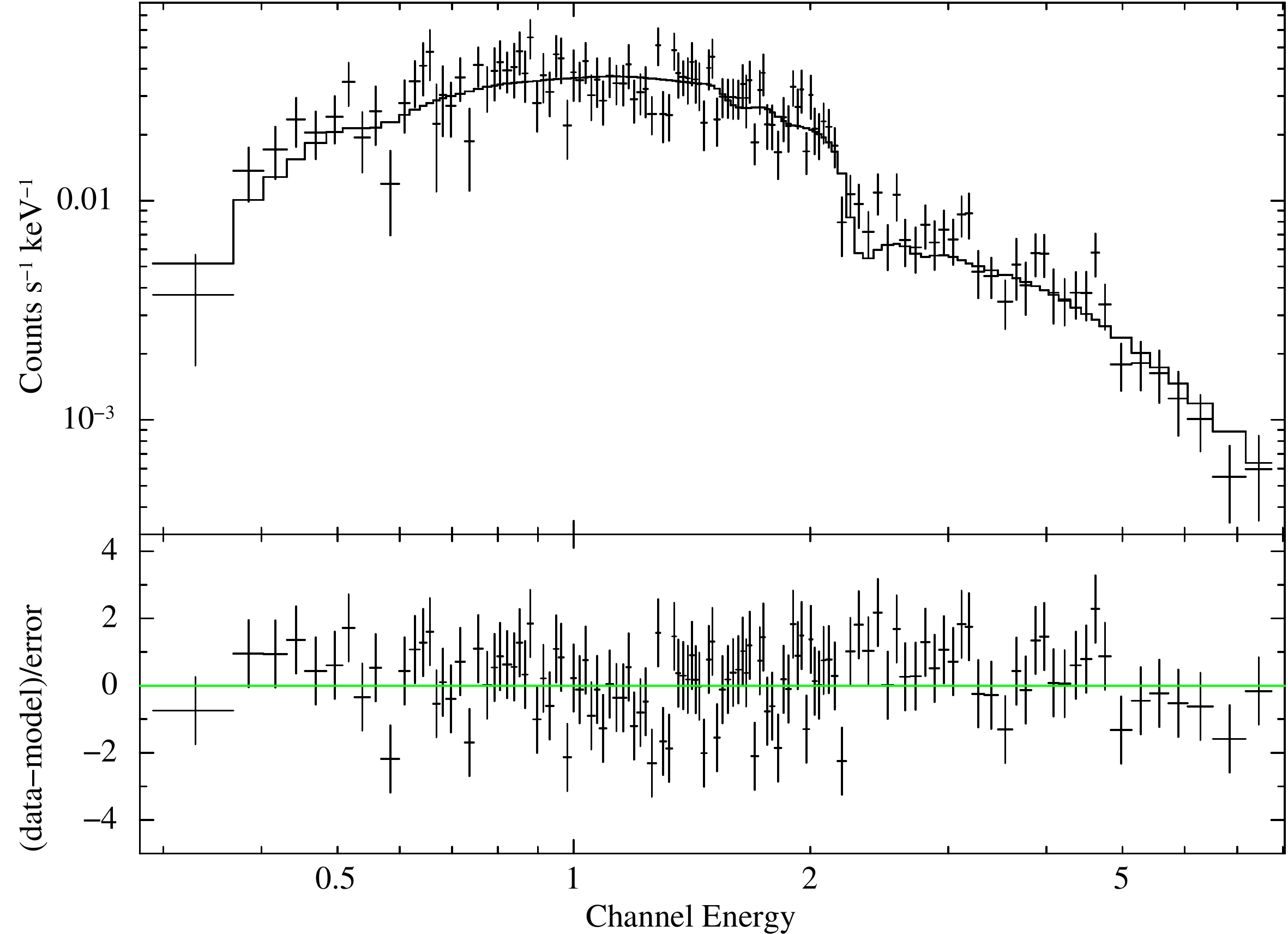}}
  \end{center}
  \caption{Top: Simultaneous spectral fit of an absorbed power law to the spectra of \sxp{726} 
           using observations 700001--5 (marked in black, red, green, blue, and cyan respectively).
           Bottom: The spectrum of \sxp{726} from observation 710000.
           All spectra are obtained by combining data from TM1--4 and 6.
           The histograms trace the best-fit model and the residuals are plotted in the narrow  panels.}
  \label{figsxp726spec}
\end{figure}

Considering the large spin-period deviation of \sxp{726} inferred from \ero data with respect to published values, we re-visited all available archival \xmm observations of the SMC covering the pulsar to investigate its evolution in more detail. In particular, we found 31 observations with counting statistics sufficient to search for periodicities. We re-processed the data from those observations using XMMSAS v.19 applying standard screening criteria described in the  documentation\footnote{\url{https://www.cosmos.esa.int/web/xmm-newton/sas-threads}}, and correcting photon arrival times to the solar system barycenter assuming the position reported in \cite{2008A&A...485..807E}. After that, source photons in the 0.2--10\,keV energy band from a source-centered region with radius of 30\arcsec\ were extracted for all EPIC cameras, and periodicity searched in the 700-900\,s range, using $Z^2_2$ statistics \citep{1983A&A...128..245B}. The uncertainties for the strongest peak were estimated as width of the peak at a level corresponding to 1$\sigma$ uncertainty which was estimated assuming that $Z^2_2$ statistics follows the $\chi^2$ distribution for 4 dof  \citep{1983A&A...128..245B}. 
 
The results, along with the values measured by \ero and values reported by \cite{2008A&A...485..807E} for some of the observations also reported here are presented in Fig.\,\ref{fig:sxp726xmmpspin}. It is evident that the \ero measurements follow a general spin-down trend. To estimate the spin-down rate we performed a linear fit to all data points obtained in this work (i.e. including \xmm points) which resulted in  $\dot{P}=0.01178(6)$\,s\,day$^{-1}$ (4.3\,s\,yr$^{-1}$), i.e. highly significant and slightly higher in value than reported by \cite{2008A&A...485..807E}. 
 
To estimate the source flux observed by \ero, we fitted all Nov. 2019 observations simultaneously with a simple absorbed power-law model. The absorption component was consistent with the Galactic foreground absorption and no additional absorbing material was required. The foreground column density was fixed at 6\hcm{20}, the power-law normalization was left free. The inferred power-law index was $0.63\pm0.13$ and the average observed flux (0.2--10\,keV) \red{1.14$\pm$0.21} \ergcm{-13}. This corresponds to a luminosity of 5.2\ergs{34}. The spectra with the best-fit model are shown in Fig.\,\ref{figsxp726spec}.

During observation 710000 \sxp{726} was sufficiently bright to analyse the spectrum individually.
The power-law index was $0.70\pm0.06$, consistent within errors with the value obtained from the Nov. 2019 observations.
The observed flux was 7.5$^{+0.61}_{-0.56}$ \ergcm{-13} and the luminosity 3.3\ergs{35}. 
 
\begin{figure}
     \centering
     \includegraphics[width=\columnwidth]{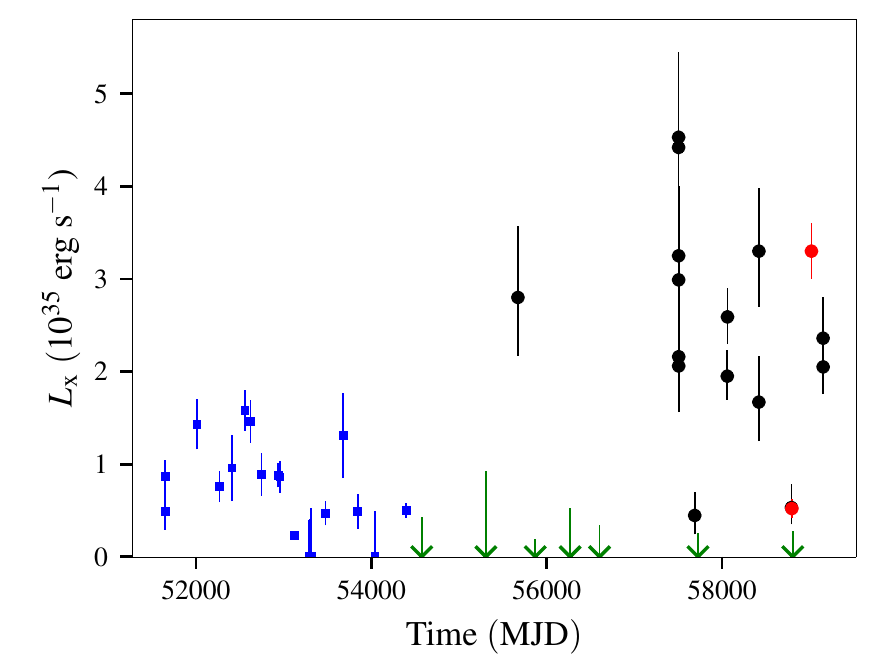}
     \caption{Variability of SXP726 as observed by \xmm and \ero (red).
     The blue points correspond to flux values reported by \cite{2008A&A...485..807E}, whereas black points are reported here for the first time. The upper limits for observations when the source was not detected are also indicated (green arrows).}
     \label{fig:sxp726xmmlc}
 \end{figure}

To put the \ero flux in context of the historical flux evolution of \sxp{726}, we also analyzed the available \xmm observations of the source. To estimate the source flux, we used the same procedure as for the \ero data, i.e. by fitting source spectra using the absorbed power-law model. For observations where the source was detected by multiple EPIC cameras, the spectra were modeled simultaneously by linking model parameters. The source flux in the  0.2--10\,keV band and its uncertainty were then estimated using the \texttt{cflux} component in \xspec. The results are presented in Fig.\,\ref{fig:sxp726xmmlc}. It is evident that the source is generally quite variable, and the \ero observations occurred at flux levels typically observed from \sxp{726} during the recent years. 

\label{sec:726}

%%%%%%%%%%%%%%%%%%%%%%%%%%%%%%%%%%%%%%%%%%%%%%%%%%%%%%%%%%%%%%%%%%%%%%%%%%%%%%%%%%%

\subsection{ RX\,J0103.6$-$7201 = \sxp{1323} = \hscat{62}}
\label{sec:1323}

\sxp{1323} has been extensively monitored by various X-ray observatories due its close angular distance to supernova remnant \snre, which is used as calibration standard by many X-ray  observatories \citep{2017A&A...597A..35P}. 
X-ray pulsations were discovered by \citet{2005A&A...438..211H} in \xmm data. 
Their long-term light curve created from \rosat, \cxo and \xmm data from 1991--2004 was extended by \citet{2017A&A...602A..81C} until the end of 2016.
Their study of the \suzaku, \cxo and \xmm data showed that the pulse period stayed relatively stable around 1323\,s from 2000 to 2006, but afterwards the source started to spin-up rapidly, reaching 1110\,s by mid 2016. Since then, the source continued to spin-up at a similar rate and the pulse period of 1005\,s, during the eROSITA CalPV phase, was already announced by \citet{2019ATel13312....1H}.

\begin{figure*}
  \centering
  \resizebox{0.9\hsize}{!}{\includegraphics[]{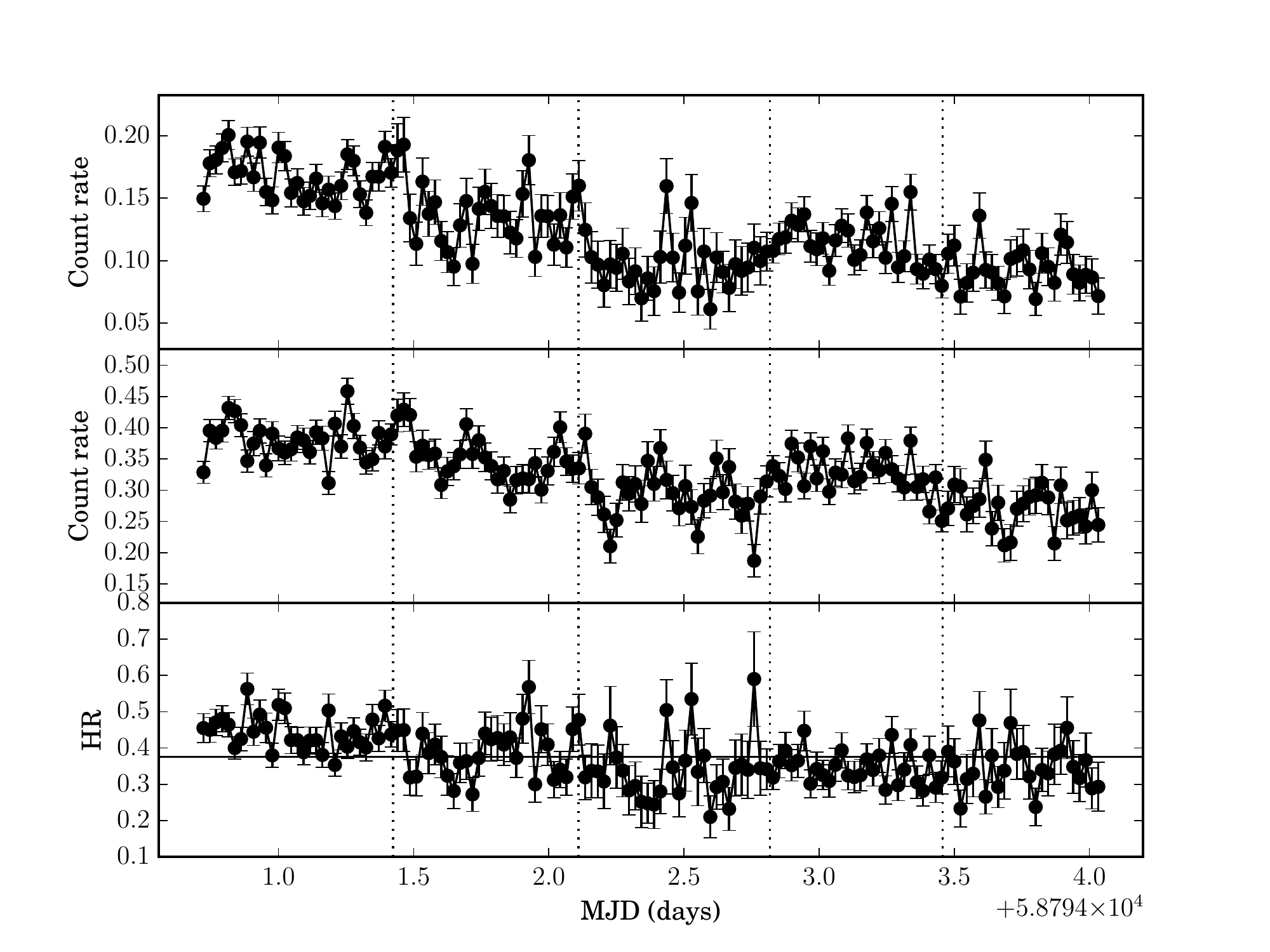}}
  \caption{Background subtracted and vignetting-corrected light curves of \sxp{1323} from observations 700001--5, binned at twice the pulse period in the energy range of 2.5--8\,keV (top) and 0.2--2.5\,keV (middle), Bottom: Hardness ratio (H/S) of \sxp{1323}.}
   \label{figperiodsxp1323lc}
\end{figure*}

\begin{figure*}
  \centering
  \resizebox{0.8\hsize}{!}{\includegraphics[angle=90]{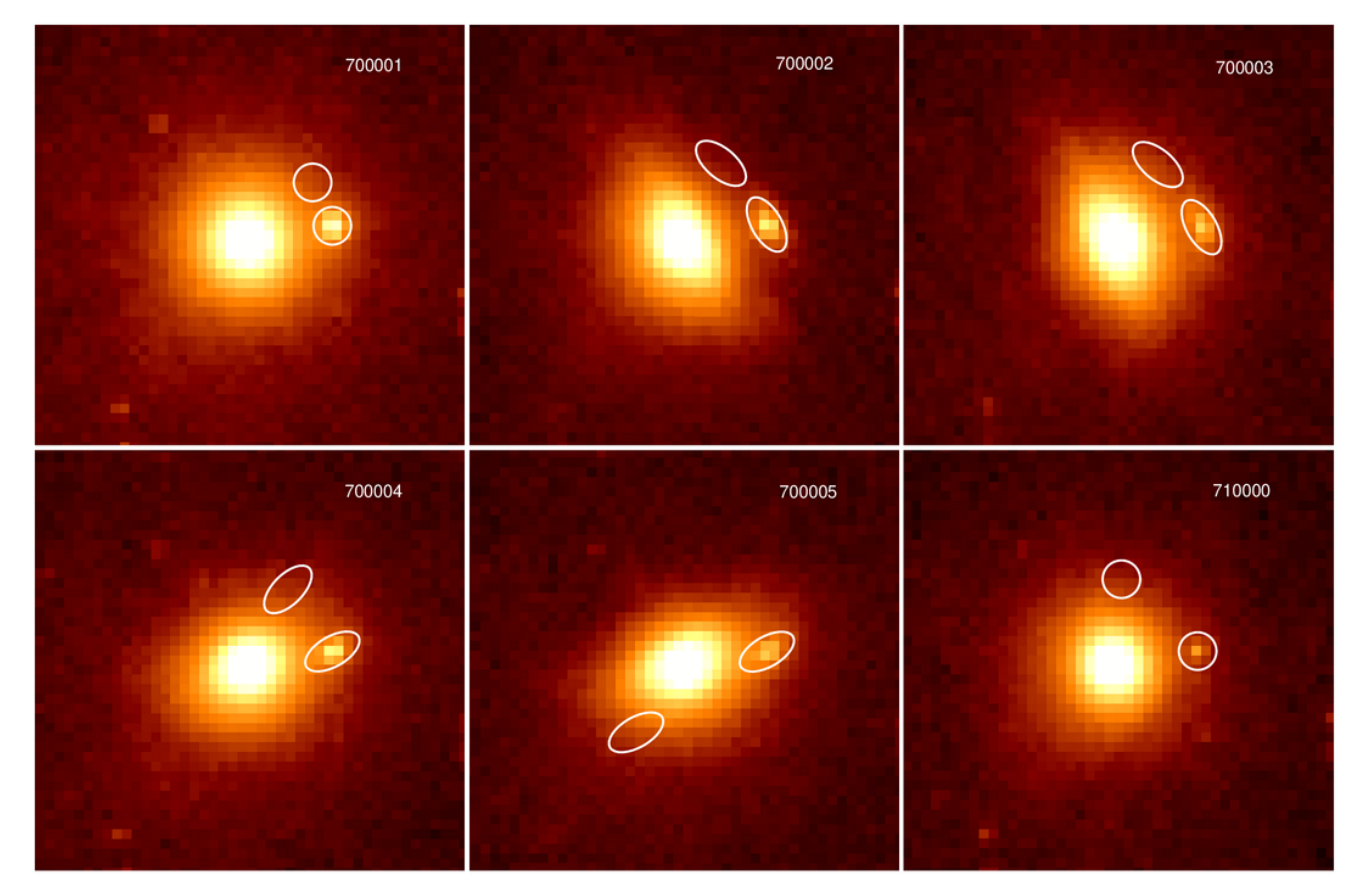}}
  
  \caption{Images of the region around \snre and \sxp{1323} from the six CalPV observations. The source and background extraction regions used for the temporal and spectral analyses of \sxp{1323} are marked.}
   \label{figsxp1323ima}
\end{figure*}

\sxp{1323} was detected in all the Nov. 2019 observations (700001--5) and the June 2020 observation (710000) performed by \ero with count rates in the range between 0.09\,\cts and 0.68\,\cts (Table\,\ref{tab:smc_rates}). 
The combined light curve from Nov. 2019 is shown in Fig.\,\ref{figperiodsxp1323lc} for two energy bands, a soft band (0.2--2\,keV) and a hard band (2.5--8\,keV). The hard band was chosen to avoid any possible residual contribution from the nearby supernova remnant (see Fig.\,\ref{figsxp1323ima}), which exhibits a soft X-ray spectrum.
Some variability by a factor of $\sim$3 is seen in the hard-band light curve of the pulsar, while in the soft band variations are reduced to a factor of $\sim$2.
The hardness ratio (ratio of count rates in the hard and soft bands) shows a slight decreasing (softening) trend during the 3.4 days, but caution is required due to the large PSF at the large off-axis angles during observations 700002--5 (see also below).

%%%%%%%%%%%%%%%%%%%%%%%%%%%
%%%%%%%%%%%%%%%%%%%%%%%%%%%

%%%%%%%%%%%%%%%%%%%%%%%%%%%
%%%%%%%%%%%%%%%%%%%%%%%%%%%

\begin{figure}
  \centering
  \resizebox{1.0\hsize}{!}{\includegraphics[]{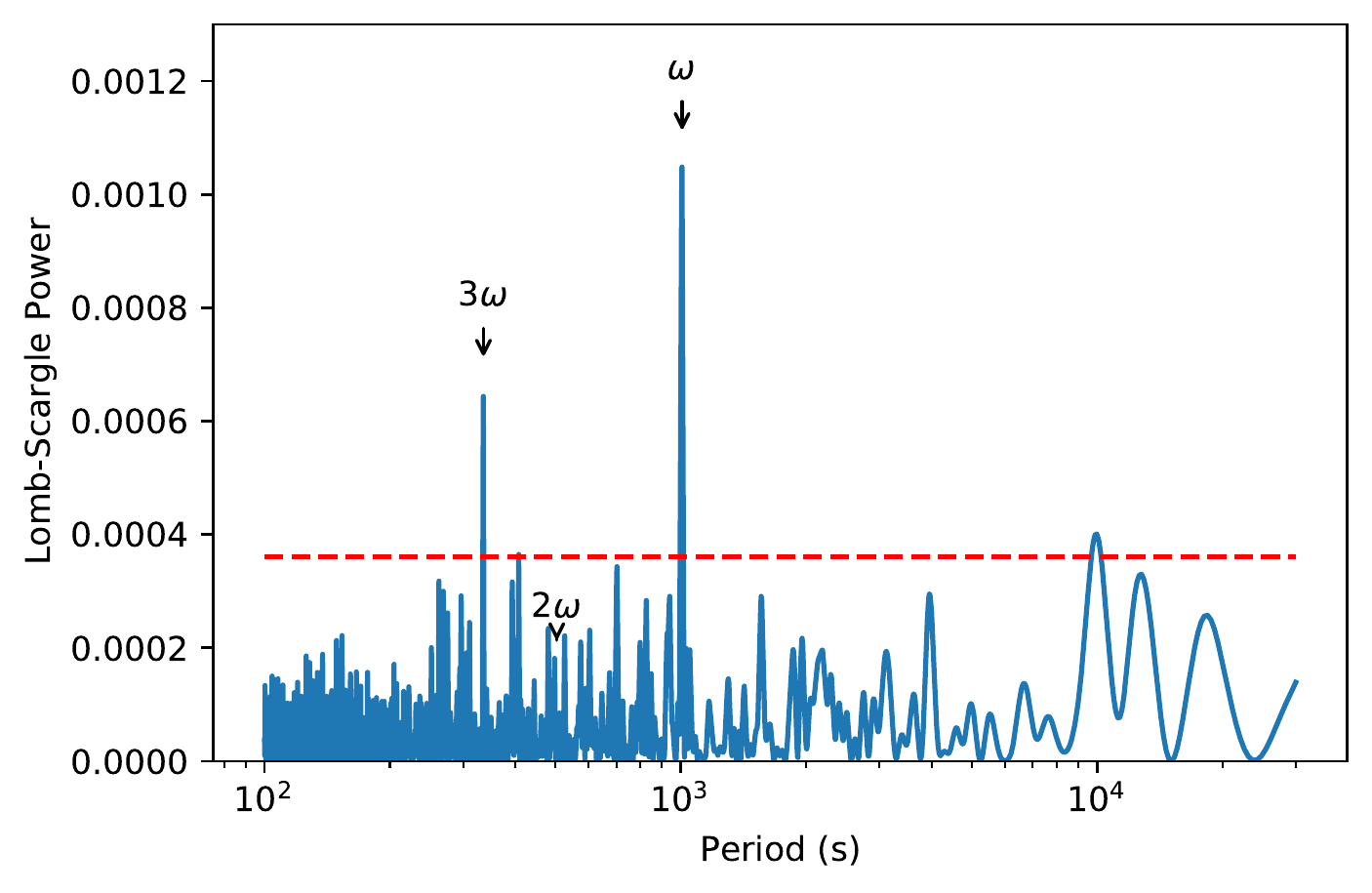}}
  \caption{LS periodogram obtained from the light curve extracted from Obsid 700001 of \sxp{1323}.
           The red dashed line indicates the 99\% confidence level.}
           
   \label{figperiodsxp1323ls}
\end{figure}

\begin{figure}
  \begin{center}
   \resizebox{0.8\hsize}{!}{\includegraphics[angle=-90]{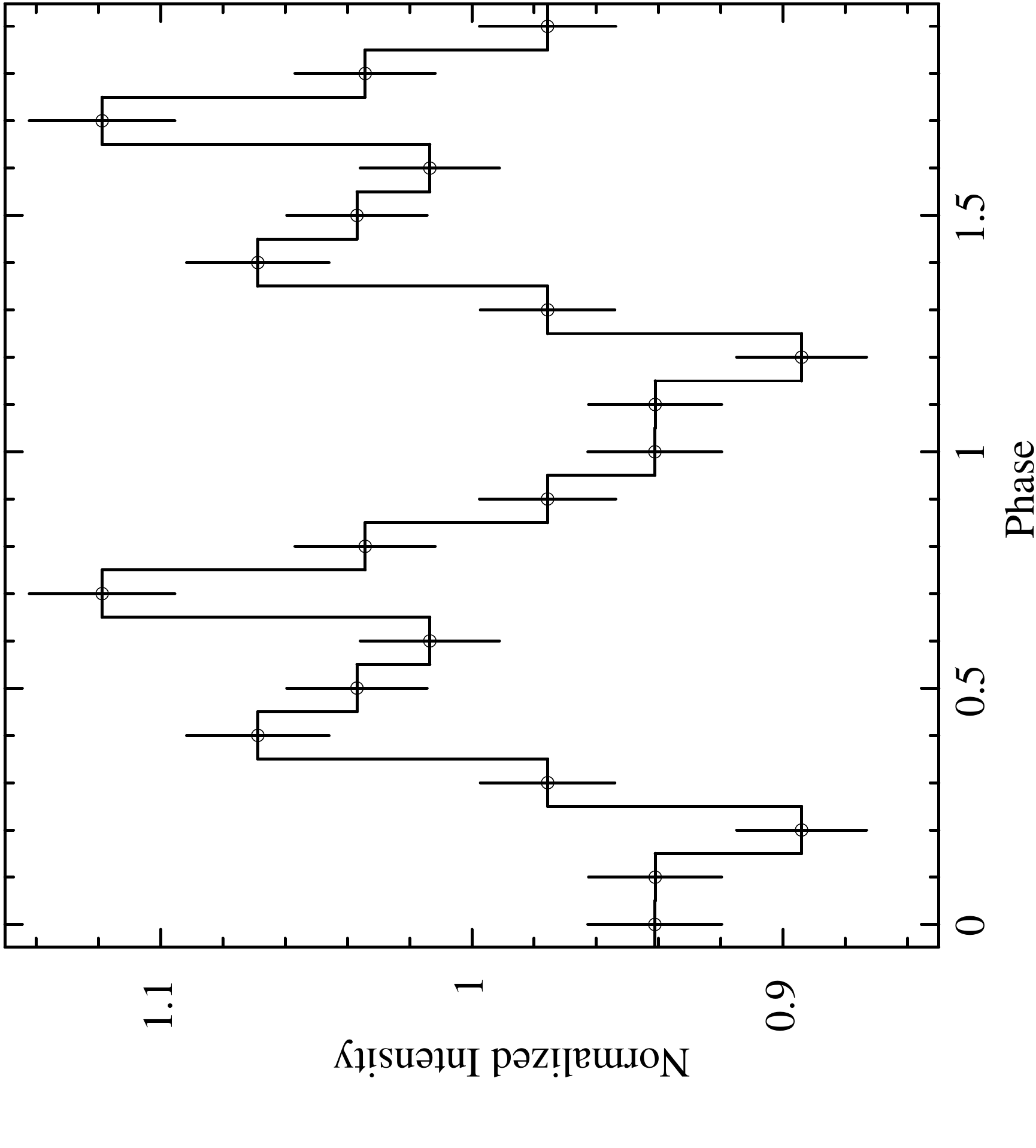}}
  \end{center}
  \caption{Background subtracted pulse profile of \sxp{1323} in the energy range of 0.2--10\,keV from 
                observation 700001 in which pulsations were detected.}
   \label{figsxp1323pulse}
\end{figure}

In the eROSITA data, pulsations are detected most  significantly during observation 700001 from 2019 November 7, when the source was brightest and close to the telescope optical axis.
Our timing analysis revealed the pulse period at 1006.3$\pm$5.6\,s.
The Lomb-Scargle periodogram is presented in Fig.\,\ref{figperiodsxp1323ls} and shows signal at the fundamental frequency and the second harmonic (3$\omega$), but not at the first harmonic (2$\omega$). This is different from periodograms obtained from \xmm observations in the first years of this century \citep{2005A&A...438..211H} when the fundamental and the first harmonic were seen. 
Correspondingly, the pulse profile from the \ero observation 700001 (presented in Fig.\,\ref{figsxp1323pulse}) shows two narrow peaks while the \xmm profiles were characterized by a structured broader peak.

%%%%%%%%%%%%%%%%%%%%%%%%%%%
%%%%%%%%%%%%%%%%%%%%%%%%%
%%%%%%%%%%%%%%%%%%%%%%%%%

\begin{figure}
\centering
\resizebox{0.9\hsize}{!}{\includegraphics{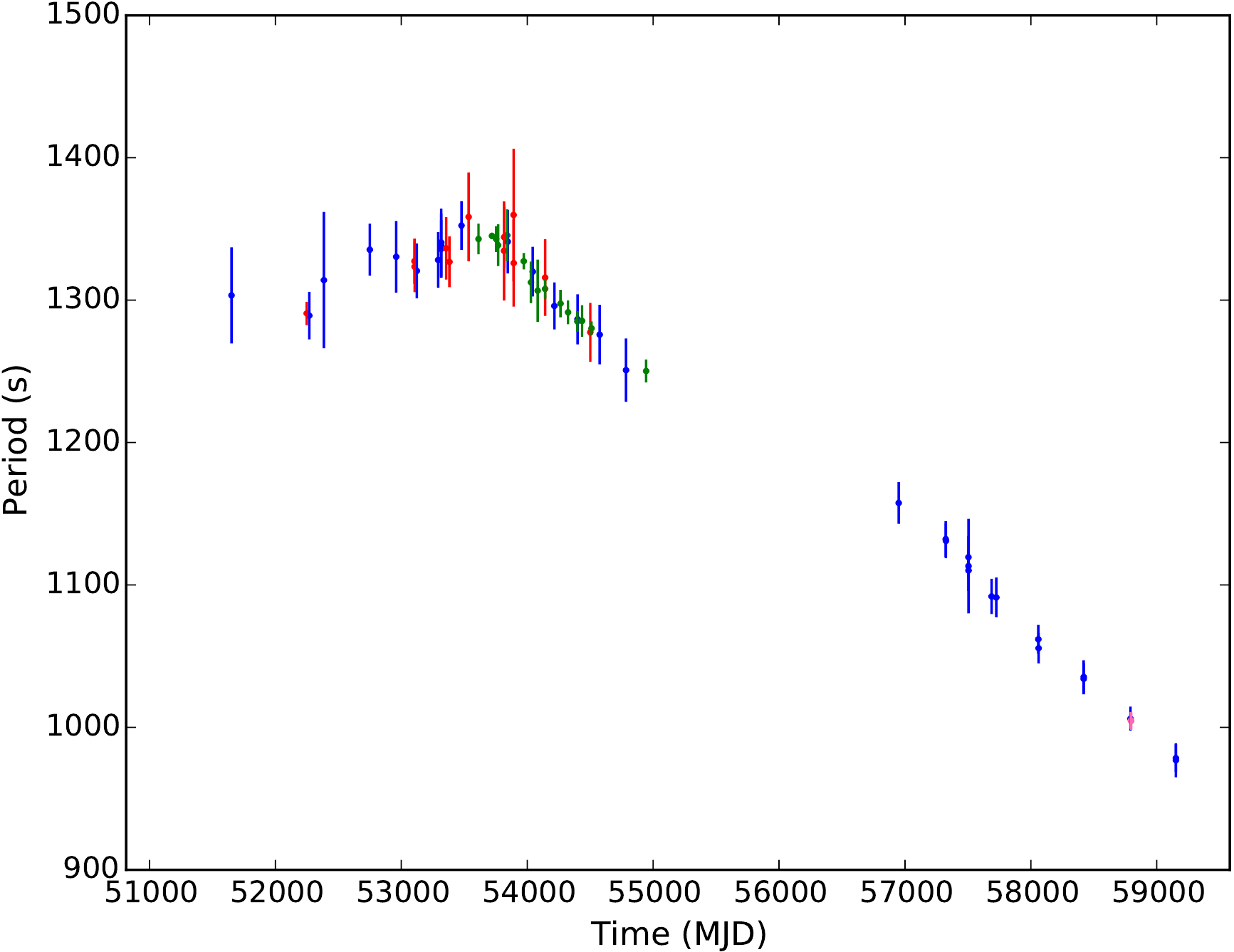}}
\caption{Long-term spin period evolution of \sxp{1323}, following the work of \cite{2017A&A...602A..81C}, which includes Suzaku (green), \xmm (blue), \cxo (red) and \ero (pink) observations.} 
\label{fig:longsxp1323}
\end{figure}

Figure\,\ref{fig:longsxp1323} shows the long-term evolution of the pulse period of \sxp{1323}, as in \citet{2017A&A...602A..81C}, but including the latest \xmm and eROSITA measurements. 
The ten new \xmm observations are spanning from 2016 October 26 (obsID: 0412983201) to 2020 October 30 (obsID: 0810880701), when the pulse period monotonically decreased from 1090.86\,s to 976.91\,s (note that pulsations are not detected significantly in all ten observations). 
Using the latest data, the spin up rate is now $\dot{P}=-23.2$\,s\,yr$^{-1}$, as inferred from a linear fit to the data in the 2005--2021 time interval. With a mean value of 1168\,s, the corresponding relative spin period change is $\lvert\dot{P}/P\lvert$=0.0199\,yr$^{-1}$.
Between 2010 and 2014 pulsations were only occasionally detected, if at all at a low confidence level \citep{2017A&A...602A..81C}. A comparison of the spin-up trend from the epochs before and after this break suggests that the spin-up rate was reduced during the break (Fig.\,\ref{fig:longsxp1323}). The first measured period after the break is about 50\,s longer than expected from the extrapolation of the data before the break. That also implies even higher spin-up rates before and after the break compared to the 16-year average.

%%%%%%%%%%%%%%%%%%%%%%%%%
%%%%%%%%%%%%%%%%%%%%%%%%%

Due to the vicinity of \sxp{1323} to \snre special care has to be taken for the definition of the extraction regions for source and background spectra. In particular for the off-axis observations of \snre the larger PSF causes a significant contribution of the supernova remnant flux to the spectra of \sxp{1323} as shown in Fig.\,\ref{figsxp1323ima}. 
The background regions were carefully selected to remove this emission, taking into account the distance of \sxp{1323} to \snre and also the asymmetric shape of the PSF at large off-axis angles.

\begin{figure}
  \begin{center}
  \resizebox{0.9\hsize}{!}{\includegraphics[angle=-90]{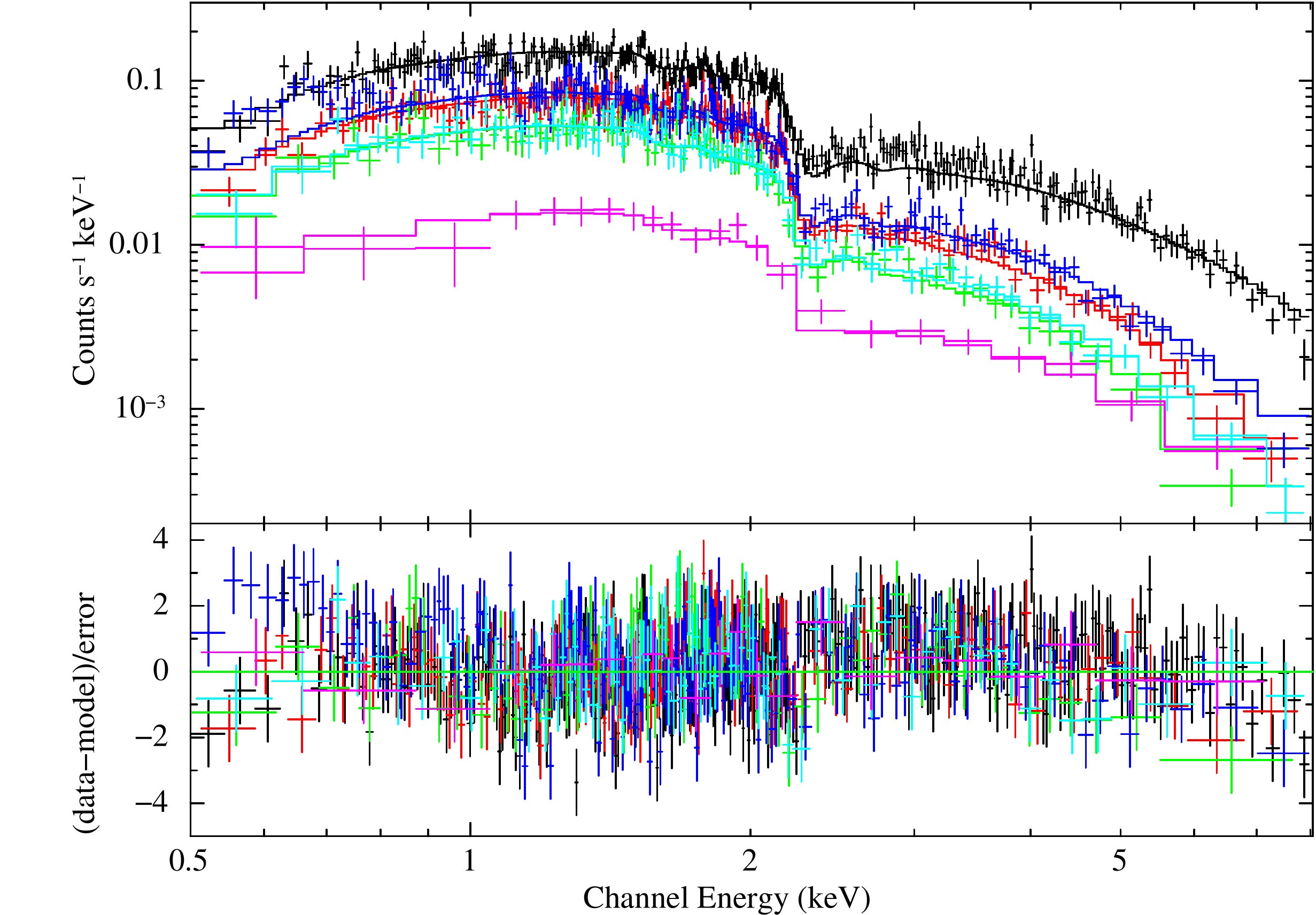}}
  \end{center}
  \caption{\ero spectra of \sxp{1323} (observations 700001--5 marked in black, red, green, blue and cyan, respectively, and 710000 in magenta) 
           together with the best-fit power-law model and the residuals from the simultaneous fit. 
           The spectrum from each observation is obtained by combining data from TM1--4 and 6.}
   \label{figsxp1323spec}
\end{figure}

The spectra extracted from observations 700001--5 and 710000 were fitted simultaneously. An absorbed power-law provided an adequate description to the spectrum. For this purpose we used two absorption components: one to describe the Galactic foreground absorption (fixed at 6\hcm{20}) and another to account for the column density of both the interstellar medium along the line of sight in the SMC and the local absorption near the source. For the latter absorption component, the abundances were set to 0.2 solar for elements heavier than helium as described before. In the fits only the power-law normalization was left free to vary between observations. The obtained power-law photon index is $0.50\pm0.02$; and \nh = $9.8\pm2.5$\hcm{20} \red{(reduced $\chi^2$ = 1.1 for 1581 dof).}
During the Nov. 2019 observations the observed flux (0.2--10\,keV) decreased from  4.5\ergcm{-12} (700001) to 3.1\ergcm{-12} (700005), as already indicated by the light curve (Fig.\,\ref{figperiodsxp1323lc}) and was even lower at 5.7\ergcm{-13} in June 2020.
This corresponds to luminosities of 2.0\ergs{36}, 1.3\ergs{36} and 2.5\ergs{35}. The result of the simultaneous spectral fit with the best-fit model is shown in Fig.\,\ref{figsxp1323spec}.

\section{Discussion}
\label{sec:discussion}

\subsection{HMXBs in the Magellanic Clouds as seen during \ero CalPV observations}

The \ero observations around SN\,1987A and N\,132D in the LMC covered 14 known (candidate) HMXBs. Six of them
were not detected in any of the observations and upper limits between  1.5\expo{-3} \cts and 1.1\expo{-2} \cts were derived. Assuming the spectral parameters derived for the new transient \srget (see below) these correspond to fluxes from 1.2\ergcm{-14} to 8.8\ergcm{-14}, or luminosities from 4.2\ergs{33} to 3.1\ergs{34} (0.2--8.0 keV).
All other sources (we exclude LMC\,X-1 from our analysis) were detected in at least one of the observations with fluxes (luminosities) in the range of 4.3\ergcm{-14} to  1.7\ergcm{-12} (1.5\ergs{34} to 5.9\ergs{35}). The brightest sources were the SFXT XMMU\,J053108.3$-$690923 \citep{2021A&A...647A...8M} and the supergiant HMXB candidate RX\,J0541.4$-$6936, the latter being undetected in one observation with an upper limit a factor of $\sim$30 below the two detections.

The SMC observations around \snre covered 23 HMXBs, most likely all BeXRBs. The SMC observations were more sensitive for point source detection than those of the LMC, resulting in a factor 2--3 smaller upper limits (5.7\expo{-4} \cts to  6.6\expo{-3} \cts, 4.5\ergcm{-15} to  5.3\ergcm{-14}, 2.2\ergs{33} to 2.6\ergs{34}). This is caused not only by the deeper exposures, but also due to the stronger diffuse X-ray emission in the LMC. Only one of the known SMC BeXRB pulsars was not detected, although in the FoV of all observations (\sxp{345}, [HS16]\,50). From the HMXBs without known spin period three were not detected ([HS16]\,114, 120 and 133). \citet{2016A&A...586A..81H} list [HS16]\,120 in their lowest confidence class for being a genuine HMXB. Nineteen (83\%) of the HMXBs covered by the \ero observations were detected, with fluxes (luminosities) in the range of 1.1\ergcm{-14} to  5.4\ergcm{-12} (5.2\ergs{33} to 2.6\ergs{36}), with \sxp{1323} the brightest in the SMC.
These moderate flux levels \citep{2016A&A...586A..81H} indicate that none of the sources was in a strong outburst during the \ero observations, as expected for long-period, low-luminosity BeXRBs.

One of the LMC calibration observations in Nov. 2019 revealed a new hard X-ray transient - \srget\ - with the typical characteristics of a BeXRB, but with an X-ray luminosity of 1.2\ergs{35} relatively faint.
The X-ray spectrum can be modeled with a power law with photon index of \red{0.66$^{+0.74}_{-0.54}$} and as optical counterpart we identified a V $\sim$ 15.7\,mag star,
which shows large brightness variations in the V and I bands as monitored by the OGLE project over 18.5 years (Figs.\,\ref{fig:oglelci} to \ref{fig:oglevi}). 
The star varies in colour as indicated by the V-I index, which is correlated with the brightness of the star. \red{There is some scatter in V-I for a given I, which is limited to a well-defined band as shown in Fig.\,\ref{fig:oglevi}.}
This behaviour of reddening with brightness was reported from several other BeXRBs in the Magellanic Clouds with large variations in brightness \citep{2021MNRAS.503.6187T,2017A&A...598A..69H,2014A&A...567A.129V,2012MNRAS.424..282C} and is interpreted as changing size of the 
circum-stellar disc, which leads to variations in the infrared excess with respect to the emission from the stellar photosphere. 

The \ero CalPV observations also covered three HMXB candidates in the LMC, which were selected based on X-ray hardness ratios and early-type star as likely counterpart, but did not show significant \Halpha emission during the spectroscopic observations performed by \citet{2018MNRAS.475.3253V}. XMMU\,J052417.1$-$692533, one of the candidates was detected in three \ero observations and the spectra can be well represented by an absorbed power law with photon index 0.83$\pm$0.17, consistent with the average value found for BeXRBs in the SMC by \citet{2008A&A...489..327H}. 
This supports the proposed HMXB nature of the source, although no \Halpha emission was observed. Simultaneous X-ray and optical observations are required to verify if X-ray luminosities of $\sim$\oergs{34} can be powered when no significant circumstellar disc is present.

In the SMC the \ero observations covered 16 BeXRB pulsars, including XMMU\,J010429.4$-$723136, from which 164\,s pulsations were discovered in \ero data \citep{2021arXiv210614536C,2019ATel13312....1H}. 
We present detailed spectral and temporal analyses for \sxp{304}, \sxp{455}, \sxp{522}, \sxp{726} and \sxp{1323}.
All five belong to a group of long-period ($\ga$100\,s) and low-luminosity ($\la$5\ergs{36}) pulsars, 
which {\red{usually do not exhibit very strong outbursts with amplitudes F$_{\rm max}$/F$_{\rm min}$ larger than $\sim$500, unlike short-period BeXRB pulsars which can reach factors larger than 10$^4$} \citep[see Fig.\,5 in][]{2016A&A...586A..81H}. 

The current pulse periods measured for \sxp{304} and \sxp{455} are within the range published previously, indicating little long-term spin changes over the last one to two decades. 
Their X-ray luminosities of $\sim$1--5\ergs{35} are also within the typical range observed in the past \citep{2008A&A...491..841E}.

\sxp{455} showed the largest flare-like flux variations of all investigated BeXRBs covered by the long Nov. 2019 observations (Fig.\,\ref{figperiodsxp455}). In particular during observation 700002 a strong flare occurred, which increased by a factor of five in flux and lasted for about 10\,hours. During the flare the pulse profile changed from double-peaked to more single-peaked (Fig.\,\ref{figsxp455pp}).
During previous \xmm observations in Apr. and Oct. 2000, the source was a factor of more than 2 fainter than in 2019 and the pulse profile showed indications for two peaks, in particular at energies below 2\,keV \citep{2003A&A...403..901S,2001A&A...369L..29S}. This indicates a dependence of the pulse profile on source luminosity, however, it should be noted that during the \ero observation the weaker peak followed the stronger one, while during the \xmm observations this order was reversed.  
A similar flaring activity and a luminosity dependence of the pulse profile was reported from the Galactic transient 2.76\,s X-ray pulsar 4U\,1901+03 \citep{2020MNRAS.493.5680J}, which the authors interpret as changes of the pulsed beaming pattern due to transitions between the sub- and super-critical accretion regimes.
However, there are some differences between the flares of \sxp{455} and 4U\,1901+03: The flares of the SMC pulsar are stronger with respect to the persistent emission and last longer (10 hr vs. a few hundred seconds). It is remarkable that the ratio of the flare durations is similar to the ratio of the spin periods of the two neutron stars.

The newly measured period of \sxp{522} indicates a small spin-up of $\sim$-6$\pm$2\,s over the last 7 years.
However, the luminosity seen by \ero in 2019 is a factor of 40--100 lower than the maximum in 2012, suggesting that the source has returned to a low state after some more active period when the pulsations were discovered.

\subsection{The long-term spin-period evolution of \sxp{726} and \sxp{1323}}

The spin history of \sxp{726} and \sxp{1323} is very different from the other three pulsars. 
While \sxp{726} shows a nearly linear spin-down rate of 4.3\,s\,yr$^{-1}$ since about 17\,years, 
\sxp{1323} is spinning up on average with -23.2\,s\,yr$^{-1}$ since $\sim$16\,years. 

The low X-ray luminosity of X-ray pulsars in the SMC ($\sim$\oexpo{35}--\oergs{36}) and long pulse periods (exceeding  $\sim$70--100\,s) suggest that they may be quasi-spherically accreting from stellar winds. In this case, the plasma entry in the neutron star magnetospheres is regulated by the plasma cooling rate \citep{1977ApJ...215..897E}. At X-ray luminosities below $\sim$4\ergs{36}, the Compton cooling of plasma above the magnetosphere is insufficient to keep pace with freely-falling fresh matter gravitationally captured from the stellar wind, and a quasi-spherical subsonic, turbulent shell is formed around the magnetosphere \citep{2012MNRAS.420..216S}. 
The accretion through the shell occurs subsonically (the settling accretion), no shock arises above the neutron star magnetosphere, and the accretion rate onto the neutron star $\dot M$ 
is smaller than the potential mass accretion rate determined by the standard Bondi-Hoyle-Littleton value.
Turbulent stresses mediate the angular momentum exchange between the shell and the magnetosphere leading to the spin-up/spin-down of the neutron star. The angular momentum balance 
in this case is written as \citep{2012MNRAS.420..216S,2017arXiv170203393S,2018ASSL..454..331S}
\begin{equation}
\label{e:torques}
\dot \omega=A(\mu,v,P_\mathrm{b})\dot M^{7/11}-B(\mu,P)\dot M^{3/11}\,.
\end{equation}  
Here $\omega=2\pi/P$, the coefficients $A$ and $B$ depend on the system parameters (the neutron star's \red{magnetic moment $\mu$}\footnote{\red{The neutron star dipole magnetic moment is defined as $\mu=BR^3/2$, where $R$ and $B$  are the neutron star surface polar field and radius. For the typical values of $B=10^{12}$~G and $R=10$~km, $\mu=0.5\times 10^{30}$~G cm$^3$.}} and moment of inertia, the stellar wind velocity $v$, the binary orbital period $P_\mathrm{b}$, and details of the shell's structure; see below).

The equilibrium pulsar period reads: 
\begin{equation}
\label{e:Peq}
    P_\mathrm{eq}\approx 1000[\mathrm{s}]~\mu_{30}^{12/11}
    \left(\frac{P_\mathrm{b}}{10\mathrm{\,d}}\right)\dot M_{16}^{-4/11}v_8^4\,.
\end{equation}
Here the parameters are normalized to the typical values; $\mu_{30}\equiv \mu$ / (10$^{30}$\,G\,cm$^3$), 
$P_\mathrm{b}$ is the binary orbital period (the orbit is assumed to be circular), 
$\dot M_{16}\equiv \dot M$ / (10$^{16}$\,g\,s$^{-1}$) is the accretion rate onto the neutron star surface corresponding to an X-ray luminosity of $L_x=0.1\dot Mc^2$, and 
$v_8\equiv v$ / (1000\,km\,s$^{-1}$) is the stellar wind velocity relative to the neutron star. 

For accretion rates below the equilibrium value, $\dot M< \dot M_{eq}=(B/A)^{11/4}$, 
the spin-up torque can be neglected.
The  spin-down torque $\dot \omega_\mathrm{sd}$ \citep{2012MNRAS.420..216S,2017arXiv170203393S,2018ASSL..454..331S} reads  \citep[see Eq.\,6.9 in][]{2017arXiv170203393S}:
\begin{equation}
    \label{e:odotsd}
    \dot \omega_\mathrm{sd}\approx -0.54\times 10^{-12} [\mathrm{rad\,s^{-2}}]~ \Pi_0~\mu_{30}^{13/11}\dot M_{16}^{3/11} \left(\frac{P}{100\mathrm{\,s}}\right)\,,
\end{equation}
where the dimensionless combination of the theory parameters $\Pi_0\sim 1$  \citep[see][for more detail]{2018ASSL..454..331S}.
This torque can be recast to the fractional period change rate as
\begin{equation}
    \label{e:PdotPsd}
    \left(\frac{\dot P}{P}\right)_\mathrm{sd}=-\frac{\dot\omega_\mathrm{sd}}{2\pi}P\approx 8.6\times 10^{-12} ~\Pi_0~\mu_{30}^{13/11}\dot M_{16}^{3/11}\,.
\end{equation}
Therefore, with known X-ray luminosity, formula
(\ref{e:PdotPsd}) can be used to estimate the neutron star's magnetic \red{moment}:
\begin{equation}
    \label{e:mu}
    \mu_{30}\approx 0.16 \left[\frac{(\dot P/P)_\mathrm{sd}}{10^{-12}\mathrm{s^{-1}}}\right]^{11/13}\Pi_0^{-11/13} \dot M_{16}^{-3/13}\,.
\end{equation}
In the case of \sxp{726}, we have $P\approx 800$\,s, $L_x\approx (3-5)$\ergs{34}, corresponding to $\dot M_{16}\approx (3-5)\times 10^{-2}$, and $(\dot P/P)_\mathrm{sd}\approx 1.7\times 10^{-10}$ s$^{-1}$, which yields the neutron star's magnetic \red{moment} estimate $\mu_{30}\simeq 2.7$ (\red{corresponding to the surface polar magnetic field $B\simeq 5.4\times 10^{12}$~G}), independent of unknown binary orbital period and stellar wind velocity.

Now turn to \sxp{1323} with a period of $P\approx 1000$\,s for which an almost steady spin-up is measured at a rate of $(\dot P/P)_\mathrm{su}\approx 0.0199\,\mathrm{yr}^{-1}=6.31\times 10^{-10}\,\mathrm{s}^{-1}$ at an X-ray luminosity of $\sim$\oergs{36}. The spin-up torque $\dot \omega_\mathrm{su}$ at the quasi-spherical settling accretion stage reads \citep[see Eq.\,4.15 in][]{2017arXiv170203393S}:
\begin{equation}
    \label{e:odotsu}
    \dot \omega_\mathrm{su}\approx 5.3\times 10^{-13}
    [\mathrm{rad\,s^{-2}}] \,\left(\frac{\Pi_0}{\zeta^{4/11}}\right)\,\mu_{30}^{1/11}\,\left(\frac{P_\mathrm{b}}{10\mathrm{\,d}}\right)^{-1}\dot M_{16}^{7/11}v_8^{-4}
\end{equation}
corresponding to a fractional period change rate of
\begin{eqnarray}
    \label{e:PdotPsu}
    &\left(\frac{\dot P}{P}\right)_\mathrm{su}=-\frac{\dot\omega_\mathrm{su}}{2\pi}P\approx \nonumber \\
    &-8.44\times 10^{-12} 
    \left(\frac{\Pi_0}{\zeta^{4/11}}\right)
    \mu_{30}^{1/11}
    \left(\frac{P/100\mathrm{\,s}}
    {P_\mathrm{b}/10\mathrm{\,d}}\right)\dot M_{16}^{7/11}v_8^{-4}\,.
\end{eqnarray}
Clearly, the strongest dependence in Eq.\,\ref{e:PdotPsu} is on the stellar wind velocity, which enables us to estimate it from measurements of the fractional period spin-up rate and X-ray luminosity:
\begin{equation}
    \label{e:v8}
    v_8\simeq 1.7
    \left(\frac{10^{-12}\mathrm{s^{-1}}}{-(\dot P/P)_\mathrm{su}}\right)^{1/4}
    \left(\frac{P/100\mathrm{\,s}}
    {P_\mathrm{b}/10\mathrm{\,d}}\right)^{1/4}
    \dot M_{16}^{7/44}\mu_{30}^{1/44}\left(\frac{\Pi_0}{\zeta^{4/11}}\right)^{1/4}
\end{equation}
Substituting parameters for \sxp{1323} ($P\approx 1000$~s, $\dot M_{16}\simeq 1$, $P_\mathrm{b}=26.2$~d \citep{2017A&A...602A..81C}) and ignoring dimensionless theory parameters $\Pi_0, \zeta$ of order of unity, we find
\begin{equation}
\label{e:mu1323}
    v_8\simeq 0.5 
    \mu_{30}^{1/44}\,.
\end{equation}
Note the virtual independence of this estimate on the unknown magnetic field of the neutron star. 
The long-term spin evolution of \sxp{1323} presented in Fig. \ref{fig:longsxp1323} suggests a torque reversal at around MJD\,53500. Therefore, the equilibrium period of this pulsar could be close to $\sim$1000\,s. Then equation (\ref{e:Peq}) with the velocity estimate (\ref{e:mu1323}) may suggest a magnetic \red{moment} of $\mu_{30}\sim 10$ for this pulsar.

Thus, the quasi-spherical subsonic accretion theory \citep{2012MNRAS.420..216S} applied to observations of the spin-up and spin-down of the slowly rotating X-ray pulsars \sxp{1323} and \sxp{726} gives reasonable (standard) estimates for the magnetic field of the neutron star and the stellar wind velocity in these systems, without making additional assumptions on the spin-up/spin-down mechanisms or involving extraordinary values of the system parameters. Further observations of such BeXRB pulsars, especially to measure their orbital periods, neutron star magnetic fields and stellar wind velocities can be used to test this theory in more detail.

\section{Conclusions}

The \ero pointed observations of the CalPV phase of fields in the LMC and SMC demonstrate the strength of the telescopes for the investigation of the HMXB population in the Magellanic Clouds, namely high sensitivity especially at energies below $\sim$2\,keV and wide field of view. 
This also holds for the \ero all-sky survey which is currently in progress and covers the whole Magellanic System; LMC, SMC and the Magellanic Bridge. 
Due to the location of the LMC in the vicinity of the South Ecliptic Pole (SEP), where the survey scans cross, sources there are scanned for a much longer time than most of the sky,
e.g. three weeks of scans in the northern part of the LMC per survey visit lead to a twenty times higher exposure and accumulates to $\sim$40\,ks over 4 years. This is similar to the exposures of the CalPV observations and opens discovery space for the faint end of the HMXB luminosity function.
The various time scales involved in the survey, from $\sim$40\,s per scan, six scans per day over many days in the LMC (depending on the distance to the SEP, two days in the SMC), to the eight half-yearly visits of the survey will allow to discover new transient HMXBs and to investigate their duty cycles.

\begin{acknowledgements}
This work is based on data from \ero, the soft X-ray instrument aboard \srg, a joint Russian-German science mission supported by the Russian Space Agency (Roskosmos), in the interests of the Russian Academy of Sciences represented by its Space Research Institute (IKI), and the Deutsches Zentrum f{\"u}r Luft- und Raumfahrt (DLR). The \srg spacecraft was built by Lavochkin Association (NPOL) and its subcontractors, and is operated by NPOL with support from the Max Planck Institute for Extraterrestrial Physics (MPE).
The development and construction of the \ero X-ray instrument was led by MPE, with contributions from the Dr. Karl Remeis Observatory Bamberg \& ECAP (FAU Erlangen-N{\"u}rnberg), the University of Hamburg Observatory, the Leibniz Institute for Astrophysics Potsdam (AIP), and the Institute for Astronomy and Astrophysics of the University of T{\"u}bingen, with the support of DLR and the Max Planck Society. The Argelander Institute for Astronomy of the University of Bonn and the Ludwig Maximilians Universit{\"a}t Munich also participated in the science preparation for \ero.
The \ero data shown here were processed using the \eSASS software system developed by the German \ero consortium. 
The OGLE project has received funding from the National Science Centre, Poland, grant MAESTRO 2014/14/A/ST9/00121 to AU.
NSh is partially supported by RSF grant 21-12-00141. KP is partially supported by the RFBR grant 19-02-00790. 
NSh and KP also acknowledge the support from the Scientific and Educational School of M.V. Lomonosov Moscow State University
'Fundamental and applied space research'.

\end{acknowledgements}

\bibliographystyle{aa} % style aa.bst
\bibliography{references,general} % your references Yourfile.bib

\begin{appendix}
\section{Electronic chopper}
To reduce telemetry load, e.g. when observing a very bright source, part of the read-out frames recorded by the \ero cameras can be discarded from transmission. For this an electronic chopper is set to a value of n, which causes only every nth frame to be processed. This results in a reduction of the net exposure time by a factor of n, which needs to be considered by the ground software. In the \eSASS pipeline version which processed the CalPV data which were publicly released\footnote{\url{https://erosita.mpe.mpg.de/edr}}, the correction for the chopper setting was applied twice, once to the good time interval extension and once to the deadtime correction extension in the event files. From the CalPV data used here three observations in the LMC (Table\,\ref{tabobs}) were affected by a chopper setting of 2 (part of the time, part of the cameras). As a workaround, we multiplied the deadtime values by 2 for the corresponding time intervals to finally obtain correct exposure times, which are used to compute count rates and fluxes. The problem has been fixed in the \eSASS software, and after a future re-processing of the data, the event files will have the correct extensions.
\end{appendix}

\end{document}